\documentclass[10pt, journal, letterpaper]{IEEEtran}

\usepackage{configuration}

\begin{document}
\bstctlcite{IEEEexample:BSTcontrol}

\title{Managing O-RAN Networks: xApp\\Development from Zero to Hero}

\newtoggle{comsoc}
\toggletrue{comsoc}

\author{
  \nottoggle{comsoc}{
    \IEEEauthorblockN{
      Joao F. Santos\IEEEauthorrefmark{1},
      Alexandre Huff\IEEEauthorrefmark{2},
      Daniel Campos\IEEEauthorrefmark{3},
      Kleber V. Cardoso\IEEEauthorrefmark{3},
      Cristiano B. Both\IEEEauthorrefmark{4},\,and
      Luiz A. DaSilva\IEEEauthorrefmark{1}
    }
    \IEEEauthorblockA{
      \IEEEauthorrefmark{1}Commonwealth Cyber Initiative -- Virginia Tech, USA,
      \{joaosantos,ldasilva\}@vt.edu}\\
    \IEEEauthorblockA{
      \IEEEauthorrefmark{2}Universidade Tecnológica Federal do Paraná, Brazil,
      alexandrehuff@utfpr.edu.br}\\
    \IEEEauthorblockA{
      \IEEEauthorrefmark{3}Universidade Federal de Goiás, Brazil,
      \{dante_campos,kleber\}@ufg.br}\\
    \IEEEauthorblockA{
      \IEEEauthorrefmark{4}Universidade do Vale do Rio dos Sinos, Brazil,
       cbboth@unisinos.br}
  }{
    Joao~F.~Santos\orcidlink{0000-0001-6439-2056}, \textit{Member, IEEE},
    Alexandre~Huff\orcidlink{0000-0003-0371-4837},
    Daniel~Campos\orcidlink{0009-0008-8900-1654},
    Kleber~V.~Cardoso\orcidlink{0000-0001-5152-5323},
    Cristiano~B.~Both\orcidlink{0000-0002-9776-4888},
    and~Luiz~A.~DaSilva\orcidlink{0000-0001-6310-6150},~\textit{Fellow,~IEEE}
    \IEEEcompsocitemizethanks{
      \IEEEcompsocthanksitem
      Manuscript received August 4th, 2024; revised December 19th, 2024;
      accepted February 4th, 2025.
      The research leading to this paper received support from the Commonwealth
      Cyber Initiative, an investment in the advancement of cyber R\&D, innovation,
      and workforce development.
      This work also received support from the OpenRAN@Brasil Program.
      \IEEEcompsocthanksitem Joao F. Santos and Luiz A. DaSilva
      are with the Commonwealth Cyber Initiative (CCI) and Virginia Tech, USA.
      Emails: \{joaosantos, ldasilva\}@vt.edu
      \IEEEcompsocthanksitem Alexandre Huff
      is with the Universidade Tecnológica Federal do Paraná, Brazil.
      Email: alexandrehuff@utfpr.edu.br
      \IEEEcompsocthanksitem Daniel Campos and Kleber V. Cardoso
      are with the Universidade Federal de Goiás, Brazil.
      Emails: dante_campos@discente.ufg.br, kleber@ufg.br
      \IEEEcompsocthanksitem Cristiano B. Both
      is with the Universidade do Vale do Rio dos Sinos, Brazil,
      Email: cbboth@unisinos.br
    }
  }
}

\maketitle

\begin{abstract}

The \ac{O-RAN} Alliance proposes an open architecture that disaggregates the
\ac{RAN} and supports executing custom control logic in near-real
time from third-party applications, the xApps.
Despite \ac{O-RAN}'s efforts, the creation of xApps remains a complex and
time-consuming endeavor, aggravated by the sometimes fragmented, outdated, or deprecated
documentation from the \ac{OSC}. These challenges hinder academia and industry
from developing and validating solutions and algorithms on O-RAN
networks. This tutorial addresses this gap by providing the first comprehensive
guide for developing xApps to manage the O-RAN ecosystem from theory to practice.
We provide a thorough theoretical foundation of the \ac{O-RAN} architecture and
detail the functionality offered by \ac{nearrtric} components. We
examine the xApp design and configuration. We
explore the xApp lifecycle and demonstrate how to deploy and manage xApps on a
\ac{nearrtric}. We address the xApps' interfaces and capabilities,
accompanied by practical examples. We provide comprehensive details
on how xApps can control the \ac{RAN}. We discuss debugging strategies and good
practices to aid the xApp developers in testing their xApps.
Finally, we review the current landscape and open challenges for creating xApps.

\end{abstract}


\IEEEpeerreviewmaketitle

\acresetall

\iftrue
\fancypagestyle{firstpage}
{
    \fancyhead[L]{This work has been submitted to the IEEE for possible
      publication.\\
      Copyright may be transferred without notice, after which this version may no longer be accessible.}
    \fancyhead[R]{}
    \pagenumbering{gobble}
}
\fi

\section{Introduction}\label{sec:intr}
 \thispagestyle{firstpage}



\acp{RAN} are transitioning from monolithic implementations using specialized
hardware in favor of more agile, innovative, and customizable solutions based on
disaggregation~\cite{kaltenberger2020openairinterface}, open
interfaces~\cite{perez20195g}, and softwarization~\cite{santos2020virtual}. An
important manifestation of this transition is embodied in the \ac{O-RAN} vision,
which has gained substantial traction through the establishment of a worldwide
consortium~\cite{ORAN-alliance} with broad industry participation
and has also attracted regulatory interest~\cite{johnson2021open}. The \ac{O-RAN}
Alliance proposes an open architecture that disaggregates the \ac{RAN} into
different functional components,
connected under a common control and management overlay that can execute
custom control logic via third-party applications
supplied by, e.g., \ac{RAN} solutions and consulting companies, \acp{MNO},
the open-source community, and new entrants in the market~\cite{polese2022colo,
kliks2023towards,garcia2021ran}.
In addition, \ac{O-RAN} provides specifications that complement and build on top
of 3GPP standards, establishing well-defined open interfaces for connecting the
disaggregated \ac{RAN} components to ensure interoperability across different
vendors~\cite{d2022orchestran}.

The \ac{O-RAN} Alliance aims to promote competition and innovation, empowering \acp{MNO}
to build more flexible and cost-effective networks, encouraging new entrants
and startups,
and facilitating collaboration among industry and academia stakeholders, with potential
benefits to \acp{MNO} and end-users alike~\cite{niknam2022intelligent}.
Through the standardization of third-party applications to manage the \ac{RAN},
the \ac{O-RAN} vision \1 fosters innovation in the telecom market, allowing
\acp{MNO} to deploy tailored applications to customize their network
operations~\cite{bonati2022intelligent}; \2 creates conditions for cost saving
through competition between app providers, but also between hardware
manufacturers; and \3 ensures the \ac{RAN} equipment is future-proof, as
\acp{MNO} can test and validate new solutions and algorithms on their
existing physical network infrastructure~\cite{polese2022understanding}.



The \ac{O-RAN} Alliance partnered with the Linux Foundation to create the
\ac{OSC}~\cite{osc}, an open-source project responsible for creating reference
implementations of \ac{O-RAN} components following the \ac{O-RAN}
specifications, serving as a starting point for prototyping \ac{O-RAN}
solutions~\cite{polese2022understanding, bimo2022osc, osc}.
The \ac{OSC} supports and distributes a number of first-party xApps,
modular applications designed to manage and optimize various aspects and
parameters of the \ac{RAN}. The xApps act as plugin-like extensions, enhancing
the capabilities of the \ac{RAN}  and providing \acp{MNO} with different
functionality, e.g., managing the admission control of \acsp{UE}, monitoring
the \acp{KPM} of base stations, detecting traffic anomalies, and performing
traffic steering~\cite{currentxapps}. In addition, both academia and industry
have developed several third-party xApps to test and demonstrate their solutions
and algorithms on real \ac{O-RAN} networks (detailed further in
Section~\ref{sec:tech}).

There is a vast literature on \ac{O-RAN}, covering aspects from the basic
understanding and new concepts~\cite{wani2024open}, to how the
\ac{O-RAN} principles are influencing the evolution of mobile networks towards
6G~\cite{polese2023empowering}, the benefits for mobile operators to adopt O-RAN
in their networks~\cite{larsen2024evolution}, and the
security vulnerabilities and threat surface introduced by the open, cloud-based
\ac{O-RAN} architecture~\cite{groen2024implementing}. However, there remains a
significant gap in the literature regarding the theoretical foundation and
technical background for developing xApps.
Despite \ac{O-RAN}'s standardization and development efforts, their creation of
an SDK to create xApps with support for different programming languages, and a
growing community of developers from academia and industry, the process of
creating xApps is still far from a straightforward endeavor.

From an implementation point of view, xApps are highly complex microservices
that interact with multiple components of the \ac{nearrtric} through widely
different APIs and protocols~\cite{polese2022colo, polese2022understanding},
making it challenging for newcomers to start prototyping their xApps.
The few existing works on the development of xApps address
specific considerations, e.g., data flows between O-RAN
entities~\cite{hoffmann2023open}, interactions with base
stations~\cite{Kouchaki-22}, or designing \ac{DRL} agents~\cite{polese2022colo},
without providing context about all
the features and capabilities available for xApp developers.
In addition, some aspects of xApps are still undergoing active standardization,
e.g., the AI/ML workflow~\cite{aimlworkflow}, while others have been left for
further studies, e.g., security inside the \ac{nearrtric}~\cite{securityric}, or
had a complete revamp in recent \ac{O-RAN} releases, e.g., subscriptions to
information from the \ac{RAN} (detailed further in Section~\ref{sec:oper}).

The documentation for creating xApps is outdated and fragmented across
the \ac{OSC}'s Wiki and Gerrit webpages, with numerous tutorials becoming
deprecated as the project evolved.
The lack of consolidated and up-to-date tutorials is a well-known issue in
the community, which prompted responses from different stakeholders, including
practitioners creating educative video series~\cite{youtube}, professional
organizations developing interactive online resources~\cite{innovationtestbed},
and private initiatives offering training courses to address the need for easy
and accessible documentation regarding \ac{O-RAN} and xApp
development~\cite{tip, intelify, rimedo}.
Ultimately, the lack of consolidated documentation and comprehensive guidelines
available to the community impose barriers for new players in the telecom
market and increase the costs for industry and academia to develop, test,
and validate their xApps. 


The purpose of this tutorial paper is to provide a thorough guide on how to
develop xApps, from theory to practice.  In particular, the contributions of
this paper are as follows:

\begin{itemize}
  \item We create the first comprehensive guide with instructions for
    developing, managing, and evaluating xApps,
    supporting xApp developers from theory to practice.

  \item We present a theoretical foundation on \ac{O-RAN} and practical
    knowledge related to the realization of \ac{O-RAN} entities and xApps,
    following the \ac{OSC}'s design choices.

  \item We provide context and detail the functionality offered by
    \ac{nearrtric} components to xApps, accompanied by practical examples to
    demonstrate their utilization.

  \item We highlight the current open challenges for developing xApps and
    testing them in end-to-end scenarios.

\end{itemize}

This paper accompanies a public online repository
containing the supporting material used throughout the tutorial, namely, the
xApp descriptor and schema files, example source codes, and Python
representations of ASN.1 documents.
For additional information, we refer the reader to~\cite{repo}.

\begin{figure}[!t]
\centering
\includegraphics[width=0.80\columnwidth]{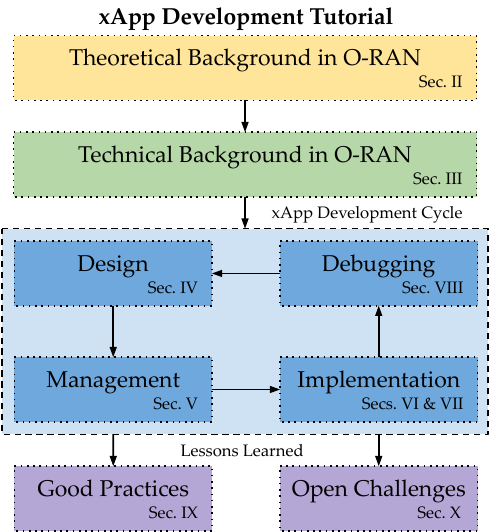}
\caption{The xApp development cycle and the tutorial paper organization.
  Sections~\ref{sec:theo}--\ref{sec:tech} provide developers with
  theoretical and technical background for getting started in \ac{O-RAN},
  Sections~\ref{sec:prep}--\ref{sec:debug} detail the xApp development cycle
  accompanied of practical examples, and Sections~\ref{sec:good}--\ref{sec:outl}
  contain lessons learned, with good practices and current
  open challenges.
}
\label{fig:structure}
\end{figure}

\textit{Paper Structure:} The remainder of this paper is organized as shown in
Fig.~\ref{fig:structure}. In Section~\ref{sec:theo},
we provide a theoretical background on \ac{O-RAN}, its architecture, and the
components of the \ac{nearrtric}. In Section~\ref{sec:tech}, we discuss
\ac{O-RAN} implementation details, review fundamentals on containers for
developing xApps, and provide an overview of current first- and third-party xApps.
In Section~\ref{sec:prep}, we detail the xApp architecture and interfaces,
and describe how to design and define xApps using configuration and schema
files.
In Section~\ref{sec:man}, we detail the xApp lifecycle and demonstrate how to
interact with the \ac{nearrtric} to manage xApps.
In Section~\ref{sec:oper}, we describe the different xApp interfaces and
functionality, providing examples of how to develop xApps capable of
communicating with one another, using persistent storage, and reacting to user
input.
In Section~\ref{sec:ran}, we detail how xApps can subscribe to information from
base stations and manage their operation.
In Section~\ref{sec:debug}, we discuss debugging strategies and methods to
validate the operation of xApps and test their interfaces.
In Section~\ref{sec:good}, we discuss good practices to facilitate the
development of xApps. 
In Section~\ref{sec:outl}, we outline ongoing xApp
standardization efforts and discuss open challenges for developing xApps.
Finally, in Section~\ref{sec:conc}, we pose our concluding remarks.
For ease of reference, we list the acronyms used throughout this paper at the end.

\section{Theoretical Background in O-RAN}\label{sec:theo}

\begin{figure}[!t]
\centering
\includegraphics[width=0.76\columnwidth]{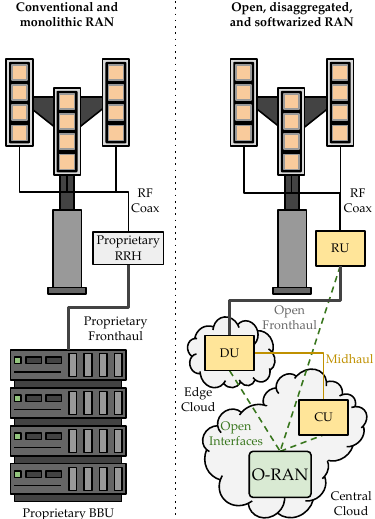}
\caption{Comparison between the conventional, monolithic
    \ac{RAN} and the \ac{O-RAN} paradigm, showing the latter's decomposition of
    the \ac{RAN} into functional components running in software, interconnected
  using   open interfaces, and orchestrated by a common control and management
  overlay.}
\label{fig:traditional-disaggregated-ran-arch}
\end{figure}

In this section, we review the \ac{O-RAN} principles and architecture, as
illustrated in Fig.~\ref{fig:traditional-disaggregated-ran-arch}.
We zoom into the \ac{nearrtric}, describe its internal components, and detail
how they provide the functionality to support the operation of xApps.

\subsection{Principles and Architecture}


\begin{figure}[t]
  \centering
  \includegraphics[width=0.90\columnwidth]{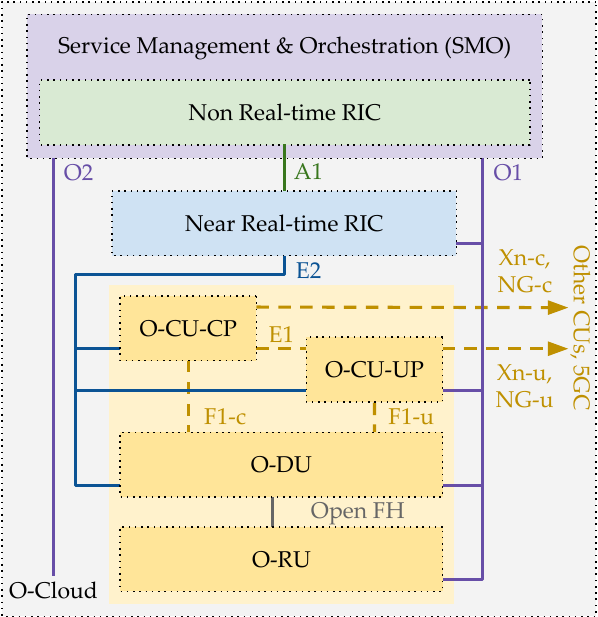}
  \caption{Logical architecture of \ac{O-RAN}, showing the \acp{RIC} and
    their interfaces for managing the \ac{RAN} and its E2 Nodes, i.e., \ac{OCU},
    \ac{ODU}, and \ac{ORU}. The dashed lines indicate interfaces
    standardized by the 3GPP, whereas the solid lines indicate new
    interfaces introduced by the \ac{O-RAN} Alliance.}
  \label{fig:oran-sc-arch}
\end{figure}

The layers of the 5G protocol stack operate at different timescales, and their computational requirements grow at different rates based on the number of users and their demand~\cite{kist2020airtime}.
As a result, the 3GPP introduced functional splits for 5G,
breaking down monolithic \ac{RAN} deployments with one-size-fits-all \acp{RRU} and \acp{BBU} into a series of discrete \ac{RAN} functions that can be placed and scaled on-demand~\cite{bonati2021intelligence}. The \ac{O-RAN} Alliance examined the split options and selected Split 7.2x to underpin their architecture, due to its balance between the simplicity of the functional components and the throughput and latency requirements for their interfaces. The Split 7.2x decomposes 5G base stations, known as gNodeBs, into \acp{CU}, \acp{DU}, and \acp{RU} that implement different functions of the 5G RAN protocol stack. Specifically, \1 the \acp{CU} implement functionalities at the higher layers that operate over larger timescales, e.g., packet processing operations;
\2 the \acp{DU} handle time-critical operations at the lower layers,
e.g., signal processing operations; and \3 the \acp{RU}
manage and interface with \ac{RF} components, e.g.,
\ac{FFT}/\ac{IFFT}~\cite{niknam2022intelligent}. In addition, \ac{O-RAN} goes one step further in disaggregation and splits the \ac{CU} into two logical components: \1 the \ac{CU}-\ac{UP}, responsible for the \ac{PDCP} layer's \ac{UP} and the \ac{SDAP} layer and \2 the \ac{CU}-\ac{CP}, responsible for the \acs{PDCP} layer's \ac{CP} and the \ac{RRC} layer. This split option allows different components to be deployed at distinct network locations and leverage different hardware accelerators, e.g., \acp{DSP} and \acp{FPGA}~\cite{polese2022understanding}.


The \ac{O-RAN} functional split promotes the decoupling between \ac{RAN} hardware and software components similar to the \ac{C-RAN} paradigm \cite{checko2014cloud}, as \ac{DU}, \ac{CU}-UP, and \ac{CU}-CP can run as software instances in a hierarchy of cloud platforms
and be autonomously deployed or scaled on demand~\cite{d2022orchestran}. It is important to note that, unlike \ac{C-RAN}, \ac{O-RAN} is not oblivious to the underlying computing and virtualization infrastructure, but instead,
\ac{O-RAN} incorporates it as a part of the ecosystem known as
\ac{OC}~\cite{polese2022understanding}. This component is an abstraction that combines \1 physical nodes, e.g., servers
and data centers; \2 software components, e.g., containers and virtual machine hypervisors; and \3  management and orchestration functionalities, e.g., \ac{FCAPS}.
\ac{O-RAN} interacts with the \ac{OC} through the \ac{SMO}, the central
component responsible for orchestration, management, and automation in the
\ac{MNO}'s infrastructure, e.g., the \ac{OSM}~\cite{etsi2016open} or the
\ac{ONAP}~\cite{onap}. A new O2
interface connects the \ac{SMO} and the \ac{OC}, enabling the programmatic
management and deployment of network functions, the definition of an inventory
of the facilities under the \ac{OC}, as well as monitoring, fault tolerance, and
update strategies~\cite{niknam2022intelligent, polese2022understanding}.

As part of the functional split, the 3GPP standardized interfaces for the
communication between \acp{CU}, \acp{DU}, and \acp{RU}. However, these standards
allowed vendors to introduce proprietary extensions between their components,
which resulted in vendor lock-in~\cite{niknam2022intelligent}. To mitigate this
issue, the \ac{O-RAN} Alliance has created more restrictive specifications
(built on top of 3GPP standards) for open interfaces between \ac{RAN} components
that ensure complete vendor interoperability~\cite{polese2022understanding}.
In addition, \ac{O-RAN} introduces new standardized APIs for controlling and
managing each E2 Node, including gNodeBs and eNodeBs (4G base stations) and
their functional splits, i.e., \acp{CU}, \acp{DU}, and \acp{RU}.
These new interfaces
contain: the \1 E2 interface for controlling different \ac{RAN} functions
exposed by each node, i.e., the control knobs supported by each node, e.g.,
handover thresholds, scheduling directives, or power-saving parameters; and the
\2 O1 interface, for operations and maintenance of each node, establishing
heartbeats, setting up alarms, and reporting \acp{KPM}. Due to these new open
interfaces and APIs, the \ac{RAN} nodes in \ac{O-RAN} are dubbed as \acf{O}
components, e.g., \acs{OCU}(-CP/UP), \acs{ODU}, and \acs{ORU}.

With the introduction of open and programmable interfaces across all
E2 Nodes, \ac{O-RAN} can orchestrate their operation under a common control and
management overlay to optimize their performance using data-driven
closed-control loops~\cite{bonati2021intelligence}.
As mentioned earlier, the \ac{O-RAN} Alliance
envisions two different \acp{RIC} for running closed-control loops in different
locations and timescales: the \1 \ac{nearrtric}, deployed closer to the edge and
\ac{RAN} nodes to perform near-real-time control loops with a periodicity
between \unit[10]{ms} and \unit[1000]{ms}, supporting xApps with custom control
logic to perform radio resource management; and the \2 \ac{nonrtric}, deployed
as a component of the \ac{MNO}'s \ac{SMO} framework to perform non-real-time
control loops longer than \unit[1]{s}, managing \ac{ML} models and supporting rApps with
custom control logic to dictate the long-term behavior of the
network~\cite{polese2022understanding}. The \acp{RIC} communicate via their A1
interface, which the \ac{nonrtric} uses to deploy policies that guide the
\ac{nearrtric} optimization goals.

Fig.~\ref{fig:oran-sc-arch} depicts the \ac{O-RAN} architecture specified by the
\ac{O-RAN} Alliance and shows the interplay between the entities discussed in
this section.
Throughout the rest of this tutorial, we have several diagrams showing logical
and practical components of the \ac{O-RAN} ecosystem, as well as their
interfaces and interactions. To make it easier for the reader to understand the
interactions between \ac{O-RAN} entities and their locations within the
\ac{O-RAN} ecosystem, we adopt a color code where: yellow refers to E2 Nodes and
the \ac{RAN}, blue refers to the \ac{nearrtric} and its components, green refers
to the \ac{nonrtric}, purple refers to the \ac{SMO}, pink refers to Docker
containers, and gray refers to the underlying software or hardware
infrastructure supporting the \ac{O-RAN} components.


\subsection{Near-RT RIC Purpose and Interactions}\label{sub:nearinter}


The \ac{nearrtric} is the \ac{O-RAN} entity responsible for providing
near-real-time \ac{RAN} orchestration and network
automation~\cite{santos2020breaking}. Its primary purpose is to host and
facilitate the operation of external applications, the xApps, for running
near-real-time closed-control loops to monitor, analyze, and optimize network
parameters for achieving desired network behavior and performance. Arguably, the
\ac{nearrtric} operates in a similar fashion to ONOS~\cite{berde2014onos}. In
the context of \ac{SDN} for transport networks. The \ac{nearrtric} has received
much attention and contributions from the members of the \ac{O-RAN} Alliance,
being one of the most complete and mature software components provided by the
\ac{OSC}. The \ac{nearrtric} interacts with other \ac{O-RAN} entities through
southbound and northbound interfaces to leverage their information and
capabilities for managing the \ac{RAN}, as shown in Fig.~\ref{fig:oran-sc-arch}.
In the following, we detail these interactions.

\begin{itemize}
  \item \textit{E2 Nodes:} The \ac{nearrtric} interacts with \ac{RAN} components, e.g., the
      disaggregated O-CU-CP, O-CU-UP, O-DU and O-RU, or the monolithic O-gNodeB
      and O-eNobeB. It collects real-time measurements and data from these nodes
      to monitor network performance, traffic load, interference levels, and
      other relevant metrics. The \ac{nearrtric} also communicates with E2
      Nodes to configure and adjust different \ac{RAN} parameters, e.g., transmit
      power, antenna settings, modulation schemes, and scheduling algorithms,
      based on the decisions made by the xApps.


    \item \textit{\ac{SMO}:} The \ac{nearrtric} interacts with the \ac{SMO} responsible
      for the overall management of the network and services.
      The \ac{SMO} provides high-level control and coordination
      functions, and the \ac{nearrtric} acts as an extension to this system by
      offering near-real-time optimization and intelligence capabilities within
      the RAN.

    \item \textit{\ac{nonrtric}:} The \ac{nearrtric} interacts with the \ac{nonrtric}
      that defines and enforces policies, quality of service requirements,
      and regulatory constraints. The Near-RT RIC exchanges information and
      aligns its decision-making process with the policies described by this
      entity to ensure that network optimizations and resource allocations are
      in compliance with the established rules.

\end{itemize}


The interactions between the \ac{nearrtric} and other \ac{O-RAN} entities to
exchange network \acp{KPM}, control information, and system settings create a
collaborative, distributed ecosystem that enables near-real-time
programmability, automation, and intelligence within the \ac{RAN}.

\subsection{Different O-RAN Flavors and \acp{nearrtric}}\label{sub:flavor}

The O-RAN specifications contain the requirements, capabilities, and
interfaces for \ac{O-RAN} entities, leaving the implementation details to
the discretion of vendors or system integrators. In addition, the \ac{O-RAN}
Alliance partnered with the Linux Foundation to create the \ac{OSC}~\cite{osc}
and provide a fully operational, open-source reference implementation of
\ac{O-RAN} entities (discussed in the following subsection) to demonstrate their
capabilities and serve as the starting point for commercial products. These
approaches fostered the creation of different \ac{O-RAN} flavors, such as the FlexRIC
from OAI~\cite{schmidt2021flexric}, the SD-RAN from ONF~\cite{SD-RAN}, and the
dRAX from AccelleRAN~\cite{durre2022disaggregated}. While possessing
interoperable external interfaces toward E2 Nodes, the different \ac{O-RAN}
flavors adopt different design choices for their \acp{RIC}, internal components,
xApps, and \acp{SM}, making them not interoperable. For example, the \ac{OSC}
and SD-RAN adopt a microservice philosophy, where each capability of the
\ac{nearrtric} is implemented through a discrete component. In contrast, the
FlexRIC and dRAX tend to be more monolithic and/or provide different
communication interfaces not standardized by the \ac{O-RAN} specifications,
such as the OAI E42 interface for direct communication between xApps and E2
Nodes~\cite{queiros2024autonomous}.
For a comprehensive comparison between \ac{O-RAN} flavors and their \acp{RIC},
we refer the reader to other works that evaluated their performance and
features~\cite{ngo2024ran}~\cite{liu2024blueprint}. 

Without loss of generality, we will focus on the \ac{O-RAN} flavor from the
\ac{OSC} throughout the rest of this tutorial. This choice is motivated by its
status as the first \ac{O-RAN} implementation available in the literature, its  
widespread adoption in academia and industry, and the extensive body of research
and solutions developed based on it. By leveraging the \ac{OSC} implementation,
we aim to benefit a broader audience and maximize the impact of this tutorial.
While other \ac{O-RAN} flavors may have different design choices and APIs, the
concepts and lessons introduced through this tutorial remain valuable and
applicable across other \ac{O-RAN} implementations.

\subsection{Near-RT RIC Components}\label{sub:nearcomp}

\begin{figure}[t]
  \centering
  \includegraphics[width=0.99\columnwidth]{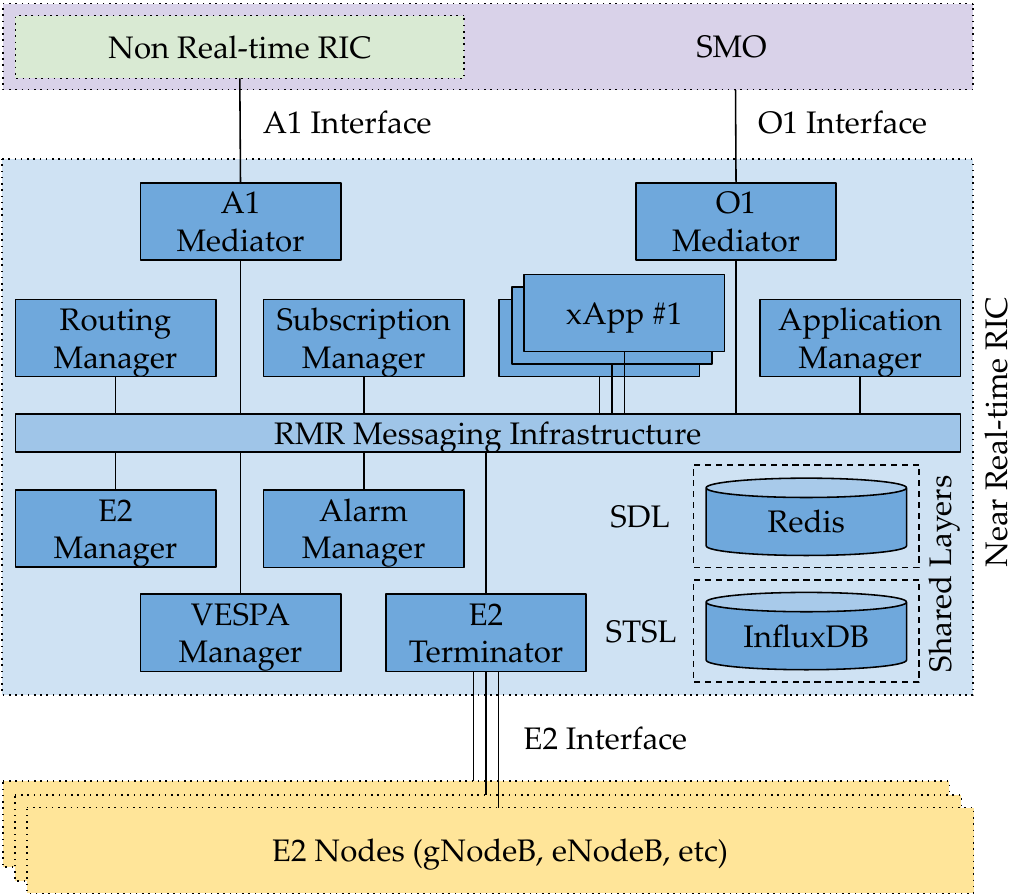}
  \caption{The internal components of the \ac{nearrtric} and how they communicate
    with other \ac{O-RAN} entities using different interfaces.
    The xApps reside inside the \ac{nearrtric} and interact with its components
    to leverage their capabilities to perform actions, e.g., subscribe
    to information from the \ac{RAN}.}
  \label{fig:nearrtric}
\end{figure}

In a similar fashion to how the \ac{O-RAN} ecosystem is composed of several
entities, the \ac{nearrtric} from the \ac{OSC} is implemented as a
collection of microservices that work in unison to allow \acp{MNO} to manage
their \acp{RAN} in near-real time. Fig.~\ref{fig:nearrtric} illustrates the
components of the \ac{nearrtric}, each of which provides specific functionality
to xApps or supports their operation inside the \ac{nearrtric} cluster. We
describe each component of the \ac{nearrtric} below.

\begin{description}[style=unboxed,leftmargin=0cm,labelindent=\parindent]

  \item[\ac{RMR}:] implements the internal messaging infrastructure for
    communication between all \ac{nearrtric} components,
    including management components, interface terminators, and xApps.
    The RMR library allows applications to send/receive messages
    to/from other applications without information about their IP,
    location, or the underlying transport mechanism.
    We detail \ac{RMR} later in Section~\ref{sub:mes}.

  \item[Routing Manager (RtMgr):] manages \ac{RMR} routes in the
    \ac{nearrtric}. It is responsible for creating and distributing
    \ac{RMR} routing policies to \ac{nearrtric} components and xApps.

  \item[Application Manager (AppMgr):] manages the xApp deployment and lifecycle
    in the \ac{nearrtric}. It is responsible for (un)installing xApps and
    notifying other \ac{nearrtric} components about the current set of xApps.
    We detail how the xApp developer can interact with it later in
    Section~\ref{sub:dms}.

\item[Subscription Manager (SubMgr):] manages subscriptions from xApps to E2
  Nodes. It is responsible for creating routes and abstracting the interaction
  with the E2 Nodes. We detail how xApps can subscribe to
  E2 Nodes later in Section~\ref{sub:sub}.

\item[E2 Manager (E2Mgr):] manages the E2 Nodes registered with the
  \ac{nearrtric}. It sets up E2 Nodes with the \ac{nearrtric}, monitors their
  health, and informs any issues to xApps.

\item[E2 Terminator (E2Term):] intermediates the communication between xApps
  (or other Near-RT RIC components) and E2 Nodes, acting as a
  translation layer between the internal \ac{RMR} messaging infrastructure used
  inside the \ac{nearrtric} and the external SCTP protocol used by E2 Nodes.

\item[Shared Layers:] provide a lightweight, high-speed interface for managing
data storage in the \ac{nearrtric}. \ac{SDL} and \ac{STSL} offer stateless
storage, abstracting the underlying database technology from the business.
\ac{SDL} stores relational data, while \ac{STSL} stores time series data. We
detail how xApps can leverage the Shared Layers for persistent storage in
Section~\ref{sub:sto}.

\item[VESPA Manager (VesMgr):] starts, configures, and uses the \ac{VES} Agent
  to adapt the collection of internal statistics using Prometheus to scrape
  metrics from \ac{nearrtric} components and xApps, and forward them to an
  \ac{SMO}, e.g., ONAP, or another \ac{VES} Collector.

\item[Alarm Manager:] manages alarms from xApps and \ac{nearrtric}
components, interfacing with the Prometheus
Alert Manager to post the alarms as alerts. It also de-duplicates,
silences, inhibits alerts, and routing them to the \ac{VES} Agent.

\item[A1 Mediator:] receives policies from the \ac{nonrtric} and forwards them
  to xApps via \ac{RMR}. A policy contains high-level directive that serves to
  steer the behavior of xApps. We detail how xApps can consume policies in
  Section~\ref{sub:pol}.

\item[O1 Mediator:] exposes metrics and information about the status of the
  \ac{nearrtric} components, registered E2 Nodes, and xApps to the management
  entities in the \ac{SMO}.

\end{description}

This section discussed the principles behind \ac{O-RAN}, its specifications, and
the concepts and motivations therein. However, to develop xApps, we must go one
step further into the \ac{O-RAN} ecosystem, exploring some implementation
details and supporting technologies underpinning the \ac{nearrtric} and the xApp
themselves.
The deployment and installation of the \ac{nearrtric}, as well as the system
requirements for running it, will differ depending on the \ac{O-RAN} flavor and
the scale of operation. We followed the system requirements from the
official \ac{OSC} documentation~\cite{ricinstallreq} to create a
development environment and the new \ac{nearrtric} installation
guidelines~\cite{newinstaller}. We refer the reader to these references
for information on system specifications and installation instructions.

\section{Technical Background in O-RAN}\label{sec:tech}


In this section, we review technical matters related to the realization of
\ac{O-RAN} entities and xApps. First, we discuss the \ac{O-RAN} implementation
and design choices taken by the \ac{OSC} that influence the development of
xApps. Then, we present the cloud technologies supporting xApps in a
\ac{nearrtric}. Next, we introduce the \ac{OSC}'s resources for facilitating xApp
development. Finally, we overview existing first- and third-party xApps to
demonstrate their capabilities and inspire readers to use them as a
starting point to develop their xApps.

\subsection{O-RAN Implementation}\label{sub:OSC}





\begin{figure}[t]
  \centering
  \includegraphics[width=0.62\columnwidth]{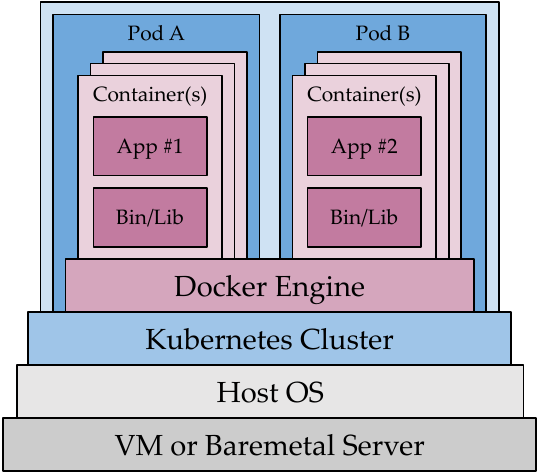}
  \caption{The layers of abstraction for running xApps, showing their Kubernetes
    pods, which include one or more Docker containers with the complete
    environment to run the xApp source code, and the underlying \ac{nearrtric}
    cluster, running on the host OS of a virtual machine of baremetal server.
}
  \label{fig:pods}
\end{figure}

Each \ac{O-RAN} flavor can take distinct
design choices and technical approaches, which leads to a lack of
interoperability across \ac{O-RAN} stacks~\cite{ngo2024ran}.
For example, the \ac{OSC} uses the \ac{RMR} messaging protocol for communication
between \ac{nearrtric} components and xApps~\cite{polese2022understanding},
whereas the SD-RAN employs the gRPC protocol~\cite{SD-RAN}. In theory, both
protocols perform the same task, but their implementations are widely different
and, hence, incompatible in practice.
These implementation differences affect xApps, as xApps developed
for a given \ac{O-RAN} flavor will not necessarily be compatible with others,
e.g., xApps for \ac{OSC} deployments are incompatible with SD-RAN and
vice-versa. Consequently, the xApp developer must decide which \ac{O-RAN}
flavor they will cater to. In the remainder of this tutorial, we will instruct
the reader on creating xApps for the \ac{O-RAN} flavor from the \ac{OSC},
providing technical background and practical examples.

\begin{figure}[t]
  \centering
  \includegraphics[width=0.95\columnwidth]{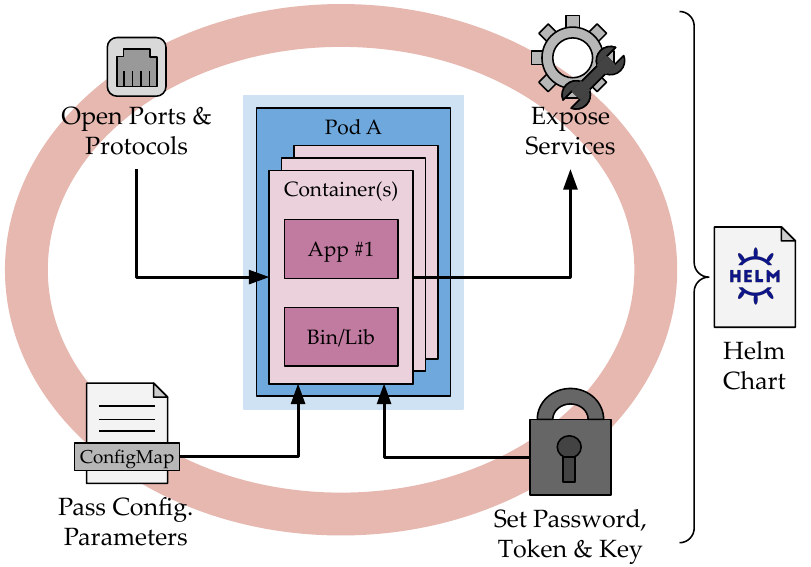}
  \caption{Examples of the different aspects that are
  automated through the use of Helm Charts, facilitating the deployment of
multiple pods and their containers in a \ac{nearrtric}.}
  \label{fig:helm}
\end{figure}

The \ac{OSC} is an open-source
community, academia, industry, and open-source developers
contribute to developing and improving an open-source \ac{O-RAN}
implementation~\cite{polese2022understanding, bimo2022osc, osc}.
It provides an open-source implementation of the \ac{nearrtric},
following a microservice architecture based on cloud
technologies~\cite{bimo2022osc}.
The \ac{nearrtric} is a specialized Kubernetes
cluster that adopts the Docker container engine~\cite{polese2022understanding}.
In this way, each \ac{nearrtric} component is an isolated Kubernetes pod running
one or more Docker containers, whose resources, ports, and interfaces are
described using Helm Charts (we detail these cloud technologies later
in this section). In the \ac{OSC}, the xApps are also cloud-native
microservices, i.e., Kubernetes pods based on Docker containers, running inside
the \ac{nearrtric} cluster, as illustrated previously in
Fig.~\ref{fig:nearrtric}.
Thus, it is essential for the xApp developer to understand the fundamentals
behind these cloud technologies and use them accordingly.
Moreover, the \ac{OSC} created an xApp SDK~\cite{oransdk} to facilitate the
development of xApps in different languages and interfacing with the
\ac{nearrtric} components, which we will leverage throughout this tutorial.


\subsection{Cloud Technologies Supporting xApps}\label{subsub:tech}



A common challenge in software distribution is ensuring that your application
will run in the environment of an interested party, such as \acp{MNO}, who may
employ widely different software and hardware platforms.
To this end, Docker containers are lightweight virtual environments that include
everything required to run a given application, i.e., source code, dependencies,
libraries, and settings~\cite{merkel2014docker}.
These applications and their virtual environments can be conveniently packaged
into a single file, a Docker container image~\cite{polese2022understanding},
making them portable and facilitating their distribution.
While sharing the same kernel with the host OS of the underlying server or
virtual machine (a common practice in cloud environments), the Docker containers
are isolated from one another, allowing them to have different OSs, libraries, and
versions, improving security for running applications from third-parties and
mitigating potential library versioning conflicts.
We detail the process for creating Docker container
images supporting xApps later in Section~\ref{sub:doc}.

A second challenge arises when managing the deployment of several containers on
a cluster of servers, such as the components of the \ac{nearrtric} and the xApps
from various developers,
which can have very different requirements in terms of computing resources,
ports, and interfaces. To this end, Kubernetes serves as an orchestration
platform for managing the lifecycle, health, communication, and storage of
container deployments~\cite{burns2022kubernetes}.
It groups one or
more containers working together to provide a service or run an application
into an independent and isolated pod.
Fig.~\ref{fig:pods} illustrates an abstract example of an xApp pod composed of
multiple containers.
Kubernetes can automatically restart pods if there are errors, scale their
computational resources according to demand, or replicate them entirely
for load balancing across servers.
The \ac{nearrtric} Kubernetes cluster separates its pods into different
namespaces: \1 \texttt{ricinfra}, containing pods that provide the supporting
infrastructure for the operation of the Kubernetes cluster; \2 \texttt{ricplt},
containing pods of the \ac{O-RAN} components in the \ac{nearrtric} platform,
e.g., \texttt{AppMgr}, \texttt{RtMgr}, \texttt{SubMgr}, etc.; and \3
\texttt{ricxapp}, containing all the installed xApp pods.
We detail the mechanism for deploying xApps pods in a \ac{nearrtric} cluster
later in Section~\ref{sub:dms}.

As the size of your Kubernetes deployment grows, so does the complexity of
managing it. Since applications based on a microservice architecture can span
many containers or pods working together, manually \1 downloading multiple 
Docker container images to create pods; \2 passing configuration parameters to
each container, \3 defining their open ports and
protocols to describe how their services are exposed, and \4 managing their
authentication credentials through passwords and access tokens can quickly
become intractable.
To this end, Helm automates the creation, description, configuration, and
deployment of Kubernetes pods~\cite{shah2019building}. It combines multiple
configuration files that define different properties and requirements of
Kubernetes pods and their containers into a single reusable package, a Helm
Chart.
Fig.~\ref{fig:helm} illustrates the Helm Chart for an abstract example of an xApp pod leveraging different features from Kubernetes.
The Helm Charts also contain a "value.yaml" file that allows
the user to customize the configuration parameters of pods before their
deployment on a Kubernetes cluster. While
the \ac{OSC} provides Helm Charts for automating the deployment of the
\ac{nearrtric} components, the xApp developer must provide their own. The
creation of Helm Charts for xApps is partially automated using the
\texttt{dms_cli} tool provided by the \ac{OSC}, detailed later in
Section~\ref{sub:dms}.


\newcolumntype{C}[1]{>{\centering\let\newline\\\arraybackslash\hspace{0pt}}m{#1}}
\definecolor{blue1}{HTML}{9FC5E8}
\definecolor{blue2}{HTML}{CFE2F3}

\begin{table*}[!t]
\centering
\caption{Qualitative classification of the current third-party xApps for the
\ac{OSC} available in the literature.}
\label{tab:surveys}
\resizebox{\textwidth}{!}{%
  \begin{tabular}{C{2.9cm}||C{1.8cm}|c||c||c||c}
\rowcolor[HTML]{9FC5E8}
 & & &  &  & \textbf{Development}  \\
\rowcolor[HTML]{9FC5E8}
  \multirow{-2}{*}{\textbf{Works}} & \multirow{-2}{*}{\textbf{Category}} &
  \multirow{-2}{*}{\textbf{Objective}} &
  \multirow{-2}{*}{\textbf{Evaluation}} & \multirow{-2}{*}{\textbf{ML Support}} & \textbf{Details}  \\ \hhline{=::==::=::=::=}

  \rowcolor[HTML]{CFE2F3} Johnson, et al. \cite{NexRAN} &   & Policy driven RAN slice control  & Emulation: OSC \& srsLTE & -- & -- \\
  \hhline{-||>{\arrayrulecolor{blue2}}->{\arrayrulecolor{black}}-||-||-||-}

  \rowcolor[HTML]{CFE2F3} Bonati, et al. \cite{Bonati-21} & RAN Slicing & RAN slice scheduling optimization & Emulation: Colosseum & DRL & -- \\
  \hhline{-||>{\arrayrulecolor{blue2}}->{\arrayrulecolor{black}}-||-||-||-}

  \rowcolor[HTML]{CFE2F3} Zhang, et al. \cite{Zhang-22} &  & RAN slice power and
  resource allocation &  Simulation: Matlab 5G Toolbox & Federated DRL & --\\
  \hhline{=::==::=::=::=}

  \rowcolor[HTML]{9FC5E8} Huff, et al. \cite{Huff-2021} & Security and & Fault-tolerance & Emulation: OSC & -- & --\\
  \hhline{-||>{\arrayrulecolor{blue1}}->{\arrayrulecolor{black}}-||-||-||-}
  \rowcolor[HTML]{9FC5E8} Wen, et al. \cite{Haohuang-22} & Fault-tolerance  & Telemetry for security analysis & Theoretical & -- & -- \\
  \hhline{=::==::=::=::=}

  \rowcolor[HTML]{CFE2F3} Lee, et al. \cite{Hoejoo-2021} & ML &  Online training environment & Emulation: OSC \& srsLTE & RL & -- \\
  \hhline{=::==::=::=::=}


  \rowcolor[HTML]{9FC5E8} Iturria-Rivera, et al. \cite{Rivera-22} & Resource & Power and radio resource allocation & Theoretical & Multi-agent DRL & -- \\
  \hhline{-||>{\arrayrulecolor{blue1}}->{\arrayrulecolor{black}}-||-||-||-}
  \rowcolor[HTML]{9FC5E8} Mungari \cite{Mungari-21} & Allocation & Radio resource management & Emulation: OSC \& OAI & RL & -- \\
  \hhline{=::==::=::=::=}

  \rowcolor[HTML]{CFE2F3} Rego, et al. \cite{Rego-22} & \multicolumn{2}{c||}{Spectrum Sensing} & Emulation: OSC & -- & \cmark\\
  \hhline{=::==::=::=::=}

  \rowcolor[HTML]{9FC5E8} Orhan, et al. \cite{Oner-21} & & Connection management optimization & Theoretical & DRL/GNN & --\\
  \hhline{-||>{\arrayrulecolor{blue1}}->{\arrayrulecolor{black}}-||-||-||-}


  \rowcolor[HTML]{9FC5E8} Kouchaki, et al. \cite{Kouchaki-22} & & QoE maximization & Emulation: OSC & RL & \cmark \\
  \hhline{-||>{\arrayrulecolor{blue1}}->{\arrayrulecolor{black}}-||-||-||-}

  \rowcolor[HTML]{9FC5E8} Huang, et al. \cite{Huang-22} & Data Traffic & Throughput maximization & Theoretical & DRL & --\\
  \hhline{-||>{\arrayrulecolor{blue1}}->{\arrayrulecolor{black}}-||-||-||-}

  \rowcolor[HTML]{9FC5E8} Agarwal, et al. \cite{Agarwal-23} & Management & QoE enhancement function & Simulation & -- & -- \\
  \hhline{-||>{\arrayrulecolor{blue1}}->{\arrayrulecolor{black}}-||-||-||-}

  \rowcolor[HTML]{9FC5E8} Lacava, et al. \cite{Lacava-23} & & Traffic Steering intelligent handover & Simulation: ns-O-RAN & DRL & \cmark \\
  \hhline{-||>{\arrayrulecolor{blue1}}->{\arrayrulecolor{black}}-||-||-||-}

  \rowcolor[HTML]{9FC5E8} Alavirad, et al. \cite{Alavirad-23} & & Admission control of UEs & Simulation: ns3 LTE & DRL & --

\end{tabular}}
\end{table*}

\subsection{\ac{OSC}'s xApp Development Resources}\label{subsub:sdk}


The \ac{OSC} provides open-source reference implementations of \1
\ac{O-RAN} entities and their internal components, e.g., the \ac{SMO}, the Non-
and Near-RT \acp{RIC}; \2 endpoints for \ac{O-RAN} interfaces, e.g., A1, O1, and
E2; and \3 simulators for modeling and testing the behavior of nodes using the
interfaces mentioned above, e.g., \texttt{A1Sim}, \texttt{O1Sim}, and 
the \texttt{E2Sim}.
In addition, to facilitate the development of xApps, the \ac{OSC} provides
an xApp SDK~\cite{oransdk}, which contains libraries to facilitate and abstract
the communication with \ac{nearrtric} components for leveraging their
capabilities as part of the xApp's business logic to manage \acp{RAN} and tools
to help the xApp development cycle.
These tools include \1 the ASN.1 Compiler, which automatically generates C++
bindings for E2 Nodes based on \acp{SM} (detailed in
Section~\ref{sec:ran}); \2 the \texttt{dms_cli}, for managing the lifecycle of
xApps in a \ac{nearrtric} cluster (detailed in Section~\ref{sec:man}); and \3
the xApp Frameworks, which contain the set of libraries listed above, as well as
their bindings in different programming languages.

The xApp Frameworks streamline
the xApp development process, allowing the xApp developers to use the same
libraries, APIs, and design philosophy to create xApps in Python, C++, Go, or
Rust. In addition, they abstract a series of tasks
related to the (de-)registration of xApp with the \ac{nearrtric} components
(detailed in Section~\ref{sec:oper}) and the subscription to E2 Nodes
(detailed in Section~\ref{sec:ran}), which considerably simplifies  the
xApp development process.
Throughout the remainder of this tutorial, we will use the Python xApp
Framework~\cite{pythonframework} to demonstrate the development of xApps due to
Python's widespread adoption and its smooth learning curve for new software
developers. We introduce the libraries available to xApps
in Section~\ref{sec:oper} and explain how to use the \texttt{E2Sim} to
test the interactions between xApps and E2 Nodes in
Section~\ref{sec:ran}.



\subsection{Existing First- and Third-party xApps}\label{sub:table}

The number of first- and third-party xApps has increased considerably since the
first \ac{OSC} release in 2019, and it is expected to grow even more as the
interest and investments in \ac{O-RAN} continue to
rise~\cite{Weissberger_2022}. In this context, first-party refers to
xApps provided and supported by \ac{OSC}, FlexRIC, and SD-RAN initiatives. For
example, the \ac{OSC} provides a few xApps out of the box, e.g.,
the \ac{AC} for controlling the maximum number of \acp{UE} admitted on the
\ac{RAN}, the \ac{KPM} Monitoring, for obtaining period reports on \acp{UE}
and gNodeB metrics, 
the \ac{AD}, for inspecting stored performance metrics and identifying anomalous
\acp{UE}, and the \ac{TS}, for managing the handover of \acp{UE} between
different gNodeBs to optimize network performance.
Meanwhile, the SD-RAN provides their version of a \ac{KPM} Monitoring xApp, as
well as the
\ac{MHO}, also for managing handovers,
the \ac{MLB}, for controlling the gNodeB's cell individual offset
according to its load, and the \ac{RSM}, for creating \ac{RAN} slices and
configuring their resource allocation.
For more detailed information about the specific use cases supported by O-RAN
flavors and the minimal running setups for the first-party xApps mentioned
above, we refer the reader to the official \ac{OSC}~\cite{ricxapp} and the
SD-RAN documentation~\cite{SD-RAN}.
Conversely, third-party refers to xApps designed by the growing community of
open-source developers from industry and academia. To understand the
capabilities of existing xApps, we conducted a literature review to identify
examples of third-party xApps for the \ac{OSC}. We classified them according to
their category, objective, evaluation approach, support for \ac{ML}, and whether
they provide implementation details to help xApp developers use or re-create
them. Table~\ref{tab:surveys} summarizes the results of our review.


We grouped these works into distinct categories related to their operation:
\ac{RAN} Slicing, addressing aspects related to the control and optimization of
\ac{RAN} slices~\cite{NexRAN,Bonati-21,Zhang-22}; Security, considering
fault-tolerance~\cite{Huff-2021} and the streaming telemetry
information~\cite{Haohuang-22}; \ac{ML} with the creation of an online
training reference workflow\cite{Hoejoo-2021}; Resource allocation, considering
radio\cite{Rivera-22}, and power resources~\cite{Mungari-21}; Spectral Sensing
based on \ac{ML}~\cite{Rego-22}; and Data
Traffic Management, with several works managing the data traffic of
for \acp{UE}~\cite{Oner-21, Kouchaki-22, Huang-22, Agarwal-23,
Lacava-23, Alavirad-23}.
Regarding their evaluation strategies, of the 15 works we identified in our
literature review, only four provided robust analytical models. However, these
theoretical works provided limited numerical results without an empirical
evaluation to demonstrate their solutions. Four other works have developed and
evaluated prototypes in simulated environments, such as ns-3 and Matlab 5G
Toolbox. The remaining seven works presented prototypes and evaluated their
proposals using emulated environments, leveraging testbeds, such as
Colosseum, and open-source radio stacks, e.g., srsLTE and \ac{OAI}. While these
works effectively demonstrated their contributions experimentally, the vast
majority did not make their xApps available or provide any development
guidelines, inhibiting the reproduction of their results by the broader research
community.

The official support of \ac{ML} on xApps was only recently introduced by the
\ac{OSC} in December 2022, with the initial release of the
\ac{AIMLFW}~\cite{grelease}, discussed later in Section~\ref{sec:outl}.
However, there is a vast number of works in the literature that predate the
release of the \ac{AIMLFW} and present xApps with support for \ac{ML}. We can
observe a wide range of \ac{ML} solutions in Table~\ref{tab:surveys}, from
\ac{DRL}, to multi-agent \ac{DRL}, Federated \ac{DRL}, and \ac{GNN}.
These works accomplish this feat by proposing a myriad of custom solutions using
homebrewed software that incorporates \ac{ML} into their xApps. While effective
in their specific use cases, it is very challenging to support, distribute, and
extend these non-standard solutions, which limits their applicability.
Finally, it is worth mentioning that only three works provided development
details about their solutions, giving the reader some understanding of the inner
workings of their solutions. 
The lack of papers with implementation details related to the development of
xApps hinders the reproduction of results, the extension of existing prototypes,
and the creation of new solutions. This tutorial paper addresses this gap by
providing comprehensive guidelines, from theory to practice, aiding xApp
developers to design, create, and evaluate their xApps in realistic end-to-end
environments.

\section{xApp Design: Defining your Application}\label{sec:prep}

In this section, we overview the prerequisites for developing xApps, and
delve on their architecture and interfaces. We examine the xApp descriptor
and schema files, providing examples of how to define xApps pods and ports,
pass control parameters, and configure its different interfaces.

\begin{figure}[t]
  \centering
  \includegraphics[width=0.99\columnwidth]{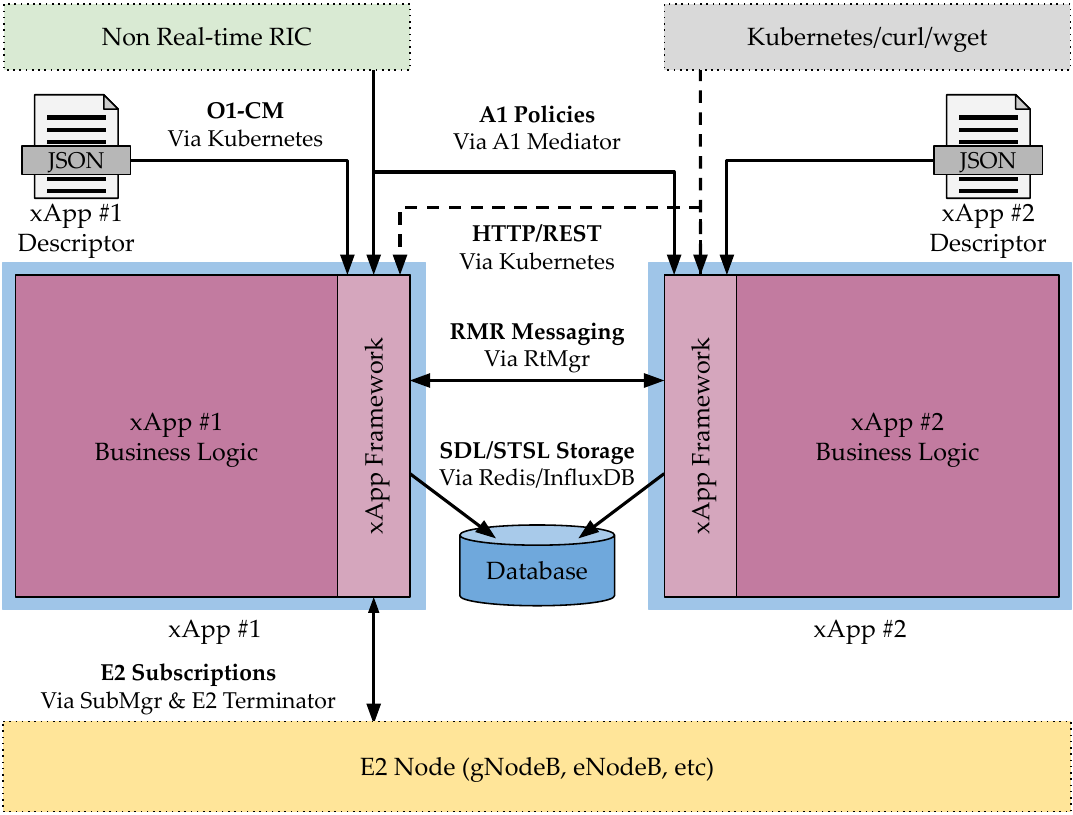}
  \caption{The xApps can leverage capabilities from and interact with several
    entities of the \ac{O-RAN} ecosystem and components of the \ac{nearrtric}
    through different APIs and interfaces.}
  \label{fig:xapp_arch}
\end{figure}

\subsection{Prerequisites}


Developing xApps differs from the traditional implementation of programs
for general-purpose operational systems, such as Linux and Windows. Unlike
traditional programs, which often run standalone 
interacting with a single kernel via system calls, 
the xApps reside
within the \ac{nearrtric} and operate as part of a
distributed system composed of \ac{nearrtric} components and other xApps.
This unique execution environment makes an xApp inseparable from the
\ac{nearrtric} and requires it to interact with \ac{nearrtric} components for
leveraging their capabilities to perform actions through well-defined APIs and
protocols, e.g., subscribe to information from the \ac{RAN} via the E2 interface.
Consequently, the xApp developer not only needs to be knowledgeable in a programming
language (preferably one with an xApp Framework library, e.g., Python, C++, Go, or
Rust) but also in microservices and cloud technologies, e.g., Docker, Kubernetes, and Helm.
Familiarity with the \ac{nearrtric} as the execution environment for the xApps 
is also required, which demands an understanding of the \ac{O-RAN} architecture 
and the capabilities, APIs, and interfaces of \ac{nearrtric} components.

In addition to the technical background and implementation skills,
the xApp developer must have a solid understanding of mobile
networking concepts and the intended use case for their xApps,
e.g., network optimization or resource management. This initial
conceptualization is essential for determining the xApp's objectives, defining
the scope of its operation, and identifying the interfaces and APIs it will
leverage ahead of the implementation of its business logic. Furthermore, it is
strongly recommended that xApp developers create or utilize an \ac{O-RAN}
development environment to test how their xApp will operate in
conjunction with other \ac{O-RAN} entities. For more information about
interactive online resources and remote access testbeds, we refer the reader
to~\cite{innovationtestbed}.

\subsection{Architecture}


From a functional perspective, xApps are discrete microservices that implement
well-defined business logic to manage \acp{RAN}~\cite{writersguide}. This logic
can involve collecting and processing data from E2 Nodes, calculating metrics to
generate reports or trigger alarms, and controlling different aspects of the E2
Nodes according to a given algorithm~\cite{polese2022understanding}. From an
implementation perspective, xApps are Kubernetes pods running inside the
\ac{nearrtric} cluster, each of which may contain one or more Docker containers,
as discussed in Section~\ref{sub:OSC}. The application running in the Docker
containers implements the business logic of the xApp, leveraging an xApp
Framework library available in Python, C++, Go, or Rust programming languages,
which provides xApps with a number of APIs to exploit the capabilities offered
by the different components of the \ac{nearrtric}, as discussed in
Section~\ref{sub:nearcomp}.

Fig.~\ref{fig:xapp_arch} illustrates the xApp architecture and its interactions
with other xApps and entities of the \ac{O-RAN} ecosystem, intermediated through
different components of the \ac{nearrtric}. In the following, we briefly
introduce the different APIs available to the xApps to give the reader a
high-level overview of the xApps' functionality. We will discuss each API
in-depth later in Section~\ref{sec:oper}, where we detail the purpose and
concepts behind their operation and exemplify their utilization.

\begin{description}
\item[O1-CM Configuration:] During startup, the xApp descriptor provides xApps with
  initial configuration parameters to interact with the \ac{nearrtric} and
  optional parameters that can be used to parameterize their
  operation~\cite{writersguide}. The content of the xApp descriptor is loaded in
  the Kubernetes pod as a ConfigMap, allowing \ac{nearrtric} users, i.e.,
  system administrators and network operators, to modify the
  optional   parameters during runtime.

\item[\acs{RMR} Messaging:] It allows xApps to communicate with one another and with
  components of the \ac{nearrtric} through a low-latency messaging library.
  \ac{RMR} uses a publish-subscribe paradigm, enabling xApps to be oblivious to
  the IP addresses of other Kubernetes pods and exchange messages based on
  message types~\cite{rmrguide}. It also serves as the API in which the xApps
  receive A1 policies from the \ac{nonrtric} through the \texttt{A1 Mediator}.

\item[\acs{SDL}/\ac{STSL} Storage:] It provides xApps access to the shared
  database within the \ac{nearrtric}, facilitating read/write
  operations to persistent storage while abstracting the specific implementation
  details of the underlying database solution.
  It also handles the authentication and authorization processes
  to access the database, ensuring that xApps remain portable and
  stateless~\cite{sdl}.

\item[HTTP/\acs{REST}:] It provides xApps with REST callbacks for handling HTTP
  requests, allowing xApp developers to customize the response to Kubernetes'
  liveness and readiness probes according to their xApp's
  requirements~\cite{writersguide}. One can also create
  REST callbacks to expose internal information about the xApp's business
  logic and respond to external commands and parameters, allowing users
  to interact with   xApps directly via HTTP.

\item[E2 Subscription:] It allows xApps to obtain information from the RAN and
  control its operation. The xApps can subscribe to metrics and updates from a
  given set of E2 Nodes for post-processing or data analytics and control of the E2
  Nodes according to their business logic by triggering or passing parameters
  to the supported \ac{RAN} functions exposed via their \acp{SM}~\cite{e2sub}.

\end{description}

An xApp only requires a valid descriptor provided via the O1-CM
to operate, as it contains the required
initial configuration to enable the xApp's deployment and
interaction with additional \ac{nearrtric} components, if demanded. In addition,
the xApp descriptor instructs the \texttt{AppMgr} on how to
install the xApp, which involves fetching
the Docker images from an accessible Docker registry, configuring the Kubernetes
pod, and notifying other \ac{nearrtric} components of the creation of a new
xApp, e.g., \texttt{RtMgr} and \texttt{SubMgr}, to allow the new xApp to
leverage their capabilities~\cite{writersguide}.
We will detail the xApp lifecycle and how the \texttt{AppMgr} uses the xApp
descriptor further in Section~\ref{sub:life}.
The other APIs listed previously in this section are optional, e.g.,
\ac{SDL}/\ac{STSL}, \ac{RMR}, E2 Subscriptions, etc., meaning that xApp
developers can focus on learning and implementing only the interfaces  needed to
accomplish their intended business logic. In the following, we
detail how to design and define xApps, specifying container images, opening ports, and
configuring the APIs mentioned above.

\begin{lstlisting}[linewidth=\columnwidth,language=json,float,
  caption={xApp Descriptor Template.},label={lst:descriptor}]
{
  "name": "example_xapp",
  "version": "1.0.0",
  "vendor": "example_vendor",
  "containers": [
    // Configures Containers and Images.
    // Detailed in %Sec.~\ref{sub:conf_cont}%.
  ],
  "rmr": {
    // Configures RMR Messages.
    // Detailed in %Sec.~\ref{sub:conf_rmr}%.
  },
  "messaging": {
    "ports": [
      // Configures Ports per Container.
      // Detailed in %Sec.~\ref{sub:conf_port}%.
    ]
  },
  "controls": {
    // Optional Control Parameters.
    // Detailed in %Sec.~\ref{sub:conf_ctl}%.
  }
}
\end{lstlisting}

\subsection{Configuration}\label{sub:conf}

As part of the xApp development cycle, the xApp developers must design their
applications according to the intended business logic, and define them through the
creation of xApp descriptor and schema files. The former is a JSON file that
instructs the \ac{nearrtric} to deploy the given xApp, specifying \1 what is the name and version of the xApp, \2 what Docker container images it requires, and
the locations of their Docker Registries, \3 which ports must be open in each container, \4 what \ac{RMR} messages the xApp will publish and subscribe, \5 what A1 policies it will consume, and \6 what optional parameters the user can control.
The latter is a
JSON schema file that the \ac{nearrtric} uses to verify and validate the content
of the xApp descriptor before triggering the xApp deployment process, a process
which we will detail further in Section~\ref{sub:life}. Assuming all the
required Docker Registries and images are reachable, an xApp developer only needs
to share their xApp descriptor and schema files to distribute their
application~\cite{writersguide}.
However, there are current discussions and research efforts toward developing a store or marketplace to distribute xApps~\cite{kliks2023towards}. Therefore, the
xApp distribution process might change in the future.

The xApp descriptor follows the structure shown in Listing~\ref{lst:descriptor},
which contains the name, version, and vendor of the xApp to be deployed in the
\ac{nearrtric}. It is worth mentioning that the name and version are required
parameters that serve to identify the xApp inside the \ac{nearrtric} and
generate a unique name for the xApp's Kubernetes pod, detailed further in
Section~\ref{sub:dms}. In the following, we first detail the subsequent sections
of the xApp descriptor, shown in Listing~\ref{lst:descriptor},
and then examine the xApp schema.



\subsubsection{Containers and Images}\label{sub:conf_cont}

This mandatory xApp descriptor section defines the containers that compose the xApp
Kubernetes pod, as shown in Listing~\ref{lst:cont}. Each xApp contains at least
one Docker container, which can come from different images in different Docker
Registries.
For each container,
the xApp developer must specify \1 its name, which will serve as a unique
identifier used internally for opening ports and routing \ac{RMR} messages to
containers, and \2 the location of its Docker image, i.e., the URL of the
Docker Registry, the image name and its tag, which the \texttt{AppMgr} will use
to pull the image locally and instantiate the container. In addition to the
aforementioned required parameters, the xApp developer can specify the minimum
computing and memory resources each container requires to run and limit the maximum resource utilization. These optional
parameters ensure the xApp has access to the resources it requires
to run and cap the resource utilization on the \ac{nearrtric} cluster.

\begin{lstlisting}[linewidth=\columnwidth,language=json,float,
  caption={Section for configuring containers and images.},
  label={lst:cont}]
  ...
  "containers": [
    {
      "name": "example_container_1",
      "image": {
        "registry": "example.registry.com",
        "name": "example_image_1",
        "tag": "1.0.0"
      }
    },
    {
      "name": "example_container_2",
      "image": {
        "registry": "example.registry.com",
        "name": "example_image_2",
        "tag": "1.0.0"
      },
      "resources": {
        "requests": {
          "cpu": "1",
          "memory": "64Mi"
        },
        "limits": {
          "cpu": "2",
          "memory": "128Mi"
        }
      }
    }
  ],
  ...
\end{lstlisting}

\subsubsection{RMR Routing and Configuration}\label{sub:conf_rmr}

This optional xApp descriptor section defines the \ac{RMR} messages that the xApp
Kubernetes pod will transmit and receive, and the A1 policies it will consume
(all of which are optional), as shown in Listing~\ref{lst:rmr}.
The \ac{RMR} messaging operates in a publish-subscribe
paradigm. If leveraging the RMR interface or consuming A1 policies, the xApp
developer must specify the message types that their xApps
will consume and the message types they will produce to use \ac{RMR}. The
\texttt{RtMgr} will use this information for creating routing tables and
propagating them to other \ac{nearrtric} components and xApps.
Some message types are required to avail from certain functionality from the
\ac{nearrtric} components, e.g.,
the \texttt{RIC_HEALTH_CHECK_REQ} and \texttt{RIC_HEALTH_CHECK_RESP} are
required for reacting to \ac{RMR} health checks from the \texttt{RtMgr}. We will
discuss in-depth the \ac{RMR} functionality and explain the essential message
types for the operation of xApps later in Section~\ref{sub:mes}. If required for
a particular use case or deployment, the xApp developer can customize
the transport protocol and port for \ac{RMR}'s operation, the maximum
message buffer size, and the number of threads listening to incoming messages.
Furthermore, the xApp developer can specify a list of
policy IDs their xApp will consume. We will discuss the A1 policies and how xApps
can avail from them later in Section~\ref{sec:oper}.

\begin{lstlisting}[linewidth=\columnwidth,language=json,float,
  caption={Section for configuring RMR message routing.},
  label={lst:rmr}]
  ...
  "rmr": {
    "txMessages": [
      "A1_POLICY_RESP",
      "A1_POLICY_QUERY",
      "RIC_HEALTH_CHECK_RESP"
    ],
    "rxMessages": [
      "RIC_INDICATION",
      "A1_POLICY_REQ",
      "RIC_HEALTH_CHECK_REQ"
    ],
    "protPort": "tcp:4560",
    "maxSize": 2072,
    "numWorkers": 1
  "policies": [1]
  },
  ...
\end{lstlisting}

\subsubsection{Ports and Services}\label{sub:conf_port}

This optional xApp descriptor section defines the open ports and messages routed to each
container that composes the xApp Kubernetes pods, as shown in
Listing~\ref{lst:port}. Each container comprising the xApp Kubernetes pod can
have different communication interfaces and avail of distinct functionality from
\ac{nearrtric} components, which the xApp developer can specify by creating port
definitions that must contain an identifying name, the target container's name,
the port number, and a brief description. The \texttt{AppMgr} uses this
information to trigger Kubernetes for creating a service port mapped to the
corresponding container. For containers using HTTP/REST, port 8080 must be open
to support reacting to external input from users of the \ac{nearrtric} and
Kubernetes' liveness and readiness probes.
For a \ac{nearrtric} cluster using the default configuration,
each container leveraging \ac{RMR} or consuming A1 policies must
open \1 port 4061 to receive dynamic routing tables from the \texttt{RtMgr} and
learn where to route messages, and \2 port  4060 to receive messages from
\ac{nearrtric} components and other xApps.
The xApp developer has fine-grained control over the \ac{RMR} message routing
inside their xApp pod and must specify which messages will be produced and
consumed per container.

\begin{lstlisting}[linewidth=\columnwidth,language=json,float,
  caption={Section for configuring ports and services.},
  label={lst:port}]
  ...
  "ports": [
    {
      "name": "http",
      "container": "example_container_1",
      "port": 8080,
      "description": "HTTP service port"
    },
    {
      "name": "rmrroute",
      "container": "example_container_2",
      "port": 4561,
      "description": "RMR route port"
    },
    {
      "name": "rmrdata",
      "container": "example_container_2",
      "port": 4560,
      "rxMessages": [
        "RIC_INDICATION",
        "A1_POLICY_REQ",
      ],
      "txMessages": [
        "A1_POLICY_RESP",
        "A1_POLICY_QUERY",
      ],
      "policies": [1],
      "description": "RMR data port"
    }
  ]
  ...
\end{lstlisting}

\subsubsection{Optional Control Parameters}\label{sub:conf_ctl}

\begin{lstlisting}[linewidth=\columnwidth,language=json,float,
  caption={Section for configuring optional control parameters.},
  label={lst:controls}]
  ...
  "controls": {
    "rmr_routing_needed": false,
    "meid": "gnb123456",
    "ran_function_id": 1231,
    "action_definition": [
      11, 12, 13, 14, 15
    ],
    "action_id": 1,
    "action_type": "policy",
    "subsequent_action": {
      "subsequent_action_type": "continue",
      "time_to_wait": "w10ms"
      }
  },
  ...
\end{lstlisting}

This optional xApp descriptor section defines additional control parameters to
customize the operation of the xApp, as shown in Listing~\ref{lst:controls}. The
xApp developer can include an arbitrary number of xApp-specific parameters,
ranging from boolean values, strings, integer (or float) numbers, and arrays to
more complex JSON objects comprised of a combination of the data types listed
above.
The control parameters listed in Listing~\ref{lst:controls} serve to
parameterize the subscription to E2 Nodes, which we will explain in detail
further in Section~\ref{sec:ran}.
The xApp descriptor file is loaded into the
Kubernetes pod as a ConfigMap, which mounts the xApp descriptor as a JSON file
inside the containers' directory tree. The location of the xApp descriptor is
defined on the \texttt{XAPP_DESCRIPTOR_PATH} environment variable, which the
xApp can use to locate and load its content accordingly. However, the
xApp Frameworks automate these tasks and make the content of the xApp descriptor
readily available for the xApp developer to use as part of their business logic.
The parameters specified in this xApp descriptor section can be updated by the user during
runtime by editing the ConfigMap of the xApp Kubernetes pod (detailed in the next
section), which the xApp developer can use to customize certain aspects of their xApp.

\subsubsection{xApp Schema File}\label{sub:conf_schema}

\begin{lstlisting}[linewidth=\columnwidth,language=json,float,
  caption={xApp Schema Template.},
  label={lst:schema}]
{
  "$schema": "http:// json-schema.org/draft-07/schema#",
  "$id": "#/controls",
  "type": "object",
  "title": "Controls Section Schema",
  "required": [
  // List of required control parameters
  ],
  "properties": {
  // Properties of the required parameters
  }
}\end{lstlisting}

It is a JSON schema that annotates and validates the xApp
descriptor JSON file. Before triggering Kubernetes to instantiate the xApp pod,
the \texttt{AppMgr} verifies the content of the xApp descriptor against the xApp
schema to ensure the descriptor contains all the required parameters for
deploying the given xApp. The \texttt{AppMgr} comes preloaded with JSON schemas
to verify most of the required and optional sections of the xApp descriptor,
e.g., \textit{containers}, \textit{rmr}, and \textit{messaging} sections. The
only exception is the \textit{controls} section, which contains customized
optional control parameters. If the xApp descriptor has an empty
\textit{controls} section, the xApp schema is entirely optional.
However, if the xApp descriptor contains a non-empty \textit{controls} section,
the xApp developer must provide the \texttt{AppMgr} with a custom schema file to
verify and validate this section, as shown in Listing~\ref{lst:schema}, or that
will cause the xApp deployment to fail. The xApp schema contains the IETF JSON
Schema version, the ID of the xApp descriptor section that this schema
will verify (\textit{controls}), a list of the required control parameters that
the xApp descriptor must contain, followed by a list of their properties.
Should the xApp developer decide to include any required control parameters, they
must define their properties, as shown in Listing~\ref{lst:prop}, specifying
their ID (the URI from the root of the descriptor), their data type, their
default values, descriptive titles, and optionally, examples of possible
values.
The xApp schema can also contain no required control parameters and properties,
making all parameters optional, as shown in Listing~\ref{lst:schema}.
For completeness, we refer the reader to our online repository~\cite{repo},
where we include the entire xApp descriptor and schema files used as
examples in this section.

\begin{lstlisting}[linewidth=\columnwidth,language=json,float,
  caption={Schema for defining required control parameters.},
  label={lst:prop}]
  ...
  "required": [
    "meid",
    "ran_function_id",
    ...
  ],
  "properties": {
    "meid": {
      "$id": "#/properties/controls/items/ properties/meid",
      "type": "string",
      "default": "gnb123456",
      "title": "E2 Node Managed Entity ID",
      "examples": [
        "gnbABCDEF", "enbMNOPQR"
      ]
    },
    "ran_function_id": {
      "$id": "#/properties/controls/items/ properties/ran_function_id",
      "type": "integer",
      "title": "E2 Node RAN Function ID",
      "default": "1231"
    }
  }
  ...
\end{lstlisting}

\section{xApp Management: Controlling its Lifecycle}\label{sec:man}

In this section, we detail the xApp lifecycle, how to create and publish Docker
Images from the xApp's source code, and how to interact
with the \ac{nearrtric} to manage xApps, teaching xApp developers to
onboard, install, query, and uninstall xApps in their \ac{O-RAN} development environment.

\begin{figure*}[t]
  \centering
  \includegraphics[width=0.99\textwidth]{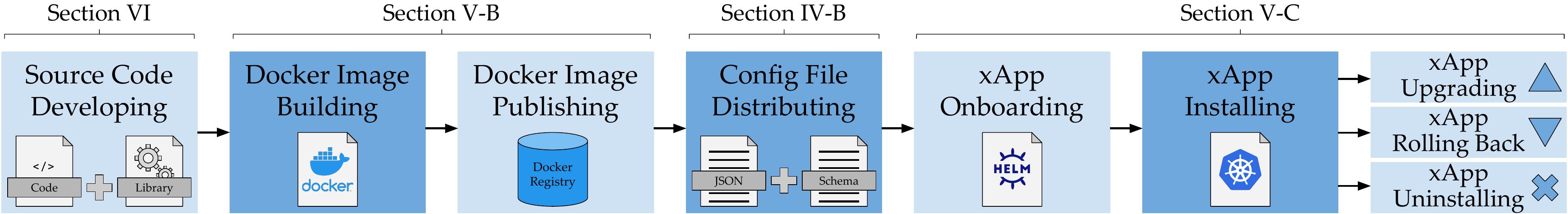}
  \caption{The stages of the xApp lifecycle, from
    software development to container generation, configuration file sharing,
    and pod execution in a testing or production \ac{nearrtric} cluster.
    The xApp developer may need to repeat these steps several times during the
    xApp development cycle to test, debug, and validate their xApp, before their
    xApp is ready for public release.
}
  \label{fig:lifecycle}
\end{figure*}

\subsection{The xApp Lifecycle}\label{sub:life}

From an implementation perspective, xApps are specialized applications
leveraging xApp Framework libraries, distributed to \acp{MNO} as universal
Docker images, and instantiated as Kubernetes pods running on a \ac{nearrtric}
cluster~\cite{writersguide}. As such, there are several steps in the xApp
lifecycle, from the initial ideation and software development to container
creation and publishing, and finally, to xApp deployment and execution inside a
\ac{nearrtric}~\cite{polese2022understanding}, as shown in
Fig.~\ref{fig:lifecycle}. In theory, the role of the xApp developer would end after
publishing their xApp in a Docker Registry and distributing their xApp's
descriptor and schema files publicly or to the intended \acp{MNO}. However, in
a practical setting, the xApp developer will most likely need to test, debug,
and validate the xApp in their own \ac{O-RAN} development environment, which can
include, but is not limited to, a testing \ac{nearrtric} cluster for testing the
deployment and operation of xApps, as well as a real or a
simulated~\cite{e2sim} E2 Node to evaluate the xApp's interaction with E2
Nodes for validating its business logic. Therefore, the xApp developer must know
how to interact with the \ac{nearrtric} to manage xApps. In this tutorial, we
cover the entire xApp development process and instruct the xApp developer
throughout all steps of the xApp lifecycle, introduced below.

\begin{description}
\item[Source Code Developing:]
The xApp developer writes the source code that implements its intended business
logic. We will go into detail about the implementation of xApps and
their available APIs further in Section~\ref{sec:oper}.

\item[Docker Image Building:]
The xApp developer prepares a Dockerfile with instructions to build Docker
image(s), specifying the complete environment to run the xApp source
code, including directories and dependencies.

\item[Docker Image Publishing:]
After creating Docker Image(s), the xApp developer pushes them
to a (local or remote) Docker Registry so the \ac{nearrtric} can fetch
the image(s) to create container(s) and instantiate the xApp.

\item[Config File Sharing:]
 With the location of the Docker image(s), i.e., the Docker Registry's URL, the
 image name, and its tag, the xApp developer includes them in the xApp descriptor,
 and shares both the descriptor and schema files with the intended \acp{MNO} to
 distribute the xApp. 

\item[xApp Onboarding:]
In possession of the xApp descriptor and schema files, the xApp developer (or
the \ac{nearrtric}'s users) onboards the xApp into the \ac{nearrtric},
generating Helm Charts stored in the Local Helm Chart
Repository~\cite{onboarding}, which we detail later in this section.

\item[xApp Installing:]
After the xApp is onboarded and its charts are in the  Local Helm Chart
Repository, the xApp developer (or the \ac{nearrtric}'s users) can install the
xApp, triggering the creation of its Docker container(s) and instantiation of the
xApp pod, 
as well as the registration of the xApp with
the \texttt{AppMgr} discussed later in Section~\ref{sec:oper}.

\item[xApp Upgrading/Rolling Back:]
Once the xApp pod is running, the xApp developer (or the \ac{nearrtric}'s
users) may upgrade it or roll it back to a different version.
After onboarding the different xApp descriptor and schema files, upgrading (or rolling back) the xApp
will uninstall its pod and subsequently install the newer (or previous) version
of the xApp.

\item[xApp Uninstalling:]
Once the xApp pod is running, the xApp developer
(or the \ac{nearrtric}'s users) may uninstall it, releasing its
computational resources and de-registering it with the \texttt{AppMgr}, which
terminates the xApp's subscriptions and RMR endpoints, discussed later in
Section~\ref{sec:oper}.

\end{description}

We discussed the design and distribution of xApp descriptor and schema files
in the previous section, and will dive deep into the xApp source code development
later in Section~\ref{sec:oper}.
In the next subsections, we overview the remaining steps of the xApp lifecycle,
detailing the Docker image building and publishing, as
well as the xApp management operations to onboard, install, query, and uninstall
xApps inside the \ac{nearrtric} cluster through interactions with the
\texttt{AppMgr}.


\subsection{Creating and Publishing xApp Docker Containers}\label{sub:doc}

The xApp developer prepares a Dockerfile
during the xApp development process, a text file that specifies the complete
environment for running the xApp source code. The Dockerfile serves to build a
Docker image file, a read-only snapshot containing the several layers that
constitute a live Docker container~\cite{merkel2014docker}. For the xApp
developer, creating an xApp Docker image serves two purposes. First, it
aggregates all the source code, packages, and directory structures required to
run the xApp into a single file, which can be uploaded into an accessible Docker
Registry, facilitating the xApp developer to distribute its xApp to interested
parties, e.g., \acp{MNO}, in a scalable and stateless manner. Second, it allows
the xApp developer to instantiate a container from the said image, either
standalone or in an xApp pod inside a \ac{nearrtric} cluster, and interact with
a live instance of their compiled source code for development and debugging. For
simplicity, we focus on the latter throughout this tutorial, i.e., instantiating
xApp containers in pods inside a \ac{nearrtric}.

We outline the structure of a Dockerfile in
Listing~\ref{lst:dockerfile_min_example}. It contains a series of basic commands that
the xApp developer can use to create the container environment, detailed below.
For the complete list of Dockerfile commands and
syntax, we refer the reader to the official Docker documentation
on~\cite{dockerfilereference}.

\begin{description}
  \item[\texttt{FROM}:] Defines the base image that we will modify in this
    Dockerfile, e.g., the certain release of a Linux distribution or the
    development environment of a Python version~\cite{dockerhubofficial}.
\item[\texttt{ARG}:] Creates a temporary variable for use in
  the Dockerfile, useful for scripting and controlling parameters used in
  multiple commands, e.g., file paths and package versions.
\item[\texttt{RUN}:] Runs a Linux shell command in the Docker image file
  system, serves to modify system settings, installs required dependencies,
  and compiles libraries and binaries.
\item[\texttt{COPY}:] Copies files and directories from the host machine of the
  xApp developer to the container image, useful to copy their repositories,
  source code files, and datasets.
\item[\texttt{ENV}:] Defines a Linux environment variable that will persist when
  the Docker container is instantiated from the resulting image; it serves to
  specify configuration file locations and pass parameters to the xApp running
  in this container.
\item[\texttt{CMD}:] The last command in a Dockerfile; it specifies the Linux
  shell command that will be run when the Docker container starts,
  we use it to start our xApp binary.

\end{description}

The xApp developer will likely need to customize their Dockerfiles using the
aforementioned commands according to
the requirements, business logic, and dependencies of their xApps.
For completeness, we refer the reader to our online repository~\cite{repo},
where we include the entire Dockerfile used to create the
Docker image containers for running the Python xApps used throughout
this tutorial.


\begin{lstlisting}[linewidth=\columnwidth,language=dockerfile,float,
caption={Basic structure of a Dockerfile with the steps to build a Docker image for
running an xApp written in Python.},
label={lst:dockerfile_min_example}]
# Start by building from a base image
FROM python:3.8-alpine

# Create temporary variable with a path
ARG dir=/tmp

# Run shell command to install dependencies
RUN apk update && apk add gcc musl-dev bash

# Copy files from host machine to the image
COPY src/ ${dir}/src
COPY init/ ${dir}/init
COPY setup.py ${dir}/

# Install the Python xApp
RUN pip3 install ${dir}

# Set location of xApp configuration file
ENV CONFIG_FILE=${dir}/config_file.json

# Starting the container running our xApp
CMD run-xapp
\end{lstlisting}

In possession of a Dockerfile, the xApp developer can use its location as an
argument to create a Docker image, as shown in Listing~\ref{lst:docker_build_push}.
The \texttt{docker build} command sequentially executes the instructions in the
Dockerfile and, when completed, generates a single file containing the snapshot
of the container. In addition to the Dockerfile, the \texttt{docker
build} command requires \1 the hostname and port of a private Docker
Registry, either local or remote~\cite{dockertag}, \2 a name for
the xApp container image, and \3 an associated tag, i.e., a custom
human-readable identifier that typically refers to the version or variant of an
image. For the container to gain access to the host network, e.g., to
clone repositories or install packages, the xApp developer may need to include
the "\texttt{-{}-network host}" argument. We refer the reader to the
official Docker Build documentation~\cite{dockerbuild} for additional
information on the \texttt{docker build} command.
After the xApp developer builds a Docker image, the next step is to publish it
to a Docker Registry, so that the
container image can be fetched by the \ac{nearrtric} and deployed as an xApp.
The \texttt{docker push} command uploads the local Docker image into
the chosen private Docker Registry, either local or remote, using the Docker
Registry's hostname and port, as well as the xApp container name and tag used
during the build phase.

\begin{lstlisting}[linewidth=\columnwidth,language=bash,float,
  caption={Command for building and pushing an xApp image.},
  label={lst:docker_build_push}]
# Build xApp image with a name and tag
docker build <DOCKERFILE_PATH> -t \
<REGISTRY_HOSTNAME>:<REGISTRY_PORT>/ <XAPP_NAME>:<XAPP_TAG> \
--network host

# Example using a local Docker Registry
docker build . -t \
localhost:5001/test_xapp:1.0.0 \
--network host

# Push the xApp image to Docker Registry
docker push \
<REGISTRY_HOSTNAME>:<REGISTRY_PORT>/ <XAPP_NAME>:<XAPP_TAG>

# Example using a local Docker Registry
docker push \
localhost:5001/example_xapp:1.0.0
\end{lstlisting}

Should the xApp developer decide to set up their own local Docker Registry
inside the \ac{nearrtric} cluster of their \ac{O-RAN} development environment for
testing and debugging xApps, they can use Docker's official open-source
registry, which on itself runs as a Docker container, as shown in
Listing~\ref{lst:docker_registry}.
The \texttt{docker run} command instantiates the Docker Registry image
(\texttt{registry:2}) as a container,
where \1 the "\texttt{-d}" flag indicates the registry will run as a daemon in
the background, \2 the \texttt{-p} flag maps an internal port from the container
(which, in this case, listens to port 5000) to an arbitrary host port, and \3
the \texttt{-{}-restart} flag specifies the conditions in which the container
will restart automatically. The xApp developer will likely want their local
Docker Registry to restart automatically upon system restarts or failures,
hence, the "\texttt{unless-stopped}" option.
In addition, the xApp developer can specify a name for its new Docker Registry
container.
By default, this Docker Registry is publicly accessible locally, but we can make
it remotely accessible and restrict access using passwords or certificates. We refer the reader to the official Docker Registry
documentation~\cite{dockerregistry} for additional information.

After the xApp developer pushes the Docker image(s) of their xApp to a Docker
Registry, and updates the xApp configuration file to include the image
location, i.e., the Docker Registry's URL, the image name and its tag,
they are ready to onboard the xApp into the \ac{nearrtric}, which we
will discuss in the next subsection. The xApp developer can
also use the commands shown in Listing~\ref{lst:docker_check} to inspect
the Docker images stored locally or available from a Docker Registry.

\begin{lstlisting}[linewidth=\columnwidth,language=bash,float,floatplacement=tbp,
  caption={Command for creating a local Docker Registry.},
  label={lst:docker_registry}]
# Run a self-restarting Docker Registry
docker run -d -p <REGISTRY_PORT>:5000 \
--restart unless-stopped \
--name <REGISTRY_NAME> registry:2

# Example using port 5001
docker run -d -p 5001:5000 \
--restart unless-stopped \
--name registry registry:2
\end{lstlisting}

\begin{lstlisting}[linewidth=\columnwidth,language=bash,float,floatplacement=tbp,
  caption={Commands for querying available Docker images.},
  label={lst:docker_check}]
# Check Docker images stored locally
docker image ls

# Query the available images in a Registry
curl -X GET http://<REGISTRY_HOSTNAME>: <REGISTRY_PORT>/v2/_catalog

# Example of query to a local Registry
curl -X GET http://localhost:5001/v2/_catalog
\end{lstlisting}



\begin{figure}[t]
  \centering
  \includegraphics[width=0.99\columnwidth]{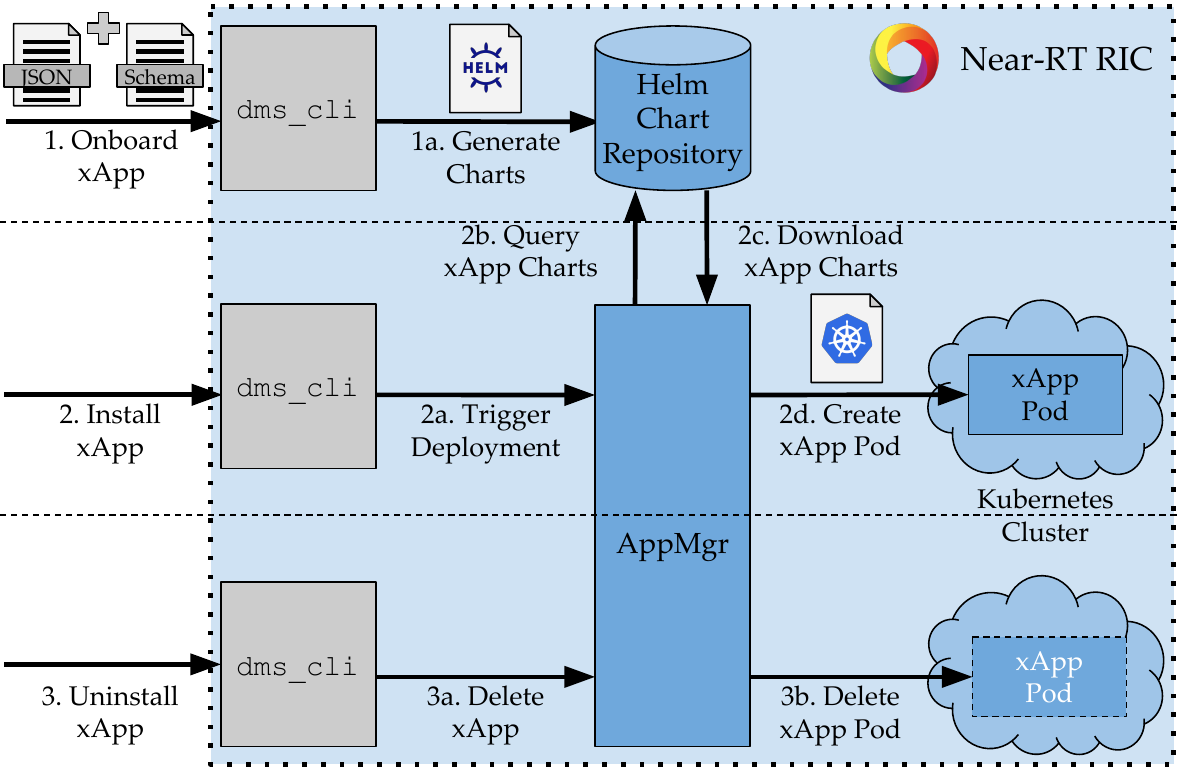}
  \caption{
    Interactions between components of the \ac{nearrtric} to
    perform operations related to the management of xApps inside the
    \ac{nearrtric},
    intermediated by the \texttt{dms_cli}.
}
  \label{fig:dms_cli}
\end{figure}

\subsection{Interfacing with the AppMgr via the \texttt{dms_cli}}\label{sub:dms}

In possession of an xApp configuration and schema files, including the
location of the Docker Image(s), the xApp developer is ready to onboard their
xApp, or any of the publicly available first- and third-party xApp listed in
Section~\ref{sub:table}, into a \ac{nearrtric}.
The onboarding process, as well as the other operations related to the
management of xApps by the \texttt{AppMgr}, are intermediated by an application
provided by the \ac{OSC} called \texttt{dms_cli}~\cite{ricinstallationguide}, which stands for \ac{DMS}
\ac{CLI}. The \texttt{dms_cli} is a command line tool for
deploying xApps and managing their lifecycle in a \ac{nearrtric}, as
illustrated in Fig.~\ref{fig:dms_cli}. We detail below the operations related to
xApp management intermediated by the \texttt{dms_cli}.

\subsubsection{xApp Onboarding}

This process collects the information required
to deploy an xApp and stores it locally as Helm Charts for later use.
First, the \texttt{dms_cli} validates the descriptor file against the schema
file, as shown in Listing~\ref{lst:dms_onboarding}, and throws errors if there
are missing parameters or invalid formatting. If the validation succeeds, then
the \texttt{dms_cli} uses the content of the descriptor file to generate Helm
Charts, which define the xApp pod's open ports, computational resources, and
environment variables (as discussed in Section~\ref{subsub:tech}), and are
stored in a Local Helm Chart Repository available to the \texttt{AppMgr}.
After onboarding is complete, the \texttt{name} and
\texttt{version} contained in the descriptor file serve as identifiers for the
Helm Charts stored in the local repository.

\subsubsection{xApp Installing}
This process triggers the \texttt{AppMgr} to deploy an xApp Kubernetes pod using
the Helm Charts stored locally during the xApp onboarding.
The \texttt{dms_cli} passes the name and version of the xApp Helm Chart,
alongside the Kubernetes namespace for xApps (defined during the \ac{nearrtric}
installation, defaulting to "ricxapp") to the \texttt{AppMgr}, as shown in
Listing~\ref{lst:dms_install}. Then, the \texttt{AppMgr} queries the Local Helm
Chart Repository
to download the xApp Helm Charts and use
the information therein for creating the xApp pod. The \texttt{dms_cli}
throws errors if the \texttt{AppMgr} cannot locate the Helm Chart or the
correct version.
Next, during the instantiation of the xApp pod, Kubernetes uses the location of
the xApp Containers in the xApp Helm Chart
to fetch the xApp images and instantiate the xApp Kubernetes pod.
If the xApp Kubernetes pod is successfully instantiated, there is an additional
step where the xApp registers with the \texttt{AppMgr} to avail of the features
and capabilities of the \ac{nearrtric}; we detail this further in
Section~\ref{sec:oper}.
There might be issues preventing pod instantiation, e.g., Kubernetes cannot
fetch the images, reach the Docker Registry location, or the cluster lacks
computational resources. However, these errors are not automatically shown to
the user as the result of running this command.
Instead, the users must debug the Kubernetes
deployment themselves to understand the reason for failure, e.g.,
using the \textit{health_check} command detailed later in this section, or other
alternatives described further in Section~\ref{sec:debug}.

\begin{lstlisting}[linewidth=\columnwidth,language=bash,float,floatplacement=tbp,
  caption={\texttt{dms_cli} command for onboarding an xApp.},
  label={lst:dms_onboarding}]
# Onboard xApp to generate its chart
dms_cli onboard <CONFIG_JSON> <SCHEMA_JSON>
# Or
dms_cli onboard \
--config_file_path=<CONFIG_JSON> \
--schema_file_path=<SCHEMA_JSON>

# Example of an onboarding command
dms_cli onboard \
xapp_path/init/config_file.json \
xapp_path/init/schema_file.json
\end{lstlisting}

\begin{lstlisting}[linewidth=\columnwidth,language=bash,float,
  caption={\texttt{dms_cli} command for installing xApps.},
  label={lst:dms_install}]
# Install an xApp on the Near-RT RIC
dms_cli install <XAPP_CHART_NAME> \
<VERSION> <NAMESPACE>
# Or
dms_cli install \
--xapp_chart_name=<XAPP_CHART_NAME> \
--version=<VERSION> \
--namespace=<NAMESPACE>

# Example of an install command
dms_cli install example_xapp 1.0.0 ricxapp
\end{lstlisting}

\begin{lstlisting}[linewidth=\columnwidth,language=bash,float,
  caption={\texttt{dms_cli} command for uninstalling xApps.},
  label={lst:dms_uninstall}]
# Uninstall an xApp from the Near-RT RIC
dms_cli uninstall <XAPP_CHART_NAME> \
<NAMESPACE>
# Or
dms_cli uninstall \
--xapp_chart_name=<XAPP_CHART_NAME> \
--namespace=<NAMESPACE>

# Example of an uninstall command
dms_cli uninstall example_xapp ricxapp
\end{lstlisting}

\subsubsection{xApp Uninstalling}
This process triggers the \texttt{AppMgr} to stop the execution of a given xApp
pod and release all of its resources, as shown in
Listing~\ref{lst:dms_uninstall}. The \texttt{dms_cli} throws errors if the
\texttt{AppMgr} cannot locate the given xApp or if it is not running. First, the \texttt{AppMgr} instructs Kubernetes to delete the xApp pod,
which sends a terminating signal (SIGTERM) to the pod and puts it in a
\textit{Terminating} state.
Then, Kubernetes grants the xApp pod 30 seconds (by default) to exit gracefully,
after which the pod is forcefully deleted. During this period, the xApp must
de-register with the \texttt{AppMgr}, which informs the \ac{nearrtric} components
to remove or release resources associated with it; we detail this further in
Section~\ref{sec:oper}. The xApp pod can also use this period to perform
additional operations before stopping, e.g., saving cached information to the
\ac{SDL}/\ac{STSL}. After the grace period, the xApp pod is deleted and its
resources are released.

\subsubsection{xApp Upgrading and Rolling Back}

This pair of operations, upgrading and rolling back, allow the xApp developer or
the user of the \ac{nearrtric} to change the version of a running xApp. They can
be useful for deploying new bug fixes or reverting to a previous stable version
of an xApp, respectively. The \texttt{dms_cli} commands for upgrading and
rolling back xApps that combine the previous uninstall and
install commands. They use the name of the xApp, its old current version, the
new intended version, and the xApp namespace, as shown in
Listing~\ref{lst:dms_up_down}, to trigger AppMgr to carry uninstall and install
operations in succession.
In that regard, one could use the \texttt{dms_cli} to perform these operations
manually, but these commands allow these processes to be partially automated.
Similar to the install and uninstall commands, the \texttt{dms_cli} will throw
errors if it cannot find the given xApp, if the xApp is not running, or if it
cannot locate its name or intended new version in the Local Helm Chart Repository.

\begin{lstlisting}[linewidth=\columnwidth,language=bash,float,
  caption={\texttt{dms_cli} commands to up/downgrade xApps.},
  label={lst:dms_up_down}]
# Upgrade an xApp to a new version
dms_cli upgrade \
--xapp_chart_name=<XAPP_CHART_NAME> \
--old_version=<OLD_VERSION> \
--new_version=<NEW_VERSION> \
--namespace=<NAMESPACE>

# Example of an upgrade command
dms_cli upgrade \
--xapp_chart_name=example_xapp \
--old_version=1.0.0 --new_version=1.1.0 \
--namespace=ricxapp

# Roll back an xApp to a previous version
dms_cli rollback \
--xapp_chart_name=<XAPPI_CHART_NAME> \
--new_version=<NEW_VERSION> \
--old_version=<OLD_VERSION> \
--namespace=<NAMESPACE>

# Example of a rollback command
dms_cli rollback \
--xapp_chart_name=example_xapp \
---old_version=1.1.0 --new_version=1.0.0 \
--namespace=ricxapp
\end{lstlisting}

\begin{lstlisting}[linewidth=\columnwidth,language=bash,float,
  caption={\texttt{dms_cli} commands to check useful information.},
  label={lst:dms_aux}]
# Check health of Helm Chart Repository
dms_cli health

# Query list of onboarded xApps
dms_cli get_charts_list

# Check the health of an xApp pod
dms_cli health_check \
--xapp_chart_name=<XAPP_CHART_NAME> \
--namespace=<NAMESPACE>

# Download the xApp Helm Charts
dms_cli download_values_yaml \
--xapp_chart_name=<XAPP_NAME> \
--version=<VERSION> \
--output_path=<OUTPUT_PATH>

# Override xApp Helm Chart's values.yaml
dms_cli install <XAPP_CHART_NAME> \
<VERSION> <NAMESPACE> \
--overridefile <VALUES_PATH>

\end{lstlisting}

In addition to the operations related to the management of xApps listed
above, the xApp developer, or \ac{nearrtric}'s users, can leverage the
\texttt{dms_cli} to perform a number of other useful operations for
querying the status of the onboarded and installed xApps, as
well as checking the health of the Local Helm Chart Repository or
xApp pods, as shown in Listing~\ref{lst:dms_aux}. We detail these additional
operations below.

\subsubsection{Checking the Health of the Local Helm Chart Repository}
This operation checks whether the \texttt{dms_cli} can successfully communicate
with the Local Helm Chart Repository, whose location is defined by the
\texttt{CHART_REPO_URL} environment variable in the \ac{nearrtric}
cluster~\cite{writersguide}. This operation is useful to ensure the
\ac{nearrtric} cluster works as it should and that the Local Helm Repository
is operational.

\subsubsection{Querying Onboarded xApps}
This operation lists all the onboarded xApps whose charts are stored in the
Local Helm Chart Repository. The \texttt{dms_cli} lists the xApp charts' names,
API versions, creation times, descriptions, hashes for validating their
integrity, and the location of their Charts, displayed as JSON strings.
This is helpful for identifying missing versions or misspelled
names in case xApp installations fail.

\subsubsection{Checking the Health of xApp Pods}
This operation checks the deployment status of an xApp, serving as an approach
to verify whether the instantiation was successful. The \texttt{dms_cli} uses
the xApp's chart name and its namespace to check whether all the containers are ready
and initialized, and whether the pod is scheduled and initialized, throwing
errors if the pod is not running correctly. We discuss other strategies to
assess the deployment of xApps later in Section~\ref{sec:debug}.

\subsubsection{Downloading and Modifying xApps Helm Charts}
This operation allows one to override the Helm Chart used to
instantiate the xApp Kubernetes pod before its deployment. This operation is useful for
customizing internal parameters according to the \ac{MNO}'s requirements or
performing quick tests without the need to modify the xApp's descriptor file and
onboarding them again.
First, the \texttt{dms_cli} downloads the "values.yaml" file of the Helm Chart,
as discussed in Section~\ref{sub:OSC}, using the name and version of the xApp
chart, as well as an output path to save the file. Then, the xApp developer or
the \ac{nearrtric} users can modify "values.yaml" file saved locally
according to their requirements. Finally,
one can use the \textit{install} command with an optional flag that loads the
modified "values.yaml" and overrides the Helm Chart stored in the
\ac{nearrtric}.


With these commands at their disposal, the xApp developer or the
\ac{nearrtric}'s are ready to manage xApps throughout their entire lifecycle.
In addition, they can perform a number of operations for testing
and debugging the deployment of xApps on a \ac{nearrtric} cluster, which will
be very useful during the xApp development process discussed in the next section.

\section{xApp Implementation: Realizing your Ideas}\label{sec:oper}


The Python xApp Framework~\cite{pythonframework} provides two types of xApp
implementations that differ regarding their approach to treating \acf{RMR}
messages:
\1 the reactive xApp, known as \texttt{RMRXapp}, is passive and only acts in
response to incoming \ac{RMR} messages, and \2 the general xApp, known as
\texttt{Xapp}, can implement any business logic and act upon any desired
criteria.
Both xApp implementations 
import
libraries for using the \ac{nearrtric} interfaces, e.g., \ac{RMR},
\ac{SDL}, and REST, provide methods for abstracting interactions with
\ac{nearrtric} components, and automatically register xApps with the
\texttt{AppMgr}, simplifying the xApp development.

In this section, we dive deep into the interfaces and functionality
available for the xApp implementations, such as messaging, 
policies, data storage, 
and external input,
accompanied by examples leveraging the Python xApp
Framework.

\subsection{Messaging}\label{sub:mes}

In the following, we explain how xApps can communicate with one another and the
components of the \ac{nearrtric}. First, we detail the operation of the \ac{RMR}
library, the \ac{RMR} routing table, and route resolution via the \texttt{RtMgr}.
Then, we introduce the two classes of xApps regarding their treatment of
\ac{RMR} messages. Next, we detail the APIs for creating callbacks to
receive, reply, and send \ac{RMR} messages.

\subsubsection{RMR Library, Routing Table, and Route Resolution}

The \ac{nearrtric}'s \ac{RMR} messaging infrastructure allows its components and
the running xApps to communicate without knowing each other's IP
addresses and open ports, which can be subject to changes as their Kubernetes
pods are scaled or redeployed. Each \ac{nearrtric} component and xApp leverages
the \ac{RMR} messaging library, which abstracts the connection establishment and
routing decisions from their business logic. The library operates by forwarding
messages to an endpoint (their destination) based on the message type
(\texttt{mtype}) and subscription ID (\texttt{subid}) contained in the message;
these fields are referred to together as the message key.
The \texttt{mtypes} are named values that identify the purpose of the message
and must be chosen according to the API of the desired endpoint.
For example, for reacting to policies, an xApp must send an A1 policy query
(\texttt{A1_POLICY_QUERY}) to the \texttt{A1 Mediator}, and later acknowledge a
response with an A1 policy response (\texttt{A1_POLICY_RESP}), as shown earlier
in Section~\ref{sub:conf_rmr} in Listing~\ref{lst:rmr}.
Each \texttt{mtype} has a numeric value, and the full list of supported
\texttt{mtypes} and their associated numeric values can be found in the \ac{RMR}
repository~\cite{message_types}.
The names of \texttt{mtypes} the xApp can transmit and receive must be
specified in its descriptor file, as shown in Listing~\ref{lst:port},
so \texttt{RtMgr} can create routes for their respective numeric values.
Conversely, the \texttt{subid} are integers generated by the
\texttt{SubMgr} during runtime when subscribing to E2 Nodes, which we detail later
in Section~\ref{sub:e2}.

\begin{lstlisting}[linewidth=\columnwidth,language=rt,float,
caption={Structure of an \ac{RMR} routing table with \texttt{mse} and
\texttt{rte} entry record types, showing their mandatory (between
chevrons) and optional fields (between brackets and chevrons).},%numbers=none,
label={lst:routing_table}
]
newrt| start|[<table_name>]
mse|<mtype>[,<sender_endpoint>]|<subid>|   <dest_endpoint>[<[,][;]>            <dest_endpoint>...] [| %meid]
rte|<mtype>[,<sender_endpoint>]|           <dest_endpoint>[<[,][;]>            <dest_endpoint>...] [| %meid]
...
newrt|end|[<route_counter>]
\end{lstlisting}

The \ac{RMR} decides how to forward outgoing messages according to the
information from the xApp's own \ac{RMR} routing table, which defines the
desired endpoints for each message key. This table can be \1 defined statically,
loaded once from a file during the xApp's instantiation, and \2 updated
dynamically, with constant updates from the \texttt{RtMgr} whenever a new xApp
or \ac{nearrtric} component starts~\cite{rmrguide}.
The xApp developer can define their static \ac{RMR} route table to specify
what \texttt{mtypes} their xApp will produce and with whom it will communicate,
i.e., which \ac{nearrtric} components and other xApps.
During the xApp instantiation, the \ac{RMR} library loads a static route table
from the path defined by the \texttt{RMR_SEED_RT} environment variable, which
can be set in the Dockerfile, as shown in Section~\ref{sub:doc}.

The \ac{RMR} routing table file possesses a standard and well-defined structure, as shown in
Listing~\ref{lst:routing_table}. It contains mandatory header and footer lines delimiting
its start and end, which can include an optional table name for identification
and a counter for the number of route entries used to parse the table's
integrity, respectively. In addition, the table can contain any number of
entries that specify the routes for each message key, known as entry records.
There are two types of entry records, the \texttt{mse} and \texttt{rte}.
The \texttt{mse} defines: an \texttt{mtype}, an optional sender application, a
\texttt{subid}, and at least one destination endpoint.
The \texttt{subid} is
only used for \ac{RMR} messages based on subscriptions, which we will detail
later in Section~\ref{sub:sub}.
For routes unrelated to subscriptions from the \texttt{SubMgr}, one must use
the \texttt{subid} $-1$.
The \texttt{rte} is a deprecated type of entry record and may be removed from
\ac{RMR} in future releases~\cite{rmrlib}. In this context, the \ac{OSC} advises xApp
developers to use only \texttt{mse} entry records for new xApps. However, we
can still find several occurrences of \texttt{rte} entries in existing
\ac{nearrtric} components and xApps, so we present it here for completeness. The
\texttt{rte} defines: an \texttt{mtype}, an optional sender application, and at
least one destination endpoint. It does not support subscriptions, and hence,
operates the same way as an \texttt{mse} entry record with the \texttt{subid}
$-1$.
Furthermore, we show in Listing~\ref{lst:example_table} a realistic example of
a static route table for an xApp that listens to policies from the \texttt{A1
Mediator} and communicates with two other xApps using custom \texttt{mtypes}.
We detail how to obtain the endpoints of existing \ac{nearrtric}
components and running xApps later in Section~\ref{sec:debug}.

\begin{lstlisting}[linewidth=\columnwidth,language=rt,float,
caption={Example of an xApp's \ac{RMR} routing table file, configured to send A1
policy query (20011) and response (20012) messages to the \texttt{A1
Mediator}, messages with custom \texttt{mtypes} (30001 and 30002)
to two other xApps, and a subscription control request message (12040) with a
\texttt{subid} 200 to the entity that owns the E2 Node, i.e., the
\texttt{E2Term}.
},%numbers=none,
label={lst:example_table}
]
newrt|start
mse|20011|-1|service-ricplt-a1mediator-rmr.ricplt
mse|20012|-1|service-ricplt-a1mediator-rmr.ricplt
mse|30001|-1|service-ricxapp-A-rmr.ricxapp
mse|30002|-1|service-ricxapp-B-rmr.ricxapp
mse|12040|200| %meid
newrt|end
\end{lstlisting}

\begin{lstlisting}[linewidth=\columnwidth,language=rt,float,
caption={Example of the different approaches for sending
messages to multiple destination endpoints. Endpoints separated by semicolons
receive copies of all messages, while endpoints separated by commas are cycled
in round robin.
},%numbers=none,
label={lst:distribution}
]
newrt|start
mse|mtype_1|subid_1|                       dest_endpoint_A;dest_endpoint_B
mse|mtype_2|subid_2|                       dest_endpoint_M,dest_endpoint_N
mse|mtype_3|subid_3|                       dest_endpoint_X;dest_endpoint_Y_1,dest_endpoint_Y_2;dest_endpoint_Z
newrt|end
\end{lstlisting}

The \ac{RMR} library also supports sending messages to a group of multiple
destination endpoints using two message distribution approaches: \1 fanout, where
each destination endpoint receives a copy of the outgoing message, which is useful
to broadcast information to multiple xApps; or \2 round-robin, where messages are
cycled to one endpoint at a time, which is useful for load balancing across
multiple xApps.
To accomplish this, the \ac{RMR} introduces the concept of endpoint groups, each
of which can contain one or more endpoints. The \ac{RMR} messages are distributed
in fanout to multiple endpoint groups, which are separated by semicolons, and each
group will receive copies of all messages. Moreover, the \ac{RMR} messages are
distributed in round-robin to the endpoints comprising an endpoint group, which are
separated by commas, and successive messages will be cycled between endpoints.
Listing~\ref{lst:distribution} shows examples of the two message distribution
approaches and how they can be combined to send messages in more complex
manners. For example, the third entry record in
Listing~\ref{lst:distribution} will fanout every message to \texttt{dest_endpoint_X} and
\texttt{dest_endpoint_Z}, and round-robin the same messages between
\texttt{dest_endpoint_Y_1} and \texttt{dest_endpoint_Y_2}.

\begin{figure}[t]
  \centering
  \vspace{-1em}
  \begin{subfigure}[t]{0.45\columnwidth}
    \includegraphics[width=\textwidth,trim={0 0.0em 0 0em},clip]{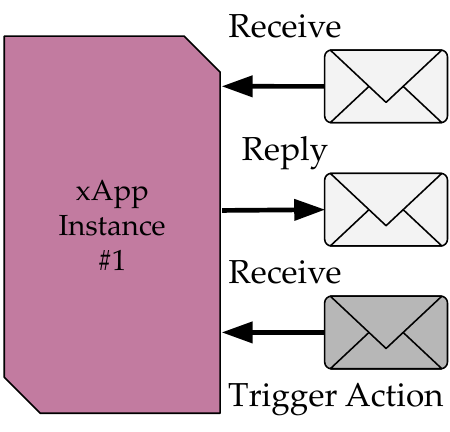}
    \caption{Reactive xApp.}
    \label{fig:xapp_reactive}
  \end{subfigure}
  \hfill
  \begin{subfigure}[t]{0.45\columnwidth}
    \includegraphics[width=\textwidth,trim={0 0.0em 0 0em},clip]{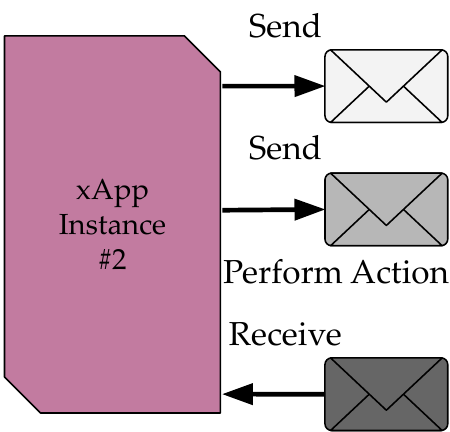}
    \caption{General xApp.}
    \label{fig:xapp_general}
  \end{subfigure}
  \caption{Example of differences between reactive and general xApps. The
    reactive xApps can only perform tasks triggered by receiving RMR messages,
    whereas the general xApps can support any desired business logic and performing
    actions in any order, including sending messages to trigger other xApps.
  }
  \label{fig:xapp_msg_types}
\end{figure}

The \ac{RMR} allows the selection of the destination endpoint based on the
\ac{MEID} contained in the \ac{RMR} message instead of selecting the
endpoint on the matching entry records~\cite{rmrlib}. When routing using
\ac{MEID}, the \ac{RMR} message
is sent to the endpoint that owns the managed entity.  To use \ac{MEID}
routing, one or more route table entry records must contain the special endpoint
name \texttt{\%meid} instead of a list of destination endpoints, as shown in
Listing~\ref{lst:example_table}.
This feature is
particularly useful in the context of subscriptions for routing messages to E2
Nodes. In this case, the \texttt{E2Term} owns the E2 Nodes and intermediates all
their communications, and the special entry records are created automatically by
the \texttt{SubMgr} in conjunction with \texttt{RtMgr}
(detailed later in Section~\ref{sec:ran}).

After the xApp initialization and RMR loading the static route
table file, the xApp's route table is updated periodically by the
\texttt{RtMgr}. These updates happen through the \texttt{rmrroute} port opened
in the xApp descriptor file, as shown in Listing~\ref{lst:port}. The
\texttt{RtMgr} populates the xApps' routing tables with information about the
accepted \texttt{mtypes} and existing endpoints of \ac{nearrtric} components and
other running xApps. Every time a new xApp is registered with the
\texttt{AppMgr} (detailed in the next section), the \texttt{AppMgr} informs the
\texttt{RtMgr} about the new endpoints. Then, the \texttt{RtMgr} propagates this
information to existing \ac{nearrtric} components and xApps.
The \ac{RMR} stashes the additional routes updated
during runtime on the same directory where the static table route file is
located, with an added \textit{.stash} extension, to facilitate debugging.

\begin{lstlisting}[linewidth=\columnwidth,language=python,float,
caption={Structure of the \texttt{RMRXapp} class constructor.},
label={lst:rmrxapp_init}
]
# Initialize the xApp
def __init__(self):
  # RMRXapp Class constructor
  self._rmr_xapp = RMRXapp(
    <default_message_handler>,
    config_handler=<config_handler>,
    post_init=<post_init_method>,
    rmr_port=<RMR_port>
  )
\end{lstlisting}

\subsubsection{Reactive and General xApps}\label{sub:rea_gen}

The Python xApp Framework contains two xApp
implementations that only differ regarding their treatment of RMR message,
as illustrated in Fig.~\ref{fig:xapp_msg_types}.
On the one hand, the \texttt{RMRXapp} provides a more straightforward starting
point for xApp developers, leveraging custom callbacks to trigger different
actions and reply to \ac{RMR} messages, e.g., controlling the E2 Nodes
based on new policies from the \texttt{A1 Mediator} or storing information on
\ac{SDL} based on messages from other xApps.
On the other hand, the
\texttt{Xapp} is more versatile and allows the development of more complex
xApps, e.g., deciding to send multiple \ac{RMR} messages to xApps and
\ac{nearrtric} components, promptly interfacing with
\ac{SDL} or the \ac{RAN} and when to listen to incoming \ac{RMR} messages, which
comes at the cost of being more involved and requiring attention to
detail from the xApp developer.

\begin{lstlisting}[linewidth=\columnwidth,language=python,float,
caption={Example of the \texttt{config_handler} function, checking for
required parameters before starting the xApp.},
label={lst:config_handler}
]
# Called when xApp descriptor file changes
def config_handler(self, rmr_xapp, config):
  # Check for missing parameters
    if "flag" not in config["controls"]:
      raise ValueError('Missing parameter')

  # Load the new configuration data
  rmr_xapp._config_data = config
\end{lstlisting}

From an implementation standpoint, the \texttt{RMRXapp} operates in a loop,
listening to and handling incoming \ac{RMR} messages with callbacks, and
checking for changes in the xApp descriptor file.
It also automatically replies to health checks and de-registers itself
with the \texttt{AppMgr} and gracefully exits when the xApp process is terminated.
The \texttt{RMRXapp} requires the xApp developer to specify: \1 a default \ac{RMR}
message callback to handle incoming messages, \2  a configuration handler to
load and sanitize the xApp configuration file, \3 a post-initialization function
that will be called after the xApp class is initialized, and \4 the port that the
\ac{RMR} library will listen to (defaults to 4060), as shown in
Listing~\ref{lst:rmrxapp_init}. The configuration  handler loads the
content of the xApp descriptor file into the xApp, as shown in
Listing~\ref{lst:config_handler}. This method is called when the xApp
starts running and whenever the configuration file is modified (either by the
\ac{nearrtric} users manually or by editing the xApp pod's ConfigMap). The xApp
developer can leverage this method to sanitize its configuration, check for
missing parameters, and log information. The post-initialization function serves
to instantiate objects and create class attributes available in the
\ac{RMR} message callbacks, e.g., logging objects (detailed later in
Section~\ref{sec:debug}) or data structures shared between callbacks, as shown
in Listing~\ref{lst:rmrxapp_post_init}.
Finally, the \texttt{RMRXapp} allows the xApp developer to create a default
\ac{RMR} message handler, serving as a catch-all for all unregistered
\texttt{mtypes}, and register specialized \ac{RMR} message handlers for
responding to specific \texttt{mtypes}, as shown in
Listing~\ref{lst:rmrxapp_register_callbacks}. We detail how to create
callbacks to receive, reply, and send \ac{RMR} messages in the next section.

\begin{lstlisting}[linewidth=\columnwidth,language=python,float,floatplacement=tbp,
caption={Example of a \texttt{post_init} function, creating class attributes and
instantiating objects shared between callbacks.},
label={lst:rmrxapp_post_init}
]
# Function called after the constructor
def _post_init(self, rmr_xapp):
  # Create a class attribute
  rmr_xapp.callback_counter = 0

  # Set the log level of the xApp
  rmr_xapp.logger.set_level(Level.DEBUG)
\end{lstlisting}

\begin{lstlisting}[linewidth=\columnwidth,language=python,float,
caption={Example on how to register RMR message callbacks
for handling different \texttt{mtypes} using the \texttt{RMRXapp}.},
label={lst:rmrxapp_register_callbacks}
]
# Register custom RMR callback handlers
self._rmr_xapp.register_callback( <custom_message_handler>, <mtype>)

# Examples of custom handlers
self._rmr_xapp.register_callback(
   self._message_handler, 30002
)
self._rmr_xapp.register_callback(
   self._policy_request_handler,
   A1_POLICY_REQ
)
\end{lstlisting}

\begin{lstlisting}[linewidth=\columnwidth,language=python,float,
caption={Structure of the \texttt{Xapp} class constructor.},
label={lst:xApp_init}
]
# Initialize the xApp
def __init__(self):
  # Xapp Class Constructor
  self._xapp = Xapp(
    <entrypoint_function>,
    rmr_port=<RMR_port>
  )
  # Potential flag to control xApp shutdown
  self.shutdown = False
\end{lstlisting}

\begin{lstlisting}[linewidth=\columnwidth,language=python,float,
caption={Example of the \texttt{entrypoint} function, opening and loading the
xApp descriptor file, and then checking if \ac{RMR} and \ac{SDL} are operational
to remain operational, sending \ac{RMR} messages, and listening to
incoming messages in a loop.},
label={lst:xApp_entrypoint}
]
# Function called after the constructor
def _entrypoint(self, xapp):
  # Set log level
  self._xapp.logger.set_level(Level.DEBUG)
  # Load configuration file
  self._xapp._config_data = load(
    open(self._xapp._config_path))

  # Loop while not set to shutdown
  while not self.shutdown:
    # Health check the RMR and SDL
    if not xapp.healthcheck():
      # Oops, something is going wrong
      xapp.logger.error(
        "Healthcheck failed. Terminating.")
      # Let us stop the xApp here
      self.shutdown = True

    # Do anything you like!
    xapp.rmr_send(<payload_1>,<mtype_1>)
    xapp.rmr_send(<payload_2>,<mtype_2>)

    # Check for incoming messages
    for (summary, msg_buf) in xapp.rmr_get_messages():
      # Log the received message
      xapp.logger.info("Msg:"+str(summary))

      # Dispatch mtypes to custom callbacks
      if summary[rmr.RMR_MS_MSG_TYPE] == 30002:
        self._message_handler(
          xapp, summary, msg_buf
        )
      elif summary[rmr.RMR_MS_MSG_TYPE] == A1_POLICY_REQ:
        self._policy_request_handler(
          xapp, summary, msg_buf
        )

    # Sleep for a while
    sleep(1)
\end{lstlisting}

The \texttt{Xapp} implementation provides only the minimal core functionality
for the operation of xApps, requiring the xApp developer to implement most of the
procedures automated and abstracted by the \texttt{RMRXapp}. Nevertheless,
it gives the xApp developer more control to implement any desired business logic.
The \texttt{Xapp} requires the xApp developer to specify \1
an \texttt{entrypoint} function that will be called after the \texttt{Xapp}
class is initialized, and \2 the port that the \ac{RMR} library will listen to
(defaults to 4060), as shown in Listing~\ref{lst:xApp_init}.
The \texttt{entrypoint} method is the only function that the \texttt{Xapp} will execute, and hence, the xApp developer must use it to implement their
business logic.
For example, setting the log level (detailed later in Section~\ref{sec:debug}),
opening and loading the xApp configuration file, and creating their own loop
with any custom actions, e.g., performing an \ac{RMR} and \ac{SDL} health check,
sending messages to two other xApp or \ac{nearrtric} components and then listening
to incoming \ac{RMR} messages, as shown in Listing~\ref{lst:xApp_entrypoint}.

It is worth mentioning that both xApp implementations automate the registration
of xApp with the \texttt{AppMgr}, a critical step for xApps to work correctly
and interface with \ac{nearrtric} components after being
instantiated~\cite{writersguide}. In
this process, the xApp \1 generates its \ac{RMR} and HTTP endpoints according to
their names, namespace, and interface types, \2 locates the
\texttt{AppMgr} exposed Kubernetes services, and \3 forwards its name, version,
namespace, \ac{RMR}, and HTTP endpoints, as well as its configuration in JSON
format to the \texttt{AppMgr}. In possession of this information, the
\texttt{AppMgr} notifies
other \ac{nearrtric} components of the new xApp, provides them
with the new endpoints for establishing communication with the xApp, and
notifies the xApp that it is ready to work.
The \texttt{RMRXapp} handles being terminated and
automatically triggers its de-registration process with the
\texttt{AppMgr}, which removes references to the given xApp and its endpoints
from all \ac{nearrtric} components. However, the \texttt{Xapp} expects the xApp
developer to handle the de-registration themselves.
\emph{Failure to de-register the xApp will leave broken references and endpoints on
the \ac{nearrtric} components, leading to undefined behavior and preventing
a new instance of that xApp from working correctly until the
\ac{nearrtric} cluster is restarted.}
We explain how the xApp developer can automatically trigger the de-registration
of their xApps in Section~\ref{sec:good}.

\begin{lstlisting}[linewidth=\columnwidth,language=python,float,
caption={Structure of the methods available to xApps for sending, receiving and
replying to RMR messages.},
label={lst:rmr_base}
]
# Returns the queue of received messages
summaries, msg_bufs = xapp.rmr_get_messages()

# Reply to received message reusing buffer
rmr_xapp.rmr_rts(<msg_buf> [,new_payload=<payload>] [,new_mtype=<mtype>] [,retries=<n_retries>])

# Send an RMR message w/ custom payload
xapp.rmr_send(<payload>, <mtype> [,retries=<n_retries>])

# Free memory allocated to message buffer
rmr_xapp.rmr_free(<msg_buf>)
\end{lstlisting}

\begin{lstlisting}[linewidth=\columnwidth,language=python,float,
caption={Information included in RMR message summary.},
label={lst:rmr_summary}
]
# Payload data
summary[rmr.RMR_MS_PAYLOAD]
# Payload length
summary[rmr.RMR_MS_PAYLOAD_LEN]
# Subscription ID
summary[rmr.RMR_MS_SUB_ID]
# Transaction id (send or reply)
summary[rmr.RMR_MS_TRN_ID]
# Status (ok or not ok)
summary[rmr.RMR_MS_MSG_STATUS]
# Error if not ok
summary[rmr.RMR_MS_ERRNO]
# Managed Entity ID
summary[rmr.RMR_MS_MEID]
\end{lstlisting}

\subsubsection{Communicating using RMR}

The Python xApp Framework provides methods for receiving, replying, and sending
\ac{RMR} messages, as shown in Listing~\ref{lst:rmr_base}. The \ac{RMR} messages
contain a payload in the form of JSON-compatible Python objects, e.g.,
dictionaries, strings, floats, etc., accompanied by a \texttt{mtype}. The
\ac{RMR} library stores the message data as \textit{bytes}, and hence, the sent payloads
must be encoded as \textit{UTF-8} strings. Conversely, the received payloads must be
decoded from \textit{UTF-8} strings. In possession of the payload and
\texttt{mtype}, the \ac{RMR} library: \1 allocates a message buffer to store the
\ac{RMR} message; \2 generates the message metadata, e.g., length,
status, etc., and stores it on the message buffer; and \3
forwards a copy of the message buffer to its destination \ac{RMR}
endpoint based on the \ac{RMR} routing table~\cite{rmrguide}.

Both xApp implementations abstract and automate the
creation of a threaded \ac{RMR} server for listening to incoming \ac{RMR}
messages and storing the received \ac{RMR} messages in a queue. Therefore, the xApp
developer only needs to
check for the presence of new messages and potentially parse them according to
their \texttt{mtypes} to select the correct callback for handling them. When
received, the \ac{RMR} messages contain a summary dictionary containing their data
and metadata, whose fields are shown in Listing~\ref{lst:rmr_summary}, and the raw message buffer where the message was stored.
After receiving an \ac{RMR} message, the xApp developer can either \1 reuse the
allocated message buffer to create a reply to the same sender, which may or may
not contain the same \texttt{mtype}, or \2 free the memory allocated for the
message buffer if they have no further use for it, which prevents memory
leaks~\cite{rmrlib}.

We can combine the methods for receiving, replying, and sending \ac{RMR}
messages to create communication protocols between xApps, as shown in
Listing~\ref{lst:rmr_example}. For example, an xApp developer may create
xApps that send information between each other, replying with an acknowledgment
confirming the reception of the messages akin to TCP, or xApps that first
manipulate the received data in some manner before returning the results to the
original sender. For more information on creating chains of xApps that
communicate via \ac{RMR}, we refer the reader to \ac{O-RAN}'s anomaly detection
use case, which employs three xApps working together
to detect anomalous \acp{UE} accessing the \ac{RAN}~\cite{anomalydetection}.

\begin{lstlisting}[linewidth=\columnwidth,language=python,float,
caption={Example on how to combine RMR methods to create a custom
communication  protocol between xApps.},
label={lst:rmr_example}
]
# All payloads must be encoded in UTF-8
xapp.rmr_send("hi".encode(), 30001, retries=5)
xapp.rmr_send(str(3.14).encode(), 30002)

# Let us iterate over the received messages
for (summary, msg_buf) in xapp.rmr_get_messages():
  # Create a new serializable payload
  new_payload = dumps({"my_key": "my_val"})

  # Reply to received msg w/ new payload
  rmr_xapp.rmr_rts(msg_buf, retries=10,
  new_payload=new_payload.encode())

  # Clear the msg_buf after we use it
  rmr_xapp.rmr_free(msg_buf)
\end{lstlisting}

\begin{lstlisting}[linewidth=\columnwidth,language=python,float,
caption={Example of the creation of callback functions for handling RMR
messages and executing desired operations.
},
label={lst:rmr_handler}
]
# Example of a default message callback
def _default_message_handler(self, xapp, summary, msg_buf):
  # Logging incoming message types
  xapp.logger.info(
  "Handler called for mtype: " + str(summary[rmr.RMR_MS_MSG_TYPE])
  )
  # Logging incoming message contents
  xapp.logger.debug(
  "Message content: " + str(summary[rmr.RMR_MS_PAYLOAD])
  )

  # Modify internal class parameter
  rmr_xapp.callback_counter += 1

  # Return an acknowledgement
  xapp.rmr_rts(msg_buf,
    new_payload="ack".enncode()
  )
  # Free allocated memory
  xapp.rmr_free(msg_buf)
\end{lstlisting}

The xApp developer can encapsulate the steps for handling messages and creating
communication protocols in callback functions, as shown in
Listing~\ref{lst:rmr_handler}. The
creation of callbacks is a requirement for the \texttt{RMRXapp}, which relies on
registered callbacks to operate. However, the creation of callbacks is an optional software design approach
for the \texttt{Xapp}, which can support the message handling directly inside
its \texttt{entrypoint} method.
For the \texttt{RMRXapp} implementation, these functions are automatically called any time the
xApp receives an \ac{RMR} message with the corresponding registered
\texttt{mtype}, as shown earlier in Listing~\ref{lst:rmrxapp_register_callbacks}.
For the \texttt{Xapp} implementation, the xApp developer must include a mechanism to parse
received messages by their \texttt{mtype} and then call the corresponding
callback, as shown earlier in Listing~\ref{lst:xApp_entrypoint}.
These callback functions receive as arguments: \1 a pointer to the class
instance where the callback was defined; \2 a pointer to the xApp
implementation being used, which is useful for accessing its internal information
and functionality; \3 the \ac{RMR} message summary dictionary, which includes
the message data and metadata; and \4 the raw \ac{RMR} message buffer. In
possession of these arguments, the xApp developer can implement any
business logic leveraging the interfaces and functionality available to the
xApp.

\subsection{Policies}\label{sub:pol}

In the following, we explain how xApps can listen to and take actions based on
policies from the \ac{nonrtric}. We detail the structure of an A1 policy and
outline the steps to enable support for policies in an xApp, describing how to
create handlers and callbacks for reacting to A1 policies.

\subsubsection{A1 Interface between Non- and Near-RT RIC}

\begin{lstlisting}[linewidth=\columnwidth,language=json,float,
  caption={Structure of the \ac{RMR} message payload from an A1 policy
that an xApp will receive from the \texttt{A1
Mediator}.},label={lst:policy_rmr}]
{
  "payload": {
    <policy_payload>
  },
  "policy_type_id": <policy_id>,
  "policy_instance_id": <policy_instance_id>,
  "operation": <operation>
}
\end{lstlisting}

The rApps in the \ac{nonrtric} can generate policies for steering the behavior of
the \ac{nearrtric} and the xApps therein. These policies contain high-level
intents, allowing xApps to decide how to interpret and act upon them.
In the \ac{nonrtric}, the A1 policies are expressed in JSON, following a specific
syntax validated through a JSON schema~\cite{polese2022understanding}. The
\texttt{A1 Mediator} serves as a
northbound interface toward the \ac{nonrtric}, translating the A1 policies
received via the A1-AP interface in JSON to the \ac{RMR} format used for
internal communication in the \ac{nearrtric}~\cite{a1mediator}.
After this translation, the \texttt{A1 Mediator} publishes policies in \texttt{RMR}
format to the xApps that have registered to receive policies of that given type.
Finally, the xApps receive and handle A1 policies in the same way they
receive other \ac{RMR} messages.

The A1 policies received via \ac{RMR} possess a predefined
\ac{RMR} message payload structure, as shown in Listing~\ref{lst:policy_rmr}.
These are Python dictionaries that contain the following fields:

\begin{description}
  \item[\texttt{payload}:]
A Python dictionary containing JSON-compatible objects. It contains the
high-level information, parameters, or flags generated by rApp in the
\ac{nonrtric} for controlling the operation of xApps in the \ac{nearrtric}.

  \item[\texttt{policy_type_id}:]
 An integer identifying the type of A1 policy that
xApps will listen to and defines the template of the policy, i.e., the fields
of the payload dictionary and its types and ranges of accepted values.

  \item[\texttt{policy_instance_id}:]
A string identifying a concrete realization of a given A1 policy, complete
with values. Instances of a given policy type will always contain the same
structure but may contain different values.

  \item[\texttt{operation}:]
A string defining the operation being performed. It can be
either "\texttt{CREATE}" when the xApp starts running or a new policy instance is
deployed; "\texttt{UPDATE}" when the policy instance is updated with new values
in its payload dictionary; or
"\texttt{DELETE}" when the policy instance is removed from the \texttt{A1 Mediator}.
\end{description}

The xApp developer may use the information about the policy instance
types, the values in its payload, and the current operation to steer the
operation of their xApps as they see fit. For example, the xApp developer
can create xApps that react to values from policies to change the signal
strength threshold for handovers~\cite{trafficsteering} or
switch the scheduling algorithm of base stations on the fly.

\begin{lstlisting}[linewidth=\columnwidth,language=json,float,
  caption={Example of an A1 Policy Handler.},label={lst:policy_json}]
def _policy_request_handler(self, xapp, summary, msg_buf):
  # Clear message buffer
  self._rmr_xapp.rmr_free(msg_buf)

  try:
    # Get JSON string encoded as bytes
    req = json.loads(
      summary[rmr.RMR_MS_PAYLOAD])

  except (json.decoder.JSONDecodeError, KeyError):
    self.logger.error("Invalid JSON")
    return

  # Check mandatory policy keys
  policy_keys = ["policy_type_id", "operation", "policy_instance_id"]
  if not all(key in policy_keys for key in req.keys()):
    self.logger.error("Invalid policy")
    return

  # Do anything you like!

  # Construct response
  req["handler_id"] = self._rmr_xapp._config_data["name"]
  req["status"] = "OK"
  del req["operation"]

  # Convert dict. to JSON string in UTF-8
  self._xapp.rmr_send(json.dumps(resp). encode(), A1_POLICY_RESP)
\end{lstlisting}


\subsubsection{Handling A1 Policies}

There are a few steps required to enable support for A1 policies on an xApp.
First, the xApp developer must edit the xApp descriptor file and include the
following \texttt{mtypes} in the \ac{RMR} configuration section: \1 The
\texttt{A1_POLICY_REQ}  on the list of received messages to
obtain \ac{RMR} messages with A1 policies, \2 the \texttt{A1_POLICY_RESP}
 on the list of transmitted messages to reply to the A1 policy
with an acknowledgment, and optionally \3 the \texttt{A1_POLICY_QUERY}
 also on the list of transmitted messages to query all existing
instances of a given policy type, as shown earlier in Listing~\ref{lst:rmr}.
Then, the xApp developer must list the policy type identifiers for all A1
policies of interest for the xApp pod on the \ac{RMR} configuration section, and the specific policy type identifiers on the ports and services section for
the containers that will handle each A1 policy,
as shown in Listing~\ref{lst:port}. Such separation allows the xApp
developer to use different containers to handle distinct A1 policies. Next, the
xApp developer must edit the static route table file of their xApp for routing
the \texttt{A1_POLICY_RESP} (and optionally the \texttt{A1_POLICY_QUERY})
messages to the \texttt{A1 Mediator}, as shown in
Listing~\ref{lst:example_table}.
Finally, the xApp is ready to receive \ac{RMR} messages containing
A1 policies from the \texttt{A1 Mediator}.

To handle A1 policies, the xApp developer must create a policy callback and
register it according to its xApp implementation
(either directly with the \texttt{RMRXapp} or manually in the \texttt{Xapp}, as
shown in Listing~\ref{lst:xApp_entrypoint}).
We show an example of such a policy callback function in
Listing~\ref{lst:policy_json}, where we first check the validity of the JSON
data structure and the integrity of the A1 policy (whether it contains the
required dictionary keys). Then, we can make any decisions according to the
consent of the A1 policy, as detailed in the previous section. Next, we must
send an acknowledgment to the \texttt{A1 Mediator} in the form
of an \texttt{A1_POLICY_RESP} message. The A1 Mediator expects a response with
the same policy type and instance identifiers from the \texttt{A1_POLICY_REQ},
as well as the name of the xApp that consumed the A1 policy and the return
status of this operation, which can be an \texttt{OK} to indicate success or
\texttt{ERROR} to indicate failure in consuming the A1 policy.
Therefore, we can reuse part of the RMR payload from the \texttt{A1_POLICY_REQ} message
and adapt it accordingly.

\begin{lstlisting}[linewidth=\columnwidth,language=python,float,
caption={The SDL API calls available for xApps in Python to leverage
the \ac{nearrtric}'s internal relational database.
% and perfom data storage, conditional and search operations.
},
label={lst:sdl_basic_commands}
]
# Reads value for a given key
xapp.sdl.get(ns, key)

# Writes a key value entry
xapp.sdl.set(ns, key, val)

# Deletes key and value entry
xapp.sdl.delete(ns, key)

# Writes key value if it does not exist
xapp.sdl.set_if_not_exists(ns, key, val)

# Updates value if old value matches search
xapp.sdl.set_if(ns, key, old_val, new_val)

# Removes entry if value matches search
xapp.sdl.delete_if(ns, key, val)

# Find keys starting with prefix
xapp.sdl.find_keys(ns, prefix)

# Find keys starting w/ prefix and get
# their associated values
xapp.sdl.find_and_get(ns, prefix)

\end{lstlisting}

\subsection{Storage}\label{sub:sto}

In the following, we explain the Shared Layer functionality that xApps can
leverage to store data within the \ac{nearrtric}, detail the APIs available for
xApps to read, write, modify, and delete information from persistent storage,
and describe the \ac{NIB} Databases.

\subsubsection{Share Layer Abstraction}

Storing data in the \ac{nearrtric} can be useful for \1 saving and retrieving
the application state, which will persist if the xApp gets updated, rolled back,
crashes, or reboots, \2 performing data analytics over long-term metrics, and \3
transferring large amounts of data between xApps.
However, each \ac{nearrtric} instance and Kubernetes deployment can have
different configurations for their data storage backends, e.g., different
credentials and authorization mechanisms, database software from various vendors
(incurring in different APIs), and distinct database architectures, e.g.,
distributed, redundant, or load-balancing databases.
Thus, the \ac{OSC} created the Shared Layers to handle the
actual data storage while providing a unified, flexible interface to xApp,
abstracting the specific implementation from the current database backend,
which allows xApps to be stateless and portable across different
\acp{nearrtric}~\cite{sdl}. There are two types of Shared Layers: the \1
\ac{SDL}, supporting structured databases, which organize data using 
keys and namespaces, and the \2 \ac{STSL}, supporting
time-series databases, which organize data sequentially with associated time
stamps. At the time of writing, \ac{STSL} support is limited and only available
for xApps implemented in Go~\cite{stsl}. While this is expected to change in the
future, we will refrain from detailing its API as it has yet to become
available in the Python xApp Framework.

\subsubsection{Storing Data using SDL}

The \ac{SDL} structures data according to keys, values, and namespaces.
Keys and values operate similarly to Python dictionaries or
JSON key-value pairs, where each data entry has a human-readable tag associated
with a value. Namespaces encapsulate the data, attributing an identifier for a
group of keys and their values~\cite{sdl}. Each xApp can use one or more namespaces to
identify its persistent data and store as many keys and values as the
underlying database backend capacity allows. By using distinct namespaces, we
can isolate data between different xApps and \ac{nearrtric} components.
Conversely, using the same namespaces enables us to share data across xApps and
\ac{nearrtric} components.

An xApp can leverage the \ac{SDL} Library, part of the Python xApp Framework,
to avail from the capabilities of this Shared Layer to read/write persistent
data on the \ac{nearrtric}.
The SDL Library offers xApps with an API for manipulating data on a given
namespace, as shown in Listing~\ref{lst:sdl_basic_commands}.
Through this API, xApps can perform \1 traditional data storage operations,
e.g., reading values associated with keys, writing new keys and values, and
deleting keys and their associated values; \2 conditional operations, e.g.,
writing new keys and values if no entries with these keys exist, updating keys
and values according to their existing values, and deleting keys-value pairs
if their keys and values match a search; and \3 search operations, e.g.,
retrieving existing keys (or keys and their associated values) that start with
a prefix (which can be an empty prefix for obtaining all the existing keys).
Moreover, we can see examples of how xApps can leverage \ac{SDL} namespaces to
isolate and share data between each other in Listing~\ref{lst:sdl_share_example}.

\begin{lstlisting}[linewidth=\columnwidth,language=python,float,
caption={Example of how xApps can use multiple SDL namespaces to isolate
and share data with each other, also showing how xApps can manipulate data
across namespaces.},
label={lst:sdl_share_example}
]
# On the First xApp:
# ------------------
# Writes an entry on its own namespace
xapp_1.sdl.set("xapp_1_ns", "gnb_meid", "gnbABCDEF")

# Writes on shared namespace if new entry
xapp_1.sdl.set_if_not_exists("shared_ns", "ue_list_240101_123836", ue_list)

# On the Second xApp:
# -------------------
# Tries to access key not in this namespace
xapp_2.sdl.get("xapp_2_ns", "gbn_meid")
# This API call will return None

# Finds key used in the shared namespace
ue_key = xapp_2.sdl.find("shared_ns", "ue_list")

# Reads value from the shared namespace
ue_list = xapp_2.sdl.get("shared_ns", ue_key)

# Deletes entry from shared namespace
xapp.sdl.delete("shared_ns", ue_key)

\end{lstlisting}


\subsubsection{User Equipment and Radio \acp{NIB}}

The \ac{nearrtric} contains two databases for storing information
about the \ac{RAN}: The \1 \ac{RNIB} database contains information about the E2
nodes, their supported \acp{SM}, and \ac{RAN} functions (detailed later in
Section~\ref{sub:sm}), and the \2 \ac{UENIB} contains information about the
associated \acp{UE}, their identity and reported
metrics~\cite{kuklinski2020ran}. The \ac{RNIB} is populated by the
\texttt{E2Mgr} whenever new base stations set up an E2 connection, serving as an
inventory of \ac{RAN} elements connected to the \ac{nearrtric}. Meanwhile, the
\ac{UENIB} contains identifying tags for associated \acp{UE}, enabling the
\ac{nearrtric} and its xApps to make user-centric decisions at the cost
of storing potentially sensitive information about
users~\cite{polese2022understanding}. At the time of writing, the \ac{UENIB}
exists in the \ac{O-RAN} specifications, but there are no current efforts to
develop it in the \ac{OSC} or enable it in the xApp frameworks.
For more information about the current implementation of the \ac{UENIB}, we refer
the reader to the SD-RAN documentation~\cite{UENIB}.


\begin{lstlisting}[linewidth=\columnwidth,language=python,float,
caption={The \ac{RNIB} API calls available for xApps in Python to find
information about the currently connected base stations.},
label={lst:rnib_api}
]
# Gets list of all base stations
xapp.GetListNodebIds()

# Gets list of all gNodeBs
xapp.get_list_enb_ids()

# Gets list of all eNodeBs
xapp.get_list_gnb_ids()

# Get detailed info about base station
xapp.GetNodeb(<inventory_name>)

# Gets definition of RAN functions
xapp.GetRanFunctionDefinition( <inventory_name>, <ran_function_oid>)
\end{lstlisting}

\begin{lstlisting}[linewidth=\columnwidth,language=python,float,
caption={Example of how to find a base stations's inventory name from the
\ac{RNIB}, and use it to find detailed information about the base station and
its supported \ac{RAN} functions.},
label={lst:rnib_example}
]
# Returns list gNodeBs
for gnb in xapp.get_list_gnb_ids():
    print("gNodeB:", gnb)

# Get the name of the last gNodeB
gnb_name = gnb.inventory_name

# Get detailed info about that gNodeb
print(xapp.GetNodeb(gnb_name))

# Get its RAN Function definition
print(xapp.GetRanFunctionDefinition( "gnb_734_733_16b8cef1", "OID123"))
\end{lstlisting}

The Python xApp Framework offers an API for accessing information stored in the
\ac{RNIB}, allowing xApps to find the list of base stations currently
connected to the \ac{nearrtric} (either eNodeBs, gNodeBs, or both), as shown in
Listing~\ref{lst:rnib_api}. The base stations are stored in the \ac{RNIB} using
an inventory name generated by the \texttt{E2Mgr}, which serves to identify
the E2 Nodes in \ac{O-RAN}. We can use the inventory
name to obtain detailed information about a particular base station, such as \1
its type and connection status, \2 the \ac{PLMN} ID and gNodeB
ID, which identify the \ac{MNO} and the base station~\cite{zhao2022multi},
respectively, \3 certain base station configurations, e.g., its
associated AMF, and \4 its supported \ac{RAN} functions (detailed later in
Section~\ref{sub:sm}). We provide an example of how to find the
inventory name of a base station and use it to obtain detailed information
about its \ac{RAN} Functions in Listing~\ref{lst:rnib_example} (detailed in
Section~\ref{sec:ran}).

\begin{lstlisting}[linewidth=\columnwidth,language=python,float,
caption={Example of the creation and configuration of a threaded HTTP server for
listening to requests inside an xApp.},
label={lst:http_server_setup}
]
# Create HTTP server to listen to requests
self.server = xapp_rest.ThreadedHTTPServer( <address>, <port>)

# Example of a server listening to
# requests from any host on port 8080
self.server = xapp_rest.ThreadedHTTPServer( "0.0.0.0", 8080)
\end{lstlisting}

\begin{lstlisting}[linewidth=\columnwidth,language=python,float,
caption={Example of the creation of a handler
to serve incoming HTTP requests and
implement a REST call.},
label={lst:http_handler}
]
# Create handler for requests on a URI
self.server.handler.add_handler( self.server.handler, <HTTP_request_type>, <REST_call_name>, <URI>, <callback_method>)

# Example of a REST method to get config
self.server.handler.add_handler( self.server.handler, "GET", "config", "/ric/v1/config", self.configGetHandler)
\end{lstlisting}

\subsection{External Input}\label{sub:ui}

In the following, we explain how xApps can respond to external input using their
REST interface. First, we detail how to enable support for REST
and handle HTTP requests. Then, we describe how to respond to
probes and user interactions.

\subsubsection{REST Interface, HTTP Server, Handlers, and Callbacks}


The xApps have an optional REST interface, allowing them to respond to external
input aside from the pods of Near-RT RIC components. It serves two purposes: \1
allows xApps to react to Kubernetes' readiness and liveness probes, indicating
their operational status; and \2 allows xApps to support interactions from the
users of the \ac{nearrtric}, which can be useful for obtaining information about
the internal state of the xApps and passing control
parameters~\cite{writersguide}. REST is a widely
popular interface for web-based applications, which maps HTTP requests acting on
exposed URI endpoints onto internal RPC calls, allowing remote hosts to query
information, execute functions, and pass parameters to a local server over HTTP.
For brevity, we refer the reader to~\cite{rodriguez2008restful} for
additional information on the RESTful paradigm and the operation of REST calls
running on top of HTTP requests.

\begin{lstlisting}[linewidth=\columnwidth,language=python,float,
caption={The structure of REST handlers and examples of how to respond to
Kubernetes probes.},
label={lst:probe_callback}
]
# Structure of a generic REST handler
def example_rest_handler(self, name, path, data, ctype):
  # Method to initiate an HTTP response
  response = xapp_rest.initResponse()

# Decode data if there was any in request
  python_data = data.decode("utf-8")

# Create resp. w/ a status code and payload
  response['status'] = <HTTP_status_code>
  response['payload'] = <desired_response>
  # Return new HTTP response
  return response

# Example of a readiness probe handler
def readiness_handler(self, name, path, data, ctype):
  # Initiate a new HTTP response
  response = xapp_rest.initResponse()

  # Check if a key was populated in SDL
  if self.xapp.sdl.get("xapp_1_ns", "gnb_meid"):
    # We are ready to start working
    response['status'] =  200
  else:
    # We are not ready yet
    response['status'] = 500

  return response
\end{lstlisting}

The Python xApp Framework contains the \texttt{xapp_rest} library, which
simplifies the process of handling HTTP requests and creating REST callbacks.
However, there are a few steps required to enable support for the REST
interface on an xApp: First, the xApp developer must edit the xApp descriptor to
open port 8080 for Kubernetes to expose the HTTP service on the desired
containers, as shown earlier in Listing~\ref{lst:port}. Then, the
xApp developer must create an HTTP server in the xApp to listen to incoming HTTP
requests. The \texttt{xapp_rest} provides a threaded HTTP server that can listen
to incoming HTTP requests without blocking the xApp's main loop, as shown in
Listing~\ref{lst:http_server_setup}. For the HTTP server to work correctly, the xApp
developer must instantiate it inside the \texttt{post_init} method for reactive
xApps, or inside the \texttt{entrypoint} method for general xApps. At this
point, the xApp is ready to receive and listen to incoming HTTP requests.

The next step is to register URI endpoints in the HTTP server, specify their
supported HTTP request types, e.g., \texttt{GET}, \texttt{POST}, \texttt{PUT},
\texttt{DELETE}, etc., and map which internal REST callback will reply to a
certain HTTP request type on given URI. This step is crucial for exposing any
internal information and functionality from the xApp through the REST interface.
The \texttt{xapp_rest} HTTP server allows us to create handlers for registering URI
endpoints and mapping an HTTP request type to an internal function that will be
called every time the server receives an HTTP request of that type on that given
URI endpoint, as shown in Listing~\ref{lst:http_handler}. The xApp developer
should consider which type of HTTP request to use when exposing internal
information and functionality via REST, as they behave in different
manners~\cite{rodriguez2008restful}. For example, \texttt{POST} and \texttt{PUT}
requests are accompanied by new resources, e.g., JSON data structures, which can
serve as input for xApps, whereas \texttt{GET} and \texttt{DELETE} requests only
contain identifiers for the resources they are operating on. Finally, we can
create REST callbacks to implement any logic for reacting to incoming
HTTP requests from a remote host. We detail how to create REST callbacks for
reacting to Kubernetes's probes and user interactions in the next subsection.

\subsubsection{Probes and Custom User Interaction}

\begin{lstlisting}[linewidth=\columnwidth,language=python,float,
caption={Examples of REST handlers, showing how xApps can
respond to different types of external input, e.g.,
retrieving the internal xApp state and passing new parameters.},
label={lst:rest_callback}
]
# Example of a GET handler
def get_config_handler(self, name, path, data, ctype):
  # Initiate a new HTTP response
  response = xapp_rest.initResponse()
  # Attribute its the OK status code
  response['status'] =  200
  # Return a JSON w/ the xApp configuration
  response['payload'] = dumps( self._xapp._config_data)

  return response

# Example of a POST hander
def set_new_parameters(self, name, path, data, ctype):
  # Initiate a new HTTP response
  response = xapp_rest.initResponse()
  # Decode new information and save it
  self.upload = data.decode("utf-8")
  # Create response w/ JSON success message
  response['payload'] = ('[{"uploaded" : "complete"}]')

  return response
\end{lstlisting}

To verify the availability and health of pods in a cluster, Kubernetes employs
two probes on each container: \1 the readiness probe checks if the container
carried out all required initialization tasks and ensures it is ready to serve
incoming traffic, and \2 the liveness probe serves as a periodic check of the
operation of the container and ensures it remains alive. During the
instantiation of pods, Kubernetes periodically probes the readiness of their
containers until they return a positive response, issuing \texttt{GET} requests
on the \textit{"/ric/v1/health/ready"} URI, and only then will Kubernetes
allow them to communicate with other pods. After the pods start
running, Kubernetes periodically probes the liveness of their containers,
issuing \texttt{GET} requests on the \textit{"/ric/v1/health/alive}" URI.
In case the liveness probe fails, Kubernetes considers the container
unhealthy (due to a crash or bug) and then tries to restart the containers
as a recovery measure~\cite{burns2022kubernetes}.

The business logic between xApps can differ vastly, and so do their conditions
for readiness and liveness. Therefore, the xApp developer must define their own
handlers for responding to Kubernetes probes, for example, waiting to create entries in SDL before the xApp is ready
to work or checking if the xApp has the necessary variables to continue
working.
We detail how to create custom REST handlers in Listing~\ref{lst:probe_callback},
which also shows how we can create HTTP responses with custom HTTP status codes
and payloads.
Any data received as an argument in the REST handler must be decoded as
\textit{UTF-8} strings before we can process it in Python, and any data we want
to return in the HTTP response must be encoded as a valid JSON string.
In addition, Kubernetes considers any response with a 2xx HTTP status code a
positive response that the probe is successful, while any response with HTTP
status codes 3xx, 4xx, or 5xx indicates a negative response that the probe failed.

The xApp developer can leverage these custom handlers to expose internal
information and functionality via REST, for example, creating custom URIs and
handlers to provide easy access to the current xApp configuration via
a \texttt{GET} request or accepting additional parameters via a \texttt{POST}
request, as shown in Listing~\ref{lst:rest_callback}.
In Section~\ref{sec:debug}, we detail how users of the \ac{nearrtric} can
find the IP addresses of xApps that enabled support for REST and explain how to
interact with them via the terminal.
Moreover, for additional information on how to trigger HTTP requests from within
the xApp itself, e.g., interfacing with \ac{nearrtric} components or other xApps
via their REST interface, we refer the reader to the documentation of the
Python Requests module~\cite{chandra2015python}.
For completeness, we refer the reader to our online repository~\cite{repo},
where we include the entire source code used on the examples in this
section.

\section{xApp Control: Managing \acp{RAN}}\label{sec:ran}

In this section, we describe how xApps can manage \acp{RAN} by interacting with
E2 Nodes through subscriptions. First, we discuss the E2 Nodes and their
interaction with the \ac{nearrtric}, which is useful for xApp developers
creating end-to-end development environments. Then, we detail the
\acp{SM}, the subscription procedure, and the interaction between xApps
and E2 Nodes. Finally, we show how xApps can subscribe to E2 Nodes, trigger
events, set up actions, and react to indication messages with information from
E2 Nodes.


\subsection{E2 Nodes, Termination, and Setup}

The E2 Nodes, whether the disaggregated O-CU, O-DU, and O-RU, or the monolithic
O-gNodeB and O-eNobeB, interact with the \ac{nearrtric} via the E2 Interface,
which exposes information and control over their internal state, enabling
near-real-time control loops to manage the \ac{RAN}. The communication over the
E2 interface occurs through the E2AP, a protocol running on top of the SCTP that
specifies a number of well-defined message types with different purposes and
goals~\cite{polese2022understanding}.
Each E2 Node can expose a number of RAN Functions related to the features and
capabilities it supports, e.g., beamforming, power control, and \ac{RAN}
slicing~\cite{e2general}.
Each \ac{RAN} Function may have widely distinct APIs involving different actions,
required parameters, and data structures. To this end, the interaction with the
\ac{RAN} Functions is structured in the form of a
\acp{SM}~\cite{irazabal2023tc}, which combines the basic \texttt{RIC
Services} provided by E2AP as building blocks to define more complex APIs for
interacting with the E2 Nodes and leveraging their
functionality (detailed in Section~\ref{sub:sub}).

At the \ac{nearrtric}, the communication with E2 Nodes is intermediated through
the \texttt{E2Term}, which serves as a translation component between
the southbound SCTP protocol and the internal \ac{RMR} messaging infrastructure,
forwarding messages between E2 Nodes and the
\texttt{E2Mgr}. Conversely, the \texttt{E2Mgr} is responsible for establishing,
maintaining, and terminating connections with E2 Nodes, as well as updating the
\ac{RNIB} inventory with information about existing E2 nodes and
their available \acp{SM}.
The \texttt{E2Term} and the \texttt{E2Mgr} play different roles in
monitoring the E2 interface: the \texttt{E2Term}
monitors the status of the SCTP connection to the E2 Nodes for identifying
sudden disconnections (and notifying the \texttt{E2Mgr}), and the \texttt{E2Mgr}
monitors the status of the \texttt{E2Term}, sending periodic probes to
for identifying errors.
When an E2 Node starts, it performs an E2 Setup procedure, where it
tries to register itself with a \ac{nearrtric}. The E2 Setup procedure creates
an entry in the \ac{RNIB} using a unique identifier for the E2
Node, known as the inventory name. Only after the E2 Node is set up with the
\ac{nearrtric} and registered in the \ac{RNIB}, the xApps can subscribe to and
communicate with it, which we detail in the Section~\ref{sub:sub}. For additional
information about registration of E2 Nodes with the \ac{nearrtric}, we refer
the reader to~\cite{hung2024security}.

\subsection{\aclp{SM}}\label{sub:sm}

The O-RAN Alliance provides several first-party \acp{SM} in their
specifications, e.g., the \texttt{\ac{RC}, the \texttt{\ac{CCC}},
and the \texttt{\acf{KPM}}
\acp{SM}}~\cite{specifications}. The specification documents for each
\ac{SM} include: \1 an overview of the \ac{SM} and the
corresponding \ac{RAN} Function, their services and capabilities; \2 the formal
description of the \ac{RAN} Function and its
supported actions (the \texttt{RIC Services} detailed in
Section~\ref{sub:sub}); \3 the formal description of the \ac{RAN} parameters,
known as \acp{IE}, i.e., the data structures and data formats for each variable
and arguments for the actions supported by the \ac{RAN} Function;
\4 the structure of how the different actions are combined to form a standard
interface descriptor, and the definition of the \ac{SM} in the form of an
ASN.1 document; and \5 their approach for handling unknown, unforeseen, and
erroneous interactions and protocol data~\cite{e2sm}.
These specifications are useful for vendors and system integrators
creating or testing E2 Nodes to ensure they abide by the standard
interfaces in the \acp{SM}.


It is worth mentioning the importance of the ASN.1 document,
which can be manually excerpted from the specifications or automatically extracted using
scripts. It provides a practical definition of the \ac{SM}, which can be
used both for understanding the operation of the \ac{RAN} Function and for
compiling the code bindings to support the
\ac{SM} on an E2 Node. 
The ASN.1 document is essential for xApp developers interested in controlling
the \ac{RAN}, as xApps only interact with E2 Nodes through the standard
interface defined in the \acp{SM}.
It is possible to develop third-party \acp{SM}
to enable custom functionality on E2 Nodes~\cite{johnson2022nexran}, but this is
outside the scope of this tutorial.
For information about custom \acp{SM}, we refer the reader to~\cite{customsm}.


Depending on the vendor, model, and version, an E2 Node may possess
multiple \ac{RAN} Functions and support their corresponding \acp{SM}
to expose different capabilities and services to
the \ac{nearrtric}. For example, an E2 Node may support the
\texttt{\ac{KPM}} and the \texttt{\ac{CCC}} \acp{SM} at the same
time, exposing \acp{KPM} and adjusting the base stations' transmit power in
near-real-time, respectively. Conversely, the \ac{nearrtric} is agnostic to
\acp{SM}, a design choice that ensures the \ac{nearrtric} architecture and
components remain general and futureproof as new \acp{SM} are developed
over time. As part of this paradigm, only the E2 Nodes and the xApps interacting
with them should be aware of the \ac{SM}'s capabilities and
parameters.
To achieve this, the data exchanged between xApps and E2 Nodes is en/decoded
according to the ASN.1 definition, and it is up to the xApp developer to
en/decode data accordingly. We discuss the ASN.1 en/decoding in Python
later in Section~\ref{sub:e2}.

The E2AP protocol has been updated regularly with bug fixes and
the inclusion of new features. However, some updates required a
significant redesign, leading to breaking changes. For example, with the E2AP
update to version 2.0, which included improved encoding and handling racing
conditions, all \acp{SM} had to be updated based on the new E2AP to remain
operational~\cite{e2general}. Accordingly, the \acp{SM} themselves have been updated
over the years to expose new capabilities and improve interoperability with
\ac{RAN} Functions from different vendors. These changes have led to differences
in supported attributes or message formats between SM versions, which can impact
compatibility with older xApps and \ac{RAN} Functions. Therefore, it is
fundamental for the xApp developer to ensure they are using the correct version
of the \ac{SM} to interface with the \ac{RAN} Functions of the intended base
stations done by checking the \ac{OID} of the \ac{SM}~\cite{e2sm}, i.e., a
universally unique string that identifies all \acp{SM} and includes their
version (detailed later in Section~\ref{sub:e2}).

\subsection{E2 Subscriptions}\label{sub:sub}

\begin{figure}[t]
  \centering
  \includegraphics[width=0.99\columnwidth]{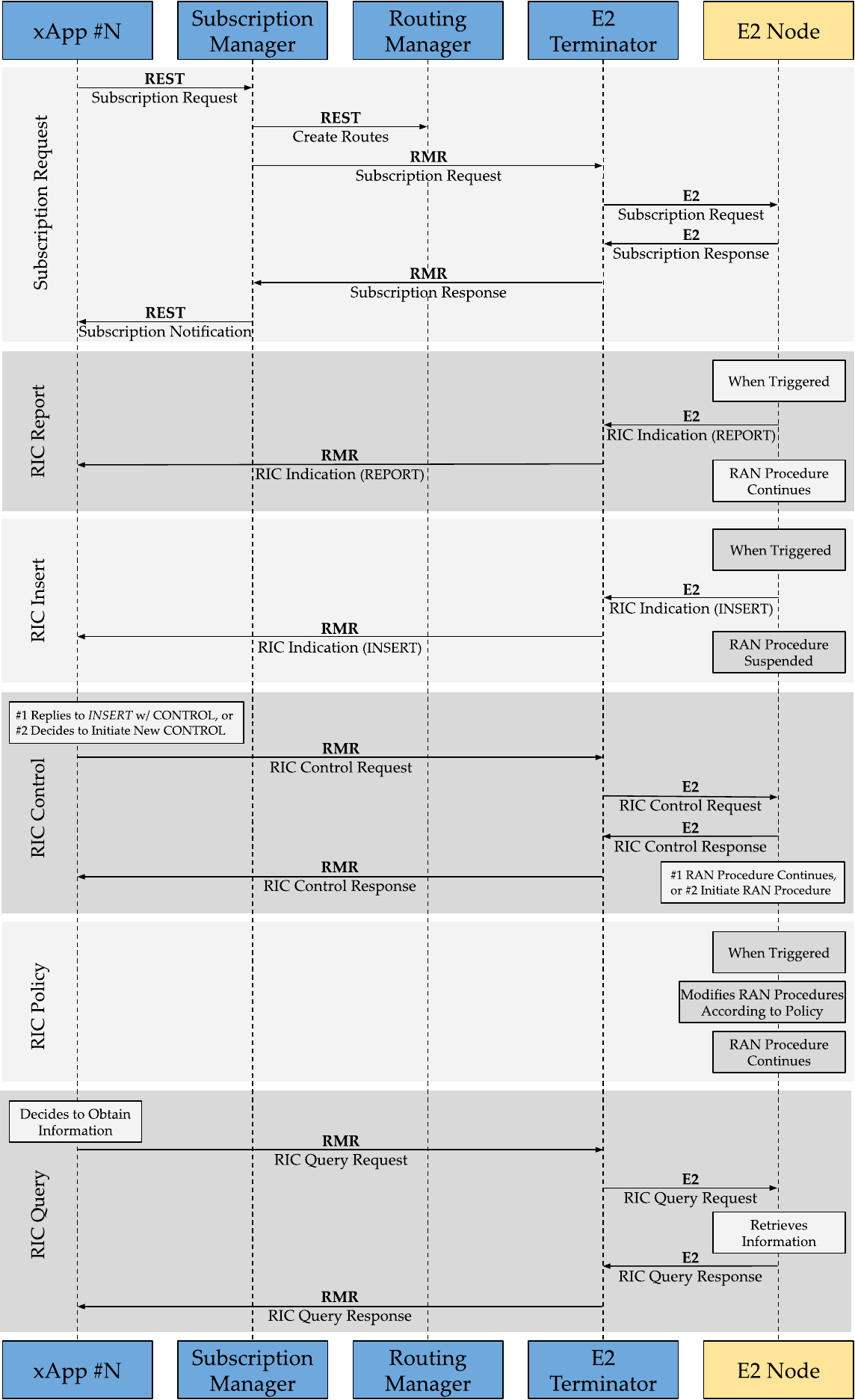}
  \caption{Communication between xApps and E2 Nodes intermediated by the
    \texttt{SubMgr}, showing the interactions with
    \ac{nearrtric} components to create subscriptions (top).
    Each type of subscription behaves differently (detailed in
    Section~\ref{subsub:services}), be it reporting information or
    requesting control decisions when events trigger, waiting
    for control decisions or autonomously handling it, or reacting to queries
    on demand.}
\label{fig:sub_proc}
\end{figure}

In an \ac{O-RAN} deployment, multiple xApps may consume data,
take control decisions, or respond to events of different \ac{RAN} Functions
on several E2 Nodes. To handle this many-to-many relationship, the interaction
between xApps and E2 Nodes follows a publish-subscribe communication pattern
intermediated by the \texttt{SubMgr}~\cite{polese2022understanding}, as
illustrated in Fig.~\ref{fig:sub_proc}.
In the following, we detail how the \texttt{SubMgr} facilitates the
communication between xApps and E2 Nodes, how xApps handle
subscriptions, and the different \texttt{RIC Services}.

\subsubsection{Communication between xApps and E2 Nodes}

The xApps interact with the \texttt{SubMgr} via REST to create subscriptions to
specific E2 Nodes. A subscription is created based on \1 the inventory name of
the specific E2 Node, referred to as the \texttt{meid} in
Section~\ref{sub:mes}, \2 the \texttt{RANFunctionID} that identifies the
\ac{RAN} Function the xApp intends to interface and the corresponding
\ac{SM}, and \3 the desired \texttt{RIC Service} that defines the intended action
to be set up at the E2 Node (detail later in Section~\ref{subsub:services}).
Moreover, we examine the structure of the
Subscription Request, its required fields, and how to create and delete
subscriptions using the Python xApp Framework in the next subsection.

Upon receiving a Subscription Request, the \texttt{SubMgr} interacts with
the \texttt{RtMgr} to create an \ac{RMR} route between the xApp and the E2 Node.
In turn, the \texttt{RtMgr} generates a new \ac{RMR} routing entry for this
subscription and distributes it to the \texttt{E2Term} and the new subscribed xApp.
The \ac{RMR} routes related to the subscription use the special \texttt{subid}
and \texttt{meid} fields, introduced earlier in Section~\ref{sub:mes}.
The \texttt{subid} is generated by the \texttt{SubMgr} to identify that
particular subscription between an xApp and E2 Node, and in the context of
subscriptions, the \texttt{meid} is the inventory name of the intended E2
Node~\cite{radhakrishnan2023detection}.
On the one hand, the routing of messages from xApps to E2 Nodes uses the
\texttt{meid}: the xApp sets the \texttt{meid} in the message that is used
by \ac{RMR} to identify the correct endpoint of the \texttt{E2Term} that is
interacting with the corresponding E2 Node. Upon receiving the \ac{RMR}
message, the \texttt{E2Term} translates it to the E2AP protocol and
forward it to the respective E2 Node~\cite{submgr}. On the other hand, the
routing of messages from E2 Node to xApps uses the \texttt{subid}: upon
receiving an E2 message from an E2 Node, the \texttt{E2Term} translates
it to \ac{RMR} and forwards it to the xApp with the corresponding
\texttt{subid}. The \ac{RMR} identifies the endpoint of the xApp based on
the \texttt{subid} populated in the message by the \texttt{E2Term}.
If the Subscription Request fails at any stage, the \texttt{SubMgr} deletes any
routes created and returns a message to the xApp indicating the reason for
failure, e.g., invalid \texttt{meid} or unsupported \texttt{RANFunctionID}, to
the xApp.

\subsubsection{Handling Multiple Subscriptions}
The xApps can use the \texttt{subid} field to operate over their active
subscriptions, such as listing active subscriptions, obtaining information about
them, updating subscription parameters, or deleting them. If multiple xApps
subscribe to the same information from the same E2 Node, e.g., using the
\texttt{\ac{KPM}} \ac{SM} to obtain \acp{KPM} from the
same E2 Node, the \texttt{SubMgr} merges the multiple subscriptions and only
appends new \texttt{subids} to the existing \texttt{RMR} routes~\cite{submgr}.
In this case, operations such as updating or deleting a particular subscription only affect the specific xApp, not the entire routes, e.g., a delete
operation removes the \texttt{subid} related to the given xApp instead of
deleting the \ac{RMR} routes entirely.
It is important to note that existing subscriptions remain active and
persist even if the xApp pod is restarted (due to crashes, updates or
rollbacks). Therefore, it is expected that the xApp either \1 gracefully
handles errors and signals to delete existing subscriptions before stopping, or
\2 stores the \texttt{subid} of its active subscriptions
on \ac{SDL} and re-sends Subscription Requests to update its
subscriptions upon booting again. The \texttt{subids} are useful
because xApps can have multiple subscriptions to \1 the same E2 Nodes for
interfacing with different \acp{SM}, and \2 different E2 nodes for
interfacing with multiple base stations. In addition, xApps are required to
generate an integer to identify each of their multiple subscriptions
locally, known as the \texttt{XappEventInstanceId}~\cite{submgr}.

\subsubsection{RIC Services}\label{subsub:services}

Each \ac{RAN} Function may support different actions, i.e., the \texttt{RIC
Services} defined in the E2AP specification, that allow xApps to instruct the E2
Nodes how to report information and/or control \ac{RAN} procedures via subscriptions~\cite{e2sm}.
Each \texttt{RIC Service}, i.e., \texttt{REPORT}, \texttt{INSERT},
\texttt{CONTROL}, \texttt{POLICY}, and \texttt{QUERY}, operates in different
manners and is better suited to cater to different use cases and applications.
Each \texttt{RIC Service} contains \ac{RAN} Function-specific data structures,
i.e., the \acp{IE} encoded according to the ASN.1 of the corresponding
\ac{SM}. Therefore, xApps managing the \ac{RAN} must understand the
particularities of the \ac{SM} of each intended \ac{RAN} Function and
en/decode data to/from ASN.1 accordingly. In the following, we detail the
\texttt{RIC Services} illustrated in Fig.~\ref{fig:sub_proc} and how they
operate.

\begin{description}
  \item[\texttt{REPORT:}] The xApp sends a subscription message instructing the
    E2 Node to report particular information according to a specified condition,
    e.g., trigger condition or periodic interval, using the \texttt{REPORT}
    message. This \texttt{RIC Service} is asynchronous and does not require a
    response. 

\item[\texttt{INSERT:}] The xApp sends a subscription message instructing the E2
  Node to suspend a particular \ac{RAN} procedure, e.g.,  handover or
  attachment, according to a specified trigger condition, and requests control
  guidance from the \ac{nearrtric} using an \texttt{INSERT} message. This
  \texttt{RIC Service} is synchronous and requires a response within a
  predefined time limit, or it executes a specific subsequent action
  if an xApp does not respond in time.

\item[\texttt{CONTROL:}] The xApp sends a \texttt{CONTROL} message instructing the
  E2 Node to initiate or resume an associated \ac{RAN} procedure,  e.g., power
  control or \ac{RAN} slicing. This \texttt{RIC Service} is synchronous and
  requires a \texttt{CONTROL} acknowledgment or failure message from the E2 Node.

\item[\texttt{POLICY:}] The xApp sends a subscription message 
  instructing the E2
  Node on a policy with specific procedures for reacting autonomously to a
  particular trigger condition, e.g., scheduling directives. This
  \texttt{RIC Service} is asynchronous and does not require a response from the
  xApp.

\item[\texttt{QUERY}:] The xApp sends a \texttt{QUERY} message to the E2
    Node to request information about the \ac{RAN} or the associated
    \acp{UE}. After each request, the E2 Node issues a single \texttt{QUERY}
    Response with the information or an error message.
\end{description}

Such diversity in how the \texttt{RIC Services} operate enables xApps to create
distinct types of subscriptions for controlling E2 Nodes in widely different
manners, for example, requesting a periodic reporting of certain \acp{KPM},
consulting the xApp on how to react to particular events, or creating rules for
reacting to particular events autonomously.
For additional information about the \texttt{RIC Services} and the steps
associated with their E2AP procedures in the \ac{nearrtric} and E2 Nodes,
we refer the reader to the E2AP documentation~\cite{e2general}.



\subsection{End-to-end Testing \& Development Environment in Python}\label{sub:e2epython}

In the following, we detail how the xApp developer can leverage simulated E2
Nodes to create an end-to-end testing and development environment. Next, we
discuss the current limitations to perform subscriptions with
the xApp Python Framework, as well as workarounds to perform
subscriptions directly with the \texttt{SubMgr} through its REST interface.

\subsubsection{Simulating E2 Nodes}


The \ac{OSC} provides a simulated E2 Node, known as the \texttt{E2Sim}, as a
tool for testing the operation of the \texttt{E2Term} and \texttt{E2Mgr} and
facilitating the xApp development process~\cite{e2sim}. The \texttt{E2Sim} is an
SCTP client that implements the E2AP protocol, which allows the testing of the
E2 Setup procedure with the \texttt{E2Term} and \texttt{E2Mgr}, the creation of
the \ac{RAN} inventory in the \ac{RNIB}, and the subscription to E2 Nodes by
xApps~\cite{lacava2023ns}. By using the \texttt{E2Sim}, xApp developers can
create an end-to-end development environment in software, including xApps,
\ac{nearrtric} components, and simulated E2 Nodes, for testing and validating
all the capabilities and interfaces used by their xApps, including \ac{RMR}
messaging, A1 Policies, \ac{SDL} storage, and E2 subscriptions.

The xApp developer can manually run the \texttt{E2Sim} as a Docker container or
as a Kubernetes pod part of its \ac{nearrtric} cluster using Helm. For
instructions on installing and setting up the \texttt{E2Sim}, we refer the
reader to its official documentation~\cite{e2sim_wiki}. The upstream
\texttt{E2Sim} provided by the \ac{OSC} supports
the \texttt{\ac{KPM}} \ac{SM}, which exposes different
metrics about the base station and its \acp{UE} using the \texttt{REPORT} type
of subscription. In this case, the \texttt{E2Sim}
streams metrics based on a trace file, the \texttt{reports.json}, which can be
edited by the xApp developer before creating the xApp's Dockerfile
to use custom or artificial data. The \texttt{E2Sim} uses the
\texttt{\ac{KPM}} \ac{SM} to encode the metrics to ASN.1
and stream \ac{RAN} telemetry to the subscribed xApps~\cite{lacava2023ns}. In
the remainder of this section, we use the E2Sim and its supported \texttt{Key
Performance Measurement} to demonstrate how xApps can perform subscriptions to
control E2 nodes in Python. For additional information about the \texttt{E2Sim},
we refer the reader to its official documentation and repository~\cite{e2sim}.

\subsubsection{Subscriptions in Python}

The different xApp Frameworks provided by the \ac{OSC} for developing xApps in
different programming languages abstract a number of interfaces and automate
many interactions with \ac{nearrtric} components to simplify the xApp
development process. However, due to the open-source nature of the \ac{OSC}, the
xApps Frameworks receive different levels of attention from the community and,
hence, possess distinct subsets of features or API versions (discussed later in
Section~\ref{sec:outl}). For example, the \ac{STSL} is currently only supported
by the Go xApp Framework.
The API for subscribing to E2 Nodes has undergone significant
changes on the F Release of the \ac{nearrtric}, with the migration of the
subscription management operations, e.g., creation, query, update, and delete,
from \ac{RMR} to REST (detailed in Section~\ref{sub:e2}). These breaking
changes across \ac{nearrtric} releases were propagated to the C++, Go, and Rust
xApp Frameworks so they could continue subscribing to E2 Nodes and managing the
\ac{RAN}.
While the Python xApp Framework has received initial support for subscriptions
via REST, at the time of writing, we observe that \1 it does not provide an
approach for en/decoding ASN.1 from/to Python data objects, and \2 its HTTP
methods for interacting with the \texttt{SubMgr} employ the snake case
convention, whereas the \texttt{SubMgr} expects HTTP requests in camel case.
Consequently, despite receiving several updates since
the F release, the subscription API of the Python xApp Framework remains
incompatible with the \texttt{SubMgr} and incapable of
subscribing to E2 Nodes.

Without loss of generality, we can leverage some of the existing
functionality provided by the Python xApp Framework and the lessons learned in
Section~\ref{sec:oper} to develop xApps in Python that can subscribe to E2 Nodes
and control the \ac{RAN} by \1 en/decoding ASN.1 data structures from/to Python
objects using external libraries, and \2 interacting with the \texttt{SubMgr} via
REST directly by creating HTTP requests. In the event that the Python xApp
Framework is updated and its subscription API is fixed, the principles therein
will remain useful to inform the reader how the subscription-related data
structures and procedures are handled under the hood.

We leverage the
Python PyCrate module to en/decode ASN.1 data structures from/to Python objects. Based on the ASN.1 documents from the E2AP and E2SM
protocols, and from the intended \ac{SM}, PyCrate can generate a Python
representation of the \ac{SM}.
This representation contains Python methods for en/decoding
data from/to ASN.1 according to the \ac{SM}'s standard interface
description.
For additional information about the utilization of the PyCrate module, we refer
the reader to its official documentation~\cite{pycrate}.
In the next subsection, we detail how we can
leverage PyCrate and manually create HTTP requests to
complement the functionality of the Python xApp Framework
to subscribe to E2 Nodes and control the \ac{RAN}.



\subsection{Controlling E2 Nodes using the Python xApp Framework}\label{sub:e2}

In the following, we discuss how the xApp developer can effectively interact with E2
Nodes using the Python xApp Framework. First, we demonstrate how to set up an
xApp to subscribe to E2 Nodes. Next, we show how to create
Subscription Requests with the \texttt{SubMgr} via REST and encode data to
ASN.1. Then, we detail how to react to subscription indication messages
and how to decode data from ASN.1. Finally, we show how xApps can
operate over their subscriptions.

\subsubsection{Setting Up the Subscriptions}

\begin{lstlisting}[linewidth=\columnwidth,language=python,float,floatplacement=tbp,
caption={Example of a \texttt{post_init} method where we register a calback for
handling Subscription Notifications, iterate over the list of registered
E2 Nodes, and subscribe to one of them according to their available \ac{RAN}
Functions.},
label={lst:sub_init}
]
# Function called after the constructor
def _post_init(self, rmr_xapp):
  ...

  # Create Subscriber Object
  self._submgr = NewSubscriber(
    uri=<SubMgr_URL>,
    local_port=<xApp_HTTP_Port>,
    rmr_port=<xApp_RMR_Route_Port>
  )

  # Register Notification Callback Handler
    self._submgr.ResponseHandler(
      responseCB=self._subscription_notif)

  # Hold active subscriptions
  self._subscriptions = []
  # Counter to identify subscriptions
  self._event_instance = 0

  # Iterate list of registered gNodeBs
  for gnb in xapp.get_list_gnb_ids():
    gnb_info = rmr_xapp.GetNodeb(
      gnb.inventory_name)

    # Iterate list of RAN Functions
    for ran_function in gnb_info.ran_functions:
      # Check for matching OID of the KPM
      if ran_function.oid == \
        "1.3.6.1.4.1.53148.1.2.2.2":

        # Subscribe to gNodeB
        self._send_subscription_request(
          gnb.inventory_name)
\end{lstlisting}

The xApp developer can leverage the \texttt{SubMgr}'s REST interface to interact
with it directly to send Subscription Requests for creating, modifying, and
deleting subscriptions to E2 Nodes.
In this case, the \texttt{SubMgr} also interacts with the xApp via REST to
send Subscription Notification messages containing the \texttt{subid} if the
Subscription Request was successful or the type and reason for errors otherwise.
To facilitate the handling of Subscription Notification messages, we can avail from
the \texttt{NewSubscriber} object from the Python xApp Framework, which
uses the \texttt{SubMgr}'s URL, as well as the xApp's HTTP and
\ac{RMR} route port, to create an HTTP server configured to
receive requests from the \texttt{SubMgr}, as shown in
Listing~\ref{lst:sub_init}. The \texttt{NewSubscriber} object allows us to
register a callback to handle Subscription Notifications, which we detail later
in Section~\ref{subsub:notif}.
It is also strongly recommended that xApps \1
store the list of \texttt{subids} of their active subscriptions, keeping them in
persistent storage via \ac{SDL} and/or in memory, e.g., using a global
\texttt{self._subscriptions} variable; and \2 create a monotonic counter to
identify each of their multiple subscriptions locally, the
\texttt{XappEventInstanceId}.

An xApp can subscribe to any E2 Node registered on the \ac{nearrtric}. However,
the xApp may decide to filter the pool of E2 Nodes to identify the subset that
supports the \ac{RAN} Functions it intends to control. The xApp developer can
accomplish this by inspecting the \ac{RNIB} to iterate over the list of E2
Nodes, parsing their \ac{RAN} Functions, and checking their vendor-specific
\texttt{RANFunctionIDs} and/or their \acp{OID}, as shown in
Listing~\ref{lst:sub_init}. Finally, we can subscribe to the matching E2 Node(s)
that support the functionality we want to control.
For information about the structure of the \ac{OID}, the meaning of each
field, and the matching list between \acp{OID} and \acp{SM}, we refer the reader
to Table 5-2 of the official documentation about the E2SM~\cite{e2sm}.

\begin{lstlisting}[linewidth=\columnwidth,language=python,float,
floatplacement=tbp, caption={Example of a custom method for creating Subscription
Requests via the \texttt{SubMgr}'s REST interface.},
label={lst:sub_req}
]
# Custom method for creating subscriptions
def _send_subscription_request(self, meid):

  # Create trigger condition ASN.1 encoded
  encoded_trigger = <Detailed %in Lst.~\ref{lst:enc_trigger}%>
  # Create action definition ASN.1 encoded
  encoded_action = <Detailed %in Lst.~\ref{lst:enc_action}%>

  # Increment counter
  self._event_instance += 1

  # Prepare Subscription Request Payload
  sub_payload = <Detailed %in Lst.~\ref{lst:json_sub_req}%>

  # Send POST request to the SubMgr
  response = requests.post(
    <SubMgr_URL> + "/ric/v1/subscriptions",
    json=sub_payload
   )

  # Handle HTTP Response
  if response.status_code == 201:
    self.logger.debug("Subscription Request Success!")

  else:
    self.logger.debug("Subscription Request Failure!")
\end{lstlisting}

\subsubsection{Creating Subscription Requests}

We can subscribe to a given E2 Node by issuing a POST request to the
\texttt{SuMgr}'s REST interface, containing a JSON  payload that defines
the Subscription Request, as shown in Listing~\ref{lst:sub_req}. This JSON
payload contains two fields encoded in ASN.1
according to the corresponding \ac{SM}: the
trigger condition and the action definition. In the following, we first discuss the
structure and content of the JSON payload, detailed in
Listing~\ref{lst:json_sub_req}. Then, we overview the details of the
trigger condition and action definition, and how to encode them in ASN.1.

\begin{lstlisting}[linewidth=\columnwidth,language=rt,float,
caption={Subscription Request payload structure.},%numbers=none,
label={lst:json_sub_req}]
{
  "SubscriptionId":"",
  "ClientEndpoint": {
    "Host": <xApp_URL>,
    "HTTPPort":8080,
    "RMRPort":4560
  },
  "Meid": <inventory_name>,
  "RANFunctionID": <RANFunctionID>,
  "E2SubscriptionDirectives":{ # Optional
    "E2TimeoutTimerValue":2,
    "E2RetryCount":2,
    "RMRRoutingNeeded":True
  },
  "SubscriptionDetails":[
    {
      "XappEventInstanceId":               self._event_instance
      "EventTriggers":[
         <ASN.1 Event Definition> ],
      "ActionToBeSetupList":[
        {
          "ActionID": 1,
          "ActionType": <RIC Service>,
          "ActionDefinition": [
            <ASN.1 Action Definition> ],
          "SubsequentAction":{
            "SubsequentActionType":        "continue",
            "TimeToWait":"w10ms"
          }
        }
      ]
    }
  ]
}
\end{lstlisting}

The Subscription Request can create or modify an existing subscription, as
specified by the \texttt{SubscriptionId} field: when creating a new
subscription, it is an empty string, whereas when modifying an existing
subscription, it uses the \texttt{subid} of the target subscription. The
\texttt{SubMgr} also needs information about the xApp's \ac{RMR} and HTTP
endpoints, i.e., the \ac{RMR} service's data port and the HTTP service's URL
and port, to \1 instruct the \texttt{RtMgr} to create or update \ac{RMR} routes
related to the subscription and \2 reply to the xApp with a Subscription
Notification via REST about the subscription result.
Next, the xApp must specify which E2 Node the xApp wants to
subscribe to, based on its inventory name, and which RAN
Function it wants to control, according to its \texttt{RANFunctionID}.
The xApp
developer can also configure optional directives related to this subscription, e.g.,
the duration of a timer to wait until the \texttt{SubMgr} receives a Subscription
Response from the E2 Node, the number of Subscription Request retries
from the \texttt{SubMgr} to the E2 Node, and whether the \texttt{RtMgr} needs to
create or update \ac{RMR} routes. 

The subscription details describe in what manner that given subscription controls the E2 Node. It contains \1
the \texttt{XappEventInstanceId} to identify that given subscription locally at
the xApp; \2  a list of trigger conditions in
ASN.1, which specifies when the given actions occur, e.g., periodically or
when a given variable reaches a threshold; and \3 a list of actions
to be set up at the E2 Node, including an ID for each action (used to notify the
xApp about the status of each action set up), the type of
\texttt{RIC Service}, e.g., \texttt{REPORT} or \texttt{INSERT}, a list of action
definitions in ASN.1, which represent  what is executed at the E2 Node, e.g.,
the metrics to be reported or the parameters to be changed. In the case
of \texttt{INSERT} and \texttt{CONTROL} actions, the xApp can specify how the E2
Node will handle the \ac{RAN} procedures (continuing or halting) if the xApp
does not reply within a given time to wait.



\begin{lstlisting}[linewidth=\columnwidth,language=python,float,floatplacement=tbp,
caption={Creating an event trigger condition for the \texttt{Key
Performance Measurement} \ac{SM} (setting up a periodic report)
and its encoding to ASN.1 using PyCrate.},
label={lst:enc_trigger}
]
event_definition = {
  "eventDefinition-formats":
    ("eventDefinition-Format1", {"reportingPeriod": 1000})
  }

trigger = E2SM_KPM_IEs. E2SM_KPM_EventTriggerDefinition
trigger.set_val(event_definition)
encoded_trigger = trigger.to_aper()
\end{lstlisting}

\begin{lstlisting}[linewidth=\columnwidth,language=python,float,floatplacement=tbp,
caption={Creating an action definition for the \texttt{Key
Performance Measurement} \ac{SM} (specifying \acp{KPM} to report)
and its encoding to ASN.1 using PyCrate.},
label={lst:enc_action}
]
action_definition = {
  "actionDefinition-formats": (
  "actionDefinition-Format1", {
    "measInfoList": [
      { "measType":                   ("measName", "DRB.PerDataVolumeDLDist.Bin"),
        "labelInfoList":              [{"measLabel": {"noLabel":"true"}}],
      },
      ...
    ],
    "granulPeriod": 1000 },
  ),
  "ric-Style-Type": 1,
}

action = E2SM_KPM_IEs. E2SM_KPM_ActionDefinition
action.set_val(action_definition)
encoded_action = action.to_aper()
\end{lstlisting}

Most parameters in the Subscription Request JSON payload are in plaintext and
independent of the \ac{SM}. However, the trigger conditions and action
definitions are ASN.1 data objects that depend on the \ac{RAN} Function and its
corresponding \ac{SM}. To find information about the supported trigger
conditions and action definitions, as well as their formats and \acp{IE}, the
xApp developer must refer to the \ac{SM}'s specifications.
The PyCrate representation of the \ac{SM} provides methods for encoding
Python data structures to ASN.1 format, as long as they abide by the strict
structure of the \acp{IE} in the ASN.1 document. We show an example of the
\texttt{\ac{KPM}} \ac{SM} in Listings~\ref{lst:enc_trigger}
and~\ref{lst:enc_action}. The former shows how to
instantiate an event definition format and create an event trigger condition to
report measurements every \unit[1000]{ms}.
The latter shows how to instantiate an action definition format, create an
action to report a list of measurements based on the names of the \acp{KPM} name
and how to label them, define a measurement granularity period of \unit[1000]{ms}, and
configure a report style that defines how to collect \acp{KPM}, e.g., per
\ac{UE}, per group of \acp{UE}, or per base station.
Both listings show the PyCrate methods for setting the values to the ASN.1
encoding and representing it using the APER format used by the 
\texttt{\ac{KPM}} \ac{SM} specification from \ac{O-RAN}.

After issuing the POST request with the Subscription Request JSON payload, the
\texttt{SubMgr} responds to the xApp with an HTTP status code 201 if the
subscription was successful or with an error status code alongside the reason
for failure.

\begin{lstlisting}[linewidth=\columnwidth,language=python,float,
caption={Example of a custom callback for handling Subscription Notification
messages from the \texttt{SubMgr}.},%numbers=none,
label={lst:sub_not}]
# Custom method to handle Notifications
def _subscription_notif(self, name, path, data, ctype):
  # Convert the JSON string to Python
  python_data = json.loads(data)

  # Extract the subid from the Notification
  subid = python_data["SubscriptionId"]
  # Store the new subscription
  self._subscriptions.append(subid)

  # Extract useful information
  sub_inst= python_data[ "SubscriptionInstances"][0]
  xapp_event_instance = sub_inst["XappEventInstanceId"]
  e2_event_instance = sub_inst["E2EventInstanceId"]
  error_cause = sub_inst["ErrorCause"]
  error_source = sub_inst["ErrorSource"]

  # Respond to the POST request
  response = initResponse()
  return response
\end{lstlisting}


\begin{lstlisting}[linewidth=\columnwidth,language=python,float,floatplacement=tbp,
caption={Example on how to register a \ac{RMR} message callback
to handle RIC Indication messages from the E2 Node.},
label={lst:rmrxapp_register_indication}
]
# Register callback to handle Indications
self._rmr_xapp.register_callback( self._indication_handler, RIC_INDICATION)
\end{lstlisting}

\subsubsection{Handling Subscription Notifications}\label{subsub:notif}

After handling a Subscription Request, the \texttt{SubMgr} returns a
Subscription Notification message to the xApp. This response message contains
the \texttt{subid} generated by the \texttt{SubMgr} to identify the
subscription, which the xApp should store in memory and/or persistent storage.
In addition, the Subscription Notification message contains information about
the result of the Subscription Request that is useful for debugging, including \1 the
\texttt{xAppInstanceEvenID}, so that the xApp knows which Subscription
Notification this refers to, \2 an \texttt{E2EventInstanceId}, to identify the
particular subscription procedure at the E2 Node, \3 the error cause, to
explain the reason for failure, and \4 the error source, i.e., the
\ac{nearrtric} component or E2 Node that raised the error. An xApp can handle a
Subscription Notification message by creating a POST request handler, as shown
in Listing~\ref{lst:sub_not}.

\subsubsection{Reacting to RIC Indications}

Depending on the type of subscription, i.e., \texttt{REPORT} or \texttt{INSERT},
the E2 Node may send a RIC Indication message to the xApp via \ac{RMR} to report
information or request a control decision, respectively. The E2 Node generates a
RIC Indication message when an event condition in the subscription is triggered,
e.g., periodically or when a variable reaches a threshold. To handle RIC
Indication messages, the xApp can register an \ac{RMR} callback to handle the
RIC Indication \texttt{mtype} (12050), as shown in
Listing~\ref{lst:rmrxapp_register_indication}.

\begin{lstlisting}[linewidth=\columnwidth,language=python,float,floatplacement=tbp,
caption={Example of a custom \ac{RMR} callback for handling RIC Indication
  messages from an E2 Node using the \texttt{\ac{KPM}} \ac{SM}.},
label={lst:sub_indi}
]
# Callback to Handle Indication Messages
def _indication_handler(self, rmrxapp, summary, msg_buf):
  # Get Message Payload
  raw_data = summary[rmr.RMR_MS_PAYLOAD]

  # Populate E2AP ASN.1 Data Structure
  e2ap_pdu.from_aper(raw_data)
  # Decode it from ASN.1 to Python
  pdu = e2ap_pdu.get_val()

  # Parse contents of the message
  if pdu[0] == 'initiatingMessage':
    # Traverse dicts to obtain protocol IEs
    ies = e2ap_pdu.get_val_at(
      ['initiatingMessage', 'value',
      'RICindication', 'protocolIEs'])
    # Iterate over protocol IEs
    for ie in ies:
      # If it is the KPM SM message header
      if ie['value'][0] == 'RICindicationHeader':
        # Populate KPM ASN.1 Data Structure
        header = E2SM_KPM_IEs. E2SM_KPM_IndicationHeader
        header.from_aper(ie['value'][1])
        data = header.get_val_at(
          ['indicationHeader-formats',
          'indicationHeader-Format1'])
        self.logger.info(f"KPM Hdr {data}")

      # If it is the KPM SM message payload
      elif ie['value'][0] == 'RICindicationMessage':
        # Populate KPM ASN.1 Data Structure
        message = E2SM_KPM_IEs. E2SM_KPM_IndicationMessage
        message.from_aper(ie['value'][1])
        data = message.get_val_at(
          ['indicationMessage-formats',
          'indicationMessage-Format1'])
        self.logger.info(f"KPM Msg {data}")

\end{lstlisting}

A RIC Indication message is encapsulated inside the \ac{RMR} payload of an E2AP
message, both of which are encoded in ASN.1. The outer E2AP message has a
generic format, encoded in ASN.1 according to the E2AP specification,
whereas the inner RIC Indication message has a very particular format, encoded
in ASN.1 according to the \ac{SM} and the type of subscription. The
PyCrate representation of ASN.1 documents also provides methods for decoding
ASN.1 data structures to Python objects, which we can leverage to decode the
E2AP Message, extract the RIC Indication message, and decode its content. In
Listing~\ref{lst:sub_indi}, we show an example of a tailored \ac{RMR} message
handler for RIC Indication messages from the \texttt{\ac{KPM}} \ac{SM}
using the \texttt{REPORT} subscription type.
This handler decodes the E2AP message using PyCrate,
traverses through its payload, extracts and decodes the RIC Indication message,
and logs its content. For additional examples of how to respond to RIC
Indication messages of the \texttt{INSERT} subscription type and issue control
messages to the E2 Nodes, we refer the reader to the \texttt{SubMgr}'s official
documentation~\cite{e2sub}.

\subsubsection{Operating over Active Subscriptions}

\begin{lstlisting}[linewidth=\columnwidth,language=python,float,
floatplacement=tbp, caption={Example of a custom method for querying the
  \texttt{SubMgr} about the all active subscriptions of an xApp.},
label={lst:sub_query}
]
# Custom method for querying subscriptions
def _query_subscriptions(self):
  # Send GET request to the SubMgr
  response = requests.get(
    <SubMgr_URL> + "/ric/v1/ get_xapp_rest_restsubscriptions/" + <xApp_URL>,
   )

  # If the query request was successful
  if response.status_code == 200:
    # List active subscriptions
    for subid in response.json():
      self.logger.info(f"Active Subscription ID: {subid}")
\end{lstlisting}

\begin{lstlisting}[linewidth=\columnwidth,language=python,float,floatplacement=tbp,
  caption={Example of a method to delete all active subscriptions and
  release resources before stopping the xApp.}, label={lst:sub_del}]
# Method to Unsubscribe from all E2 Nodes
def unsubscribe(self):
  # Iterate over the active subscriptions
  for subid in self._subscriptions:
    # Unsubscribe to each E2 Node
    data, reason, status = self._submgr.UnSubscribe(subid)

    # Handle Unsubscribe Response
    if status == 204:
      self.logger.debug("Subscription Delete Successful!")
    else:
      self.logger.debug(f"Subscription Delete Failure! {status} {reason}")
\end{lstlisting}

After the \texttt{SubMgr} returns to the xApp with a successful Subscription
Notification, the subscription is active, and the xApp can perform operations on
it, e.g., querying information about it, modifying its parameters, or deleting
it. The active subscriptions of an xApp remain active and persist across
reboots due to updates, rollbacks, or crashes. More importantly, the actions set
up on an E2 Node through a subscription stay in effect until \1 the E2
Node is removed from the \ac{nearrtric} or \2 the xApp deletes the subscription
to the E2 Node.
To query the list of active subscriptions, e.g., for recovering the previous
operational state after a crash, an xApp can issue a GET request to the
\texttt{SubMgr}'s REST interface, as shown in Listing~\ref{lst:sub_query}. In
this case, the \texttt{SubMgr} will respond with a list of subids of the active
connections, if any. To modify a subscription, the xApp can send a new
subscription request to the \texttt{SubMgr}, as shown earlier in
Listing~\ref{lst:sub_req}, using the \texttt{subid} of the active subscription in
the \texttt{SubscriptionId} field and updated subscription details, e.g.,
different action definitions or event triggers.
To delete a given subscription, e.g., for canceling actions no longer required
as part of the xApp's business logic or releasing resources before gracefully
exiting, an xApp can leverage the \texttt{UnSubscribe} method from the Python
xApp Framework, as illustrated in Listing~\ref{lst:sub_del}.

\section{xApp Debugging: Inspecting your Application}\label{sec:debug}


In this section, we discuss debugging strategies to assist the xApp developer
in identifying and fixing errors as part of the xApp development cycle.
First, we examine approaches to debug the deployment of xApps.
Then, we discuss debugging exposed services, open ports, and REST
communications.
Next, we show how to log information from xApps to debug their operation during
runtime. Finally, we detail how to debug issues with RMR communications and SDL
data storage.

\subsection{Debugging xApp Deployment}


\begin{lstlisting}[linewidth=\columnwidth,language=bash,float,
  caption={Commands for interacting with the \ac{nearrtric} cluster to obtain
  information about the xApp Kubernetes pods.},
  label={lst:kube}]
# List Kubernetes pods in all namespaces
kubectl get pods -A

# List Kubernetes pods in a given namespace
kubectl get pods -n <namespace>

# Example to list all running xApp pods
kubectl get pods -n ricxapp

# Describe information about a given pod
kubectl describe \
pod <pod_name> -n <namespace>

# Example to describe info a given xApp
kubectl describe pod ricxapp-examplexapp-6867f6c785-9pvc5 -n ricxapp

# Print the log/stdout of a given pod
kubectl logs <pod_name> -n <namespace>

# Example of the log command for an xApp
kubectl logs ricxapp-examplexapp-6867f6c785-9pvc5 -n ricxapp
\end{lstlisting}

During the xApp deployment, the \texttt{AppMgr} performs several steps to
instantiate xApps: fetching Docker images from a Docker Registry, spawning their
containers, configuring their resources, opening ports, and exposing services.
Each step is prone to errors that can prevent the xApp from
being deployed or working correctly.
In addition, there can be issues when installing or
restarting the \ac{nearrtric} Kubernetes cluster, which can prevent some of its
components from starting and, consequently, impair the operation of the
xApps. To debug the aforementioned errors, the xApp developer can use the
commands in Listing~\ref{lst:kube} to identify potential issues. We detail
these commands below:

\begin{description}
  \item[List All Pods:]
    This command lists all Kubernetes pods in the \ac{nearrtric} cluster,
    showing their namespaces, names, number of ready containers, number of
    restarts, and the time elapsed since their creation. It is useful for
    finding the names of the pods and getting a global view of the \ac{nearrtric}, as an operational cluster should have all its
    \texttt{ricinfra} and \texttt{ricplt} pods (detailed in
    Section~\ref{subsub:tech}) in the \textit{Running} state (except for the
    \texttt{tiller-secret-generator} pod as \textit{Completed}).

  \item[List Pods in a Given Namespace:] This command lists all pods that belong
    to a given namespace, which is useful to focus and inspect a particular
    aspect of the \ac{nearrtric}, such as the deployed xApps in the
    \texttt{ricxapp} namespace.

  \item[Describe a Given Pod:] This command provides in-depth information about
      a particular pod, including its container ID, Docker Registry's location,
      open ports, state, the ConfigMaps used to create its environment
      variables, the results from Kubernetes probes, the mounted
      volumes, and events that occurred during the pod's lifecycle.
      The events provide detailed information about the pods' creation,
      including when Docker images were fetched and when containers were created
      and started. The events also record information about failures, including
      why and when they occurred, if and why the pod is back-off restarting, and
      if it was evicted due to the lack of resources.
      For the complete list of events, we
      refer the reader to their comprehensive documentation
      in~\cite{dockerfilereference}.

  \item[Print Logs of a Given Pod:] This command displays the standard output
    generated by the given pod. It is useful to see the steps taken by the
    Python xApp Framework to start the xApp, including setting up the \ac{RMR}
    library, loading the \ac{RMR} route table, the content of the xApp
    configuration file loaded by the xApp, the registration with the
    \texttt{AppMgr}, the IP addresses of HTTP and RMR endpoints, and
    information about \ac{RMR} messages exchanged with components of the
    \ac{nearrtric} and other xApps. In addition, this command displays any
    information logged by the xApp developer (detailed later in this section),
    e.g., debug information from different functions, data received from
    \ac{RMR} and A1 from callbacks, or warnings and errors regarding the
    xApp's business logic.

\end{description}


These commands are helpful to debug issues related to the
implementation of new xApps, including potential errors associated with accessing
the Docker Registry, typos in the image name and tag, which prevent the xApp
from being instantiated, as well as the wrong content in the xApp descriptor,
crashes in the xApp business logic or failures to register with the
\texttt{AppMgr}, which will prevent the xApp from working correctly.

\begin{lstlisting}[linewidth=\columnwidth,language=bash,float,
caption={Commands for debugging exposed services and interacting with xApps or
\ac{nearrtric} components via REST.},
label={lst:curl}
]
# List exposed services and open ports
kubectl get services -A

# Send HTTP/REST request
curl -X GET <xapp_IP>:<xapp_port>/<path>

# Example of HTTP/REST request
curl -X GET 10.107.57.43:8080/ric/v1/health
\end{lstlisting}

\subsection{Debugging Ports, Services, and REST Communications}\label{sub:debug_rest}

The xApps and \ac{nearrtric} components can have an optional REST interface for
obtaining debug information about their internal state and passing control
parameters during runtime.
To communicate with an xApp via REST, the xApp developer must first ensure it
has exposed the HTTP service and located the associated IP address and open port,
which can be accomplished using the commands listed in Listing~\ref{lst:curl}.
The \texttt{get services} command lists all exposed services in the \ac{nearrtric}
cluster, showing their namespaces, names, types, IP addresses inside the cluster,
optionally an external IP, open ports and supported protocols, and the
time elapsed since their creation. With this information, the xApp
developer can use the \texttt{curl} command to issue custom HTTP requests
to the desired xApp, following the HTTP endpoint structure detailed in
Section~\ref{sub:ui}. We
refer the reader to the official \texttt{curl} documentation~\cite{curl} for more information about the \texttt{curl} command,
including instructions on transferring JSON objects and files.

These commands are helpful to debug issues related to connectivity
and communication
with xApps, including missing the HTTP service in the xApp
descriptor, using an old IP address due to the pods' containers restarting,
or using an incorrect port number, which would prevent the xApp developer
(or users of the \ac{nearrtric}) to interact with the xApp via REST.
There are also some circumstances in which the xApp developer may want to
interact with the \texttt{AppMgr}
to debug the onboarded xApps and their parameters.
The \texttt{AppMgr} has
a REST interface that the xApp developer (or the
\ac{nearrtric}'s users) can leverage to debug its operation during runtime. For
information about their REST interfaces and supported calls, we refer the reader
to the \texttt{AppMgr}'s documentation in~\cite{appmgr}.

\begin{lstlisting}[linewidth=\columnwidth,language=python,float,
caption={Commands for creating logger objects inside xApps, and logging
messages with different severities.},
label={lst:log}
]
# Create a logger object in a RMRXapp
def _post_init(self, rmr_xapp):
  # Set log level
  rmr_xapp.logger.set_level(<log_level>)
  ...

# Create a logger object in an Xapp
def _entrypoint(self, xapp):
  # Set log level
  xapp.logger.set_level(<log_level>)
  ...

# Example creation of a logger w/ DEBUG
  xapp.logger.set_level(level.DEBUG)

# Logging messages w/ different severities
xapp.logger.debug(<msg>)
xapp.logger.info(<msg>)
xapp.logger.warning(<msg>)
xapp.logger.error(<msg>)

# Example of a log message
xapp.logger.error("Missing input parameter:" + str(my_parameters))
\end{lstlisting}

\subsection{Logging xApp Data}


The Python xApp Framework provides a streamlined logging API, ensuring that log
entries adhere to a standardized format and are handled uniformly, which helps
the xApp developer to debug and track the execution of their xApp's business logic
and control loops~\cite{writersguide}. To leverage the logging API, the xApp
developer must initialize a logger object with a default log level in their
xApp's post-initialization function (for \texttt{RMRXApps}) or entrypoint
function (for \texttt{XApps}), as shown in Listing~\ref{lst:log}. The log levels
range from \texttt{DEBUG}, \texttt{INFO}, \texttt{WARNING}, and \texttt{ERROR},
which correspond to the growing severity levels of the logged messages.
Then, the xApp developer can start logging messages in different parts of their
xApp's business logic to log debug information, raise warnings, and throw
errors. The log entries are displayed on the standard output of the xApp pod,
and they contain the following fields:

\begin{description}

  \item[Timestamp:] When the log entry was created, in milliseconds.
  \item[Criticality:] The severity level of the log entry.
  \item[ID:] The name of the process that called the logging library.
  \item[Message:] A custom message defined by the xApp developer.
\end{description}

The logging API is useful to debug the internal state of the xApp and track its
control flow, allowing the xApp developer to easily display information about
input parameters, results from conditional expressions, calling different
methods, errors, exceptions, or any other tests and verifications. By
generating comprehensive log messages as part of their business logic and
inspecting them during execution using Kubernetes commands, the xApp developer
can find valuable information about errors and crashes during development, as
well as the users of the \ac{nearrtric} when in production.

\begin{lstlisting}[linewidth=\columnwidth,language=bash,float,
caption={Commands for interacting with the \texttt{RtMgr} to obtain
debug information related to \ac{RMR} communication.}, label={lst:rtmgr_rest}
]
# List services to search RtMgr's REST info
kubectl get svc -A

# Or obtain the RtMgr's HTTP endpoint IP
kubectl get svc -n ricplt \
service-ricplt-rtmgr-http \
-o=jsonpath='{.spec.clusterIP}'

# And the RtMgr's HTTP endpoint port
kubectl get svc -n ricplt \
service-ricplt-rtmgr-http \
-o=jsonpath='{.spec.ports[0].port}'

# Send request to RtMgr's REST interface
curl -X GET <RTMGR_IP>:<RTMGR_PORT> /ric/v1/getdebuginfo

# Example request to obtain the current
# RMR routes displayed as a formatted JSON
curl -X GET \
10.110.179.180:3800/ric/v1/getdebuginfo | jq .
\end{lstlisting}

\subsection{Debugging RMR Communications} 

While the \ac{RMR} library offers a high-speed, low-latency
communication interface between xApps and components of the \ac{nearrtric},
availing of it requires a number of steps that can be error-prone and prevent
xApps from communicating. Namely, some of those steps include: \1 adding the \texttt{mtypes} on the xApp
file descriptor, both to configure the \ac{RMR} library and to route messages to
the different containers that comprise the pod; \2 creating a static route
table file with the desired \texttt{mtypes} and \ac{RMR} endpoints; and \3
creating methods to send, receive, and reply to messages using the
\texttt{mtypes} used in the previous steps. In the face of any errors,
the xApp developer has a couple of strategies for debugging the \ac{RMR}
communication:

\subsubsection{Inspecting RMR Logs}
The \ac{RMR} library also displays debug information on the standard output of
the xApp pods, which include exchanged messages and error messages that
indicate potential issues. For example, the error message \1
"\texttt{Name does not resolve}" is displayed when the \ac{RMR} destination
endpoint cannot be resolved, which occurs when the destination xApp is not
running,  has yet to be registered with the \texttt{AppMgr}, or has crashed; and
\2 "\texttt{No route table entry for mtype=<given_mtype>}" is displayed when
the \ac{RMR} library cannot find entry record to route the given \texttt{mtype},
which occurs when the sent message used a \texttt{mtype} that has not been
registered on the static \ac{RMR} route table file, or when the table itself
cannot be found.

To obtain more information from the \ac{RMR} logs, e.g., the current \ac{RMR}
data port, the location where the RMR library expects to find a static route
table file, and the name of the \ac{RMR} endpoint, the xApp developer can
either: \1 set the \texttt{RMR_LOG_LEVEL} environment variable to 4 on the
xApp's Dockerfile and re-deploy the xApp, which makes the \ac{RMR}
library log additional debug information~\cite{rmrguide}; or \2 use
\texttt{kubectl} commands to open a shell to the xApp pod
(detailed in the next subsection) and parse its
environment variables starting with the "\texttt{RMR_}" prefix.

\subsubsection{Querying the \texttt{RtMgr}}

Another approach to debugging the \ac{RMR}
communication between xApps and \ac{nearrtric} components is to query the
\texttt{RtMgr} through REST to obtain the current \ac{RMR} routes
distributed and used inside the \ac{nearrtric}, which is useful to debug the
entire flow of messages from different sources.
To interact with the \texttt{RtMgr} via its REST interface, the xApp
developer must first find its IP address and open port (discussed in
Section~\ref{sub:debug_rest}), as shown in Listing~\ref{lst:rtmgr_rest}. In
possession of this information, the xApp developer can
issue a GET request to the \texttt{RtMgr}'s URI endpoint
\texttt{/ric/v1/getdebuginfo}, for obtaining the current
\ac{RMR} routes distributed and used in the \ac{nearrtric}
(expressed as a JSON string).
For more information about the \texttt{RtMgr}'s REST interface and other
supported REST methods, we refer the reader to its documentation~\cite{rtmgr}.

\subsection{Debugging Persistent Storage}\label{sub:debug_sdl}

The \ac{nearrtric} provides persistent storage for the xApp pods, which they can
leverage to store datasets, transfer data between each other, and save their
internal states between reboots and upgrades. However, issues with the database
infrastructure or data manipulation can prevent the xApps from performing
read/write operations, as detailed below.

\subsubsection{Optional Features}
The InfluxDB component and the persistent volume are optional
features of a \ac{nearrtric} cluster that can easily be overlooked
during the \ac{nearrtric} installation. However, the lack of an InfluxDB component to
store relational data prevents xApps from using the \ac{SDL} API.
Moreover, the lack of a persistent volume prevents the installation of the InfluxDB
during the \ac{nearrtric} cluster installation. While InfluxDB and
persistent volume are likely available in most production environments, the
xApp developer may accidentally set up their \ac{O-RAN} development
environment with a \ac{nearrtric} without one or both.

\begin{lstlisting}[linewidth=\columnwidth,language=bash,float,
caption={Commands for debugging the storage class and persistent storage
configuration on the \ac{nearrtric} cluster.},
label={lst:storage}
]
# List the configured storage classes
kubectl get storageclass -A

# List the configured persistent volumes
kubectl get pv -A
\end{lstlisting}
\begin{lstlisting}[linewidth=\columnwidth,language=bash,float,
caption={Command for openning a shell to a Kubernetes pod and example on how
to use it to acces the \texttt{DBaaS} pod.},
label={lst:shell}
]
# Open shell to a given pod in the cluster
kubectl exec -it <pod_name> \
-n <namespace> -- /bin/bash

# Example to open a shell to the DBaaS pod
kubectl exec -it \
statefulset-ricplt-dbaas-server-0 \
-n ricplt -- /bin/bash
\end{lstlisting}

To verify whether the InfluxDB component is installed and running, the xApp
developer can use the commands in Listing~\ref{lst:kube} to list running pods on
the \texttt{ricplt} namespace and confirm the existence of a pod with
"\texttt{influxdb}" on its name. To verify
whether the persistent volume is enabled, the xApp developer can use the
commands on Listing~\ref{lst:storage}, which should return an "\texttt{nfs}"
storage class and a persistent volume claim with the "\texttt{influxdb}" name.
If the InfluxDB is not installed and/or the persistent volume is not present,
the xApp developer needs to re-install their \ac{nearrtric}, making sure they
perform the additional, one-time setup for the persistent volume and include the
InfluxDB in the list of \ac{nearrtric} components to be
installed~\cite{ricinstallationguide}.

\subsubsection{Inspecting the Data}
To debug the data storage during runtime and verify whether
xApps are manipulating data as expected, the xApp developer can directly access
the SDL database abstraction layer and inspect the data stored therein. To do
so, the xApp developer must open a shell to the \ac{nearrtric}'s \texttt{DBaaS}
pod, using the commands in Listing~\ref{lst:shell}. Once the xApp developer has
shell access to the \texttt{DBaaS} pod, they can use the \texttt{sdlcli} command
to obtain statistics about the database backend, check the health of the
database backend, list database keys, and get and set values into the
database. For more information about the \texttt{sdlcli}  command and
functionality available on the \texttt{DBaaS} abstraction layer, we refer the
reader to their official documentation on~\cite{dbaas}.

\section{Good Practices: Lessons Learned}\label{sec:good}


In this section, we share some good practices related to the initialization and
registration of xApps, as well as their teardown and gracefully exiting, to
facilitate their development and ensure correct operation in the
\ac{nearrtric}.

\subsection{Initialization and Registration}

A critical step in the xApp lifecycle is the xApp registration 
for notifying the
\texttt{AppMgr} and other \ac{nearrtric} components about the existence of a new
xApp and the endpoints to communicate with it. This process requires the xApp to
locate and load its configuration file to obtain essential information for the
registration, i.e., its name, namespace, and interfaces. If the xApp cannot
locate the configuration file or if there is any missing information, it prevents
the registration with the \texttt{AppMgr} and results in errors on the xApp's
standard output, e.g., "\texttt{Cannot Read config file for xapp Registration}". However,
a registration failure leaves the xApp in an undefined state where the xApp
pod remains running, but it cannot work correctly or interact with
\ac{nearrtric} components.

\begin{lstlisting}[linewidth=\columnwidth,language=python,float,floatplacement=tbp,
caption={Example of a \texttt{post_init} function, where we wait a few seconds
and check whether the Python xApp framework was able to successfully load the
configuration file.},
label={lst:rmrxapp_register}
]
# Called after the RMRXapp constructor
def _post_init(self, rmr_xapp):
  # Set the log level of the xApp
  rmr_xapp.logger.set_level(Level.DEBUG)

  # Wait while the xApp is registered
  sleep(5)

  # Check for empty dict or False flag
  if not bool(self._config_data) or not self._keep_registration:
    # Log config file path
    rmr_xapp.logger.error(
      "Could not load config file" + str(
        self._config_path))
    # Stop the xApp
    rmr_xapp.stop()
\end{lstlisting}

To prevent an xApp from reaching this undefined state, the xApp developer has a
few options at their disposal. For example, the \texttt{RMRXapp} implementation
contains the \texttt{self._config_data} class parameter, which stores the
configuration file data (this can be replicated in the \texttt{Xapp}
implementation, as shown earlier in Listing~\ref{lst:xApp_entrypoint}).
The xApp developer can check whether this parameter is
an empty dictionary, which indicates that the Python xApp Framework could not
find or load the configuration file. Moreover, the \texttt{RMRXapp}
implementation contains the \texttt{self._keep_registration} boolean flag,
which is automatically set to \texttt{False} if any issues prevented
the registration process from starting.
Furthermore, the xApp developer can also log the content of the
\texttt{self._config_path} class parameter, which contains the configuration
file path the Python xApp framework tries to open.
After inspecting these variables, the xApp developer may decide to
debug and/or stop the xApp and avoid unintended behavior, as shown in
Listing~\ref{lst:rmrxapp_register}.

\subsection{Signals and Teardown}

The xApp pod can be terminated by \1 the users of the \ac{nearrtric} through
the \texttt{dms_cli} to uninstall a given xApp; or \2 Kubernetes itself due to
the lack of resources or when the cluster reboots (in this case, the pods are
automatically scheduled to restart after the cluster boots up). As part of this
process, Kubernetes issues a \texttt{SIGTERM} signal to inform the pods of their
impending termination so they can cease operations and gracefully exit. However,
the xApp must listen to the \texttt{SIGTERM} signal to react to it and then
manually perform procedures to exit gracefully, e.g., saving their internal
state on persistent storage and, importantly, triggering the unsubscription and de-registration
processes with the \texttt{SubMgr} and \texttt{AppMgr}, respectively.
If the pod is still running after the grace period, Kubernetes issues a
\texttt{SIGKILL} signal to terminate the pod forcefully.
As mentioned in Section~\ref{sub:rea_gen}, failure to de-register the xApp with
the \texttt{AppMgr} leaves unresolved references and communication endpoints on
the \ac{nearrtric} components, leading to undefined behavior on the system and
preventing the xApp from being installed again 
until the entire \ac{nearrtric} cluster reboots, which can disrupt
service and affect a large number of users.

Both the \texttt{RMRXapp} and \texttt{Xapp} implementations provide the
\texttt{stop()} method for abstracting the xApp de-registration, as well as
stopping its \ac{RMR} message receiving loop and any other running threads.
However, the xApp developer must ensure their xApp can catch these signals and
react accordingly to call \texttt{stop()} method, which can be accomplished
using signal handlers, as shown in Listing~\ref{lst:good_init} for the
\texttt{Xapp}. We can leverage the Python Signal module to register callbacks that
will be invoked whenever the xApp receives the respective type of signal.
In addition to the \texttt{SIGTERM} issued by Kubernetes,
we also registered a callback for the \texttt{SIGINT}, 
which can be useful for xApp developers using an open shell to their xApps for
testing and debugging in real-time. 
After these additional preliminary steps, the xApp
developer can proceed with the initialization of their xApp as usual.
The last step to catch and react to signals is to create callbacks for handling
signals, as shown in Listing~\ref{lst:xapp_signal_handler}. These methods
receive the signal type (in the form of an integer) and the stack frame (which
helps to identify which thread was interrupted) as arguments, and can be used to
toggle flags to shutdown control loops and call the stop method to exit the xApp
gracefully.  


\begin{lstlisting}[linewidth=\columnwidth,language=python,float,
caption={Example of the \texttt{Xapp} constructor that registers signal
handler callbacks for reacting to different signals.},
label={lst:good_init}
]
# Register callbacks and initialize xApp
def __init__(self):
  # Catch and react to the SIGTERM
  signal.signal(signal.SIGTERM,
    self.signal_handler)
  # Catch and react to the SIGINT
  signal.signal(signal.SIGINT,
    self.signal_handler)

  # Either instantiate the Xapp class
  self._xapp = Xapp(
    self._entrypoint,
    rmr_port=4560
  )
\end{lstlisting}

As an xApp ceases to operate and gracefully exits, it is also beneficial to
delete all active subscriptions to E2 Nodes. The benefits of this preemptive
action are twofold: \1 it
releases resources from the \texttt{SubMgr} and \texttt{RtMgr}, saving time
spent resolving routing decisions, and \2 it does not introduce unnecessary
latency on the near-RT control loops, as lingering \texttt{CONTROL} and
\texttt{INSERT} subscriptions cause \ac{RAN} procedures to wait for a
timeout until these subscriptions are deleted, even if the corresponding xApps
were uninstalled.

\begin{lstlisting}[linewidth=\columnwidth,language=python,float,
caption={Example of a signal handler callback for catching and reacting to Linux
signals and gracefully exitting xApps.},
label={lst:xapp_signal_handler}
]
# Callback that catches registered signals
def signal_handler(self, signal, frame):
  self._xapp.logger.info("signal handler called")

  # Let's first stop the entrypoint loop
  self.shutdown = True

  # Next, let's stop and de-register xApp
  self._xapp.stop()
\end{lstlisting}

At the time of writing, these additional considerations and procedures are not
covered in the documentation provided by the \ac{OSC}, despite being essential
for ensuring the correct operation of both the xApps and the \ac{nearrtric}.
Instead, we observe that existing xApp developers learn these lessons
through trial and error throughout their xApp development cycles. Therefore, these lessons serve as invaluable information and good practices to facilitate 
newcomers in starting to prototype their xApps and accelerate their development.

\section{Outlook and Open Challenges}\label{sec:outl}

In this section, we discuss the current landscape of xApp development using the
\ac{OSC} O-RAN flavor and resources, new feature capabilities and
standardization efforts, as well as open challenges for evaluating xApps in
end-to-end scenarios.

\subsection{Embedding of AI/ML to Manage \acp{RAN}}


One of the main aspects that attracted attention to \ac{O-RAN} is its
standardized platforms for \1 deploying custom control logic via third-party
applications and \2 embedding intelligence in mobile networks through
AI/ML-based control loops.
The \ac{OSC} provided a reference implementation of the \ac{nearrtric} with
support for third-party applications since its first A Release in 2019, which
led to the community developing several xApps for 
accomplishing different tasks over the years, as discussed in
Section~\ref{sub:table}. However, the \ac{OSC}'s support for AI/ML has moved at
a much slower pace.
As a result, many researchers attempted to expand on the \ac{nearrtric}
with AI/ML capabilities using a number of homebrew patches and custom
implementations~\cite{overbeck2024data, mallu2023ai,
tsampazi2023comparative}.
The \ac{OSC} only started supporting
AI/ML capabilities as part of its H Release in late 2022, with the initial
inclusion of the \ac{AIMLFW}~\cite{grelease}. This optional entity complements
the Non- and Near-RT \acp{RIC} and provides them with a complete pipeline for
data preparation, AI/ML model training, AI/ML inference, and model management.
Consequently, we expect future xApps leveraging AI/ML inference
to rely on the \ac{AIMLFW} instead of developing their own homebrewed
solutions. 
Due to a considerable change in focus,
we will detail how to leverage the \ac{AIMLFW} for performing
standard-compliant, AI/ML-based control loops to manage the \ac{RAN} in a
follow-up tutorial.

\subsection{Features Across xApp Frameworks}

The \ac{OSC} offers different xApp Frameworks in distinct programming
languages to assist developers in creating their xApps, as discussed in
Section~\ref{subsub:sdk}.
However, not all xApp Frameworks receive equal attention, maintenance,
and updates from the community due to their open-source nature and volunteer
contributions to the \ac{OSC}.
Consequently, certain xApp Frameworks may contain features
not present in others, lack support for new capabilities, or suffer from
breaking API changes that render some of their features unusable.
For example, the Python xApp Framework lacks the Alarm API present in other
xApp Frameworks~\cite{alarm}, while the \ac{STSL} API is currently only
available in the Go xApp Framework. Furthermore, since the F Release,
the \ac{OSC} transitioned its xApp subscription API from \ac{RMR}
to REST, a change that has yet to be ported to the Python xApp
Framework, as discussed in Section~\ref{sec:ran}.
Such inconsistency between features across xApp Frameworks poses significant
challenges for xApp developers, who may need to adapt their workflows
and adopt an xApp Framework based on its available features rather than their
preferred programming language. This challenge also poses barriers to new
xApp developers, who may need to learn a new programming language to avail from
particular capabilities.
We strongly believe that new features and breaking API changes should be
addressed consistently throughout all xApp Frameworks before rolling out new
\ac{OSC} releases.


\subsection{New xApps and RMR Message Types}

The xApps and \ac{nearrtric} components communicate with one another using
different \ac{RMR} mtypes, as discussed earlier in Section~\ref{sub:mes}.
Through the development of xApps, their distinct business logic, and
different control loops, new xApps will likely require new \texttt{mtypes} to
establish communication protocols between one another while avoiding message
routing conflicts with existing xApps and their \texttt{mtypes}.
However, the supported \texttt{mtypes} in the \ac{nearrtric} are hardcoded in
the \ac{RMR} source code as a lookup table of known
\texttt{mtypes}~\cite{message_types}. Consequently, we identified that including a
new \texttt{mtype} requires modifying the \ac{RMR} source code and updating (or
recompiling) the \ac{nearrtric} components and the xApps therein using the
patched \ac{RMR} library to recognize and avail from the new \texttt{mtypes}.
The \ac{OSC} performs these exact steps whenever a new  xApp becomes a
first-party, supported application on a new release, as can be  seen by the
inclusion of \texttt{mtypes} 30001, 30002, 30003, and 30010 to support the
\textit{Anomaly Detection}, \textit{QoE Predictor}, \textit{Traffic Steering},
and \textit{Measurement Campaign} xApps, respectively.
However, this process imposes significant challenges for third-party xApps, as
their developers may not have access or permission to modify the \ac{RMR}
source code used by components of a \ac{nearrtric} cluster,
or their installation may incur significant overheads to the
\ac{nearrtric} users, which can include system administrators and network operators,
but may not necessarily possess the required development skills.
We believe that the \ac{OSC} should provide an API for xApps to dynamically
register new \texttt{mtypes} as part of their deployment process, which would
considerably facilitate the deployment of new third-party xApps and lower the
barrier to entry for new developers.

\subsection{Security of the \ac{O-RAN} Ecosystem, \ac{nearrtric}, and xApps}


While current \ac{O-RAN} components are operational and capable of working in
unison for managing \acp{RAN}, several critical security considerations remain
to be addressed. For example, there is a lack of safeguards against misbehaving
or malicious xApps, which can cause conflicts and degrade network
performance~\cite{wadud2023conflict,zolghadr2024learningreconstructingconflictsoran,
del2024pacifista}, and protections against resource depletion 
and denial of service attacks, which can disrupt \ac{RAN} control
loops~\cite{radhakrishnan2023detection}. More fundamentally, there is an urgent
need for \ac{AAA} capabilities for verifying xApp identities, controlling
shared resource access, and tracing their activities~\cite{aaa}.
Addressing these  issues is essential to ensure network integrity in realistic
settings, prevent unauthorized actions, and maintain accountability, especially
in an open, multi-vendor, and multi-party environment~\cite{Haohuang-22}.
There is a strong regulatory and industry interest in addressing these
challenges, and future \ac{O-RAN} releases are expected to introduce
new security mechanisms, e.g., encrypted communication protocols, certificates,
and key management systems. These updates will likely affect the existing
xApp deployment process and development cycle, as well as their associated
xApp interfaces and APIs for conforming to the more strict security standards.

\subsection{End-to-end Testing and Validation}


For xApp developers looking into performing end-to-end testing and validation of
their xApps to manage \acp{RAN}, there are a few options to interact with E2
Nodes in different environments. Depending on their needs and expertise, xApps
developers can avail from \1 actual software radio stacks, e.g.,
srsRAN~\cite{gomez2016srslte} or OAI~\cite{kaltenberger2020openairinterface}
and their recently introduced E2 interfaces to create
softwarized \ac{O-RAN}-compatible base stations 
running either with an emulated air
interface or over the air; \2 emulated E2 Nodes, e.g., the VIAVI's RIC
Test~\cite{rictest} product or the \texttt{ns-3} \ac{O-RAN} E2
module~\cite{Lacava-23}, for emulating an arbitrary number of base stations
on a parameterizable radio environment; and \3 simulating
E2 Nodes, e.g., the \texttt{E2Sim} mentioned earlier in Section~\ref{sec:ran}, a
simple simulator for testing the communication between the \ac{nearrtric} with a
mock E2 Node, enabling the test and validation of xApps, \acp{SM}, and
their interactions with the \texttt{E2Term} and \texttt{E2Mgr}. On the one hand,
the more comprehensive software radio stacks and emulated E2 Nodes come from
third-party software suppliers, requiring some learning curve and integration
effort with the \ac{nearrtric}, and may have licensing considerations. On the
other hand, the \texttt{E2Sim} is provided by the \ac{OSC} and should work out
of the box as part of the platform.

While useful for small-scale validations and sanity checks, the \texttt{E2Sim}
possesses significant limitations that restrict the types and scope
of end-to-end testing and validation, namely: \1 the \texttt{E2Sim} can only
simulate a single base station at a time, limiting the scale of experiments and
types of xApps would benefit from the \texttt{E2Sim}, e.g., handover operations;
and \2 the \texttt{E2Sim} only supports the \ac{RAN} function ID 200, which
serves to report \acp{KPM} back to the \ac{nearrtric} and hence
makes the \texttt{\ac{KPM}} the only supported \ac{SM}.
These limitations led to research efforts developing their own
homebrewed extensions to the \texttt{E2Sim} to support other types of messages
and \acp{SM} and expand its use cases~\cite{Lacava-23}. However, these
works tend to cater to particular scenarios and \ac{O-RAN} releases, often missing
documentation or not releasing their source code. We believe it
would be beneficial to the community if the \ac{OSC} provided extensive
documentation for the utilization of the \texttt{E2Sim} to perform end-to-end
testing and validation, as well as extended it with other \ac{RAN} Functions and
\acp{SM}, e.g., the \texttt{\ac{RC}} and \texttt{\ac{CCC}}.

\section{Conclusions}\label{sec:conc}

In this tutorial, we provided the first comprehensive guide on the development
of xApps for managing \acp{RAN}, from theory to practice. This paper addresses
a significant gap in the literature and provides extensive material
for the community to expedite and accelerate the development of xApps
by academia and industry alike. First, we presented a theoretical foundation
about the \ac{O-RAN} ecosystem and its entities, as well as the \ac{nearrtric}
components and its enabling technologies. Then, we introduced the APIs available
to xApps, described how to design them through xApp descriptors, overviewed
their lifecycle, and demonstrated how to control their deployment. Next, we
addressed the functionality available to xApps and explored how to communicate
via \ac{RMR}, leverage persistent storage via \ac{SDL}, and react to external
input via REST. In addition, we detailed how xApps can interface with E2 Nodes
and their \ac{RAN} functions via \acp{SM} and subscriptions. Moreover, we
discussed debugging strategies to verify and validate the operation of xApps, as
well as good practices to ensure their correct functioning. Finally, we
discussed the current landscape of xApp developments, accompanied by new
features, open challenges for xApp development,  and suggestions for future
improvements.
It is worth mentioning that the supporting material used throughout the
tutorial, i.e., the xApp descriptor and schema files, and source codes, can be
found in our public online repository~\cite{repo}.



\DeclareAcroListStyle{extra-tabular}{extra-table}{
  table = supertabular,
  table-spec = p{.22\columnwidth}p{.7\columnwidth}ll
}
\section*{Acronyms}
\setlength{\tabcolsep}{4pt}
\acsetup{list-style=extra-tabular, list-heading=none}
\printacronyms[exclude-classes={test}]

\nottoggle{comsoc}{
\section*{Acknowledgements}

The research leading to this paper received support from the Commonwealth
Cyber Initiative, an investment in the advancement of cyber R\&D, innovation,
and workforce development. For more information, visit:
\url{www.cyberinitiative.org}.
This work was also supported by the Program OpenRAN@Brasil.
}{}


\bibliographystyle{IEEEtran}
\bibliography{IEEEabrv,bibliography}


\end{document}